\documentclass[11pt]{article}

\usepackage{amsmath}
\usepackage{graphicx}
\usepackage{enumerate}
\usepackage{natbib,bibunits}
\usepackage{url}
\usepackage{enumitem}
\usepackage{lscape}
\usepackage{amssymb,amsthm}
\usepackage{bm}
\usepackage{bbm}
\usepackage{isomath}
\usepackage{cases}
\usepackage{color}
\usepackage{comment}
\usepackage{todonotes}
\usepackage{hyperref}
\usepackage{caption}
\usepackage{subcaption}
\usepackage{adjustbox}
\usepackage[normalem]{ulem}
\usepackage{cancel}
\usepackage[para,online,flushleft]{threeparttable}
\usepackage{tabularray}
\usepackage{booktabs}
\usepackage{multirow}
\usepackage{float}
\usepackage[page,header]{appendix}
\usepackage{titletoc}
\usepackage{etoolbox}
\usepackage{pifont}
\newcommand{\cmark}{\ding{51}}
\newcommand{\xmark}{\ding{55}}
\usepackage{tcolorbox}
\usepackage{tocloft}
\usepackage{etoc}
\usepackage{setspace}

\usepackage{soul}

\usepackage{newtxtext}
\usepackage[subscriptcorrection]{newtxmath}

\usepackage[plain,noend]{algorithm2e}

\makeatletter
\renewcommand{\algocf@captiontext}[2]{#1\algocf@typo. \AlCapFnt{}#2}

\def\@algocf@capt@plain{top}
\renewcommand{\algocf@makecaption}[2]{%
  \addtolength{\hsize}{\algomargin}%
  \sbox\@tempboxa{\algocf@captiontext{#1}{#2}}%
  \ifdim\wd\@tempboxa >\hsize%
    \hskip .5\algomargin%
    \parbox[t]{\hsize}{\algocf@captiontext{#1}{#2}}%
  \else%
    \global\@minipagefalse%
    \hbox to\hsize{\box\@tempboxa}%
  \fi%
  \addtolength{\hsize}{-\algomargin}%
}
\makeatother

\graphicspath{{./art/}}

\newtheorem{definition}{Definition}
\newtheorem{theorem}{Theorem}
\newtheorem{assumption}{Assumption}
\newtheorem{corollary}{Corollary}
\newtheorem{proposition}{Proposition}
\newtheorem{lemma}{Lemma}
\newtheorem{example}{Example}
\newtheorem{remark}{Remark}

\newcommand{\dd}{\mathrm{d}}
\newcommand{\cH}{\bm{\mathcal{H}}}
\newcommand{\vvartheta}{\boldsymbol{\vartheta}}

\newcommand{\vvarepsilon}{\boldsymbol{\varepsilon}}

\newcommand{\vtheta}{\boldsymbol{\theta}}

\newcommand{\vxi}{\boldsymbol{\xi}}

\newcommand{\vupsilon}{\boldsymbol{\upsilon}}

\newcommand{\vpsi}{\boldsymbol{\psi}}
\newcommand{\vomega}{\boldsymbol{\omega}}

\newcommand{\va}{\text{\textbf{\textit{a}}}}

\newcommand{\vd}{\text{\textbf{\textit{d}}}}

\newcommand{\vf}{\text{\textbf{\textit{f}}}}

\newcommand{\vu}{\text{\textbf{\textit{u}}}}

\newcommand{\vx}{\text{\textbf{\textit{x}}}}
\newcommand{\vy}{\text{\textbf{\textit{y}}}}

\newcommand{\mA}{\mathbf{A}}
\newcommand{\mB}{\mathbf{B}}
\newcommand{\mC}{\mathbf{C}}
\newcommand{\mD}{\mathbf{D}}

\newcommand{\mH}{\mathbf{H}}
\newcommand{\mI}{\mathbf{I}}

\newcommand{\mP}{\mathbf{P}}
\newcommand{\mQ}{\mathbf{Q}}

\newcommand{\mV}{\mathbf{V}}
\newcommand{\mW}{\mathbf{W}}

\newcommand{\mZ}{\mathbf{Z}}

\newcommand{\mZeros}{\mathbf{O}}
\newcommand{\vzeros}{\text{\textbf{\textit{0}}}}

\newcommand{\calF}{\mathcal{F}}

\newcommand{\SR}{\mathbb{R}}
\newcommand{\SN}{\mathbb{N}}

\newcommand{\rN}{\mathrm{N}}

\DeclareMathOperator{\tr}{tr}
\DeclareMathOperator{\cov}{cov}
\DeclareMathOperator{\rd}{d}

\makeatletter
\newcommand{\distas}[1]{\mathbin{\overset{#1}{\kern\z@\sim}}}
\newsavebox{\mybox}\newsavebox{\mysim}
\newcommand{\distras}[1]{%
  \savebox{\mybox}{\hbox{\kern3pt$\scriptstyle#1$\kern3pt}}%
  \savebox{\mysim}{\hbox{$\sim$}}%
  \mathbin{\overset{#1}{\kern\z@\resizebox{\wd\mybox}{\ht\mysim}{$\sim$}}}%
}
\makeatother

\newcommand{\blind}{1}

\begin{document}

\def\spacingset#1{\renewcommand{\baselinestretch}{#1}\small\normalsize}
\spacingset{1.4}

\title{\bf Gradient-based filtering under misspecification: \\
Stability and error bounds}

\if1\blind
  \author{
    \textsc{Simon Donker van Heel}\textsuperscript{a,c,*},\;
    \textsc{Rutger-Jan Lange}\textsuperscript{a,c},\; \\
    \textsc{Bram van Os}\textsuperscript{b,c},\; \text{and}\;
    \textsc{Dick van Dijk}\textsuperscript{a,c}
  }
  \date{\today}
  \maketitle
  \begin{center}
    \small
    \textsuperscript{a}\textit{Econometric Institute, Erasmus University Rotterdam, The Netherlands} \\
    \textsuperscript{b}\textit{Econometrics and Data Science Department, Vrije Universiteit Amsterdam, The Netherlands} \\
    \textsuperscript{c}\textit{Tinbergen Institute, The Netherlands}
  \end{center}
  \renewcommand{\thefootnote}{\fnsymbol{footnote}}
  \footnotetext[1]{Corresponding author. \\
  E-mail addresses: \texttt{donkervanheel@ese.eur.nl} (S.W.\ Donker van Heel),
  \texttt{lange@ese.eur.nl} (R.-J.\ Lange),
  \texttt{b.van.os2@vu.nl} (B.\ van Os),
  \texttt{djvandijk@ese.eur.nl} (D.\ van Dijk).}
  \renewcommand{\thefootnote}{\arabic{footnote}}
\else
  \author{}
  \date{}
  \maketitle
\fi

\begin{abstract}
\noindent Can stochastic gradient methods track a moving target? We study the problem of tracking multidimensional time-varying parameters under noisy observations and possible model misspecification. Gradient-based filters update the time-varying parameters using the gradient of a postulated objective function. A natural filtering objective is the logarithm of the postulated observation density, which gives rise to the widely used class of score-driven filters. As in the optimization literature, these filters come in two forms: explicit filters evaluate the gradient at the predicted parameter, whereas implicit filters evaluate it at the updated parameter. For both filter types, we derive novel sufficient conditions for exponential stability of the filtered parameter path, showing that stability can be guaranteed independently of the data-generating process. Under mild additional moment conditions on the data-generating process, we also obtain finite-sample and asymptotic mean squared error bounds relative to the pseudo-true parameter path. For implicit filters, these guarantees hold under weak parameter restrictions. For explicit filters, they additionally require Lipschitz continuity of the score and a sufficiently small learning rate. Simulation studies support our theoretical findings and show that implicit gradient filters outperform explicit ones in both accuracy and stability.
\end{abstract}

\medskip
\noindent\textbf{Keywords:} Error bound; Explicit and implicit gradient method; Filtering; Pseudo-true parameter; Score-driven model; Time-varying parameter.

\bigskip


\begin{bibunit}[chicago]

\section{Introduction}
\label{sec:intro}

In a wide range of disciplines, from economics and finance to engineering and climate science, variables exhibit characteristics that change over time. Such fluctuations can be represented by unobserved states or parameters and tracked by filtering techniques that alternate between prediction and update steps. These steps yield state or parameter estimates from lagged and contemporaneous information, respectively, as in \citeauthor{kalman1960new}'s (\citeyear{kalman1960new}) classic approach and its successors in the state-space literature (e.g., \citealp{durbin2012time}).

In this paper, we study gradient-based filters for tracking vectors of unobserved parameters. These filters update the parameter estimates using the gradient of a stochastic objective function. Using the postulated observation log density as the objective yields the widely used class of score-driven (SD) filters \citep{artemova2022score, harvey2022score}. As in the optimization literature, they come in two variants: explicit methods evaluate the gradient at the predicted parameter, as in classical stochastic approximation (e.g., \citealp{robbins1951stochastic}), whereas implicit methods evaluate the gradient at the updated parameter, as in implicit gradient methods (e.g., \citealp{rockafellar1976monotone}) and recent work on proximal methods \citep{toulis2017asymptotic, toulis2021proximal}.

Our theoretical contribution is twofold. First, we establish new sufficient conditions for the exponential stability of (multivariate) implicit and explicit SD filters (Theorem~\ref{thrm:invertibility}). Exponential stability means that any two filtered parameter paths, started from different initial values but driven by the same data, converge toward each other exponentially fast. Because our conditions depend only on the postulated density, they are verifiable in practice, readily interpretable, and applicable under arbitrary misspecification, holding uniformly over all data-generating processes (DGPs). Second, for both filter types, we combine stability with mild conditions on the DGP to obtain finite-sample and asymptotic upper bounds on the mean squared errors (MSEs) of the filtered and predicted paths relative to the pseudo-true parameter path (Theorem~\ref{Th: (Non-)asymptotic MSE bounds}).

Our approach differs in several respects from the literature on SD models, also known as dynamic conditional score \citep[DCS;][]{harvey2013dynamic} and generalized autoregressive score \citep[GAS;][]{creal2013generalized} models. First, we allow the researcher’s SD filter to be misspecified, possibly severely so, rather than assuming that the true time-varying parameter follows SD dynamics. Second, while recent work allowing for misspecification focuses on  guarantees at a \emph{single} time step (e.g., \citealp{creal2024moment,gorgi2023optimality}), to our knowledge ours is the first paper to give performance guarantees for the \emph{entire} filtered or predicted parameter path. Third, while the literature has focused on explicit SD filters, we also consider the implicit SD filter recently introduced by \citet{lange2024robust} and show that it offers stronger stability and performance guarantees than the standard explicit version. For a more detailed comparison with the literature, see Section~\ref{sec:lit}.

Our first result (Theorem~\ref{thrm:invertibility}) underpins large-sample theory for the maximum-likelihood estimator (MLE; \citealp{straumann2006quasi}) of the filter's static (hyper-)parameters, including consistency and asymptotic normality. Although explicit SD filters are commonly estimated by maximum likelihood, existing stability results are often limited to the scalar case (e.g., \citealp{harvey2018modeling, blasques2022maximum}; \citealp{d2026empirical}), while in the multivariate setting available analytical conditions tend to be overly restrictive (see \citealp[sec.\ 6.4]{potscher1997dynamic}; \citealp[p.\ 5]{artemova2023order}). For implicit SD filters, the stability result of \citet{lange2024robust} relies on the assumed concavity of the postulated log observation density. By contrast, Theorem~\ref{thrm:invertibility} provides two new verifiable sufficient conditions for the stability of multivariate SD filters: one each for the implicit and explicit versions. These conditions are formulated without imposing concavity and guarantee stability of the filtered path for every data sequence, meaning that we can remain entirely agnostic about the DGP. Under correct specification, when both the filter and the DGP are score driven with the same static parameters, exponential stability implies asymptotic recovery of the true time-varying parameter. Under misspecification, exact recovery need not hold, which motivates our second contribution.

Our second result (Theorem~\ref{Th: (Non-)asymptotic MSE bounds}) provides formal guarantees for the tracking accuracy of SD filters under misspecification. The behavior of misspecified stochastic gradient methods is an important concern in statistics (e.g., \citealp{liang2019statistical}), machine learning (e.g., \citealp{cutler2023stochastic}), and time-series analysis (e.g., \citealp{brownlees2024empirical}; \citealp{lange2024bellman}; \citealp{beutner2023consistency}), yet it remains relatively underexplored. For example, \citet{koopman2016predicting} study misspecified explicit SD filters by simulation, but do not provide formal guarantees. Under misspecification, exact recovery 
is no longer possible, so the natural target becomes the \emph{pseudo}-true state. This only requires mild assumptions on the DGP, such as the existence of a pseudo-true state with finite-variance increments. For both implicit and explicit filters, Theorem~\ref{Th: (Non-)asymptotic MSE bounds} establishes finite-sample and asymptotic guarantees by bounding the MSE of the entire filtered and predicted parameter paths relative to the pseudo-true path. The asymptotic bounds can often be minimized analytically with respect to tuning parameters such as the learning rate; in a special case related to the Kalman filter, the resulting minimized bound is tight.

A central insight from both Theorems~\ref{thrm:invertibility} and~\ref{Th: (Non-)asymptotic MSE bounds} is that explicit SD filters require substantially stronger assumptions than implicit SD filters. In particular, theoretical guarantees for misspecified explicit SD filters require the score to be Lipschitz continuous in the parameter of interest. Under twice differentiability, this requires the Hessian of the postulated log density to be bounded. Although this Lipschitz assumption is standard in the optimization literature (e.g., \citealp{nesterov2018lectures}), it is often overlooked in the SD literature and rarely mentioned in the ${\sim}450$ papers listed on \href{https://www.gasmodel.com/}{gasmodel.com}. Yet it is frequently violated in practice, even in relatively simple settings such as a Poisson model with time-varying log-intensity. Our theory identifies Lipschitz continuity as a sufficient condition for stability of misspecified explicit filters, while simulation results suggest that it may even be close to necessary in practice. When it fails and the true state is sufficiently volatile, explicit filters may become unstable or even diverge, whereas implicit filters remain accurate. This mirrors well-known results for stochastic gradient methods, where implicit procedures enjoy stronger stability under weaker assumptions \citep{toulis2017asymptotic, toulis2021proximal}.

In three Monte Carlo experiments, we assess SD filters in terms of stability, theoretical MSE bounds, and empirical MSEs. First, we consider a linear setting with high-dimensional states, as in \cite{cutler2023stochastic}, where the Lipschitz condition holds and performance guarantees are available for both the explicit and implicit versions. Compared with three recent alternatives, the implicit SD filter achieves the lowest theoretical and empirical MSEs across all time steps, state dimensions, and observation dimensions, and is the only filter whose performance improves as the Lipschitz constant increases. Notably, our MSE bounds are up to three orders of magnitude smaller than those in \cite{cutler2023stochastic}. Second, we examine nine nonlinear settings studied by \cite{koopman2016predicting}. Finite MSE bounds are available for all DGPs under the implicit SD filter, whereas for the explicit SD filter such guarantees hold only in the two cases where the score is Lipschitz continuous. In the remaining seven DGPs, when the true state is sufficiently volatile, the explicit SD filter becomes unstable and sometimes diverges. This finding adds nuance to \citeauthor{koopman2016predicting}'s (\citeyear{koopman2016predicting}) conclusion that misspecified explicit SD filters accurately track unobserved states, which appears to hold only when the score is Lipschitz continuous or the state volatility is low. Third, we analyze the widely used time-varying Poisson count model, for which the Lipschitz condition fails. All examined variants of the explicit filter diverge, whereas the implicit filter remains stable.

The remainder of the article is organized as follows.
Section~\ref{sec:lit} relates our contribution to the existing literature.
Section~\ref{sec:Implicit and explicit score-driven filters} introduces implicit and explicit SD filters.
Sections~\ref{sec: Invertibility} and~\ref{sec:Error bounds for score-driven filters} present our stability and performance guarantees.
Section~\ref{sec:Simulation studies} reports the simulation studies, and Section~\ref{sec:conc} concludes.
The online supplement provides proofs of the main results (Appendix~\ref{Appendix A. Proofs}), additional theoretical results (Appendix~\ref{Appendix B. Further theoretical results}), and further numerical results and discussion (Appendix~\ref{Appendix C. Further numerical results}).

\subsection{Contribution relative to existing work}
\label{sec:lit}

Our work relates to time-varying parameter models, observation-driven time-series models, and stochastic gradient methods for tracking moving targets. A common focus in this literature is the stability and accuracy of noisy updates, especially under misspecification. We discuss three closely related strands of work on observation-driven models: (i) stability and invertibility, (ii) pseudo-true targets and optimality, and (iii) recent extensions of the score-driven framework.

First, we connect to the literature on stability and invertibility of observation-driven models. The seminal contribution of \citet{straumann2006quasi} studies invertibility as a crucial ingredient for likelihood-based inference. Many later contributions on score-driven filters, such as \citet{blasques2022maximum}, \citet{blasques2024maximum}, and \cite{d2026empirical}, formulate invertibility conditions for \emph{scalar} time-varying parameters through contraction conditions whose expectations are taken under the \emph{true} stationary law of the data, which is generally unknown. By contrast, our main stability result is stated directly in terms of the \emph{filter} itself: the conditions are verifiable from the filter's parameters and the postulated density alone, apply in any number of dimensions, and remain meaningful under arbitrary misspecification. For explicit score-driven filters, our Lipschitz-type condition implies stability uniformly over any realized data sequence without requiring assumptions on the data-generating process. Standard affine filters such as GARCH arise as special cases, and existing multivariate models such as \citet{dinnocenzo2023robust} also fit within our framework. For implicit score-driven filters, the stability guarantees are stronger still, as they hold uniformly over all realized data sequences without requiring a Lipschitz condition.

Second, we contribute to the recent literature on misspecification and optimality in score-driven models. \citet{gorgi2023optimality} show that the expected score-driven update at a given time step moves the parameter toward a pseudo-true target defined by conditional Kullback--Leibler divergence, while \citet{creal2024moment} obtain improvement results for observation-driven updates based on moment conditions. These results concern a \emph{single update} at a fixed point in time. By contrast, we analyze the \emph{entire} filtered and predicted parameter paths and derive finite-sample and asymptotic error bounds relative to a time-varying pseudo-true path. In our discrete-time setting with fixed observation intervals, exact recovery of the pseudo-true path is generally impossible, so our error bounds do not vanish asymptotically, much as in the classic discrete-time Kalman filter, albeit here for a broader class of gradient-based filters. This contrasts with \citet{beutner2023consistency}, who obtain convergence towards the pseudo-true state under \emph{infill asymptotics}, where the number of observations per unit of time diverges, making exact recovery possible. To our knowledge, the discrete-time pathwise error bounds for both explicit and implicit score-driven filters developed here are new.

Third, we relate to recent work modifying the basic score-driven setting. \citet{blasques2023quasi} propose quasi score-driven models, which leave the explicit filtering recursion unchanged but replace the log likelihood used to estimate static parameters with a quasi log likelihood. Thus, from a filtering perspective, quasi score-driven filters fall within our explicit class. By contrast, \citet{lange2024robust} use the same log likelihood for filtering and estimation, but replace the explicit gradient with an implicit one. Relative to their work, we retain the log-likelihood objective but develop theory for both explicit and implicit filters in parallel. For the explicit class, our stability and pathwise error-bound results are new. For the implicit class, our pathwise error bounds are new, and we strengthen the stability result of \citet[Thm.~1]{lange2024robust} by allowing non-concave observation log densities and deriving sharper contraction rates.

\section{Implicit and explicit score-driven filters}
\label{sec:Implicit and explicit score-driven filters}

\subsection{Problem setting}

For $t=1, \ldots, T$, the $n \times 1$ observation $\vy_t$ is drawn from a true conditional observation density $p^{0}(\vy_t \mid \vvartheta_t, \vpsi^{0}, \calF_{t-1})$, where $\vvartheta_t$ is a time-varying parameter vector taking values in some parameter space $\mathbf{\Theta}^{0}$, $\vpsi^{0}$ is a vector of static shape parameters, and $\calF_{t-1}$ is the information set at time ${t-1}$. The dependence on $\calF_{t-1}$ allows for exogenous variables and lags of $\vy_t$. For brevity, we suppress the dependence on $\vpsi^{0}$ and $\calF_{t-1}$, and for discrete observations interpret $p^{0}(\vy_t \mid \vvartheta_t)$ as a probability rather than a density. The researcher-postulated density, typically misspecified, is denoted by $p(\vy_t \mid \vtheta_t)$, where $\vtheta_t \in \mathbf{\Theta}$ is a parameter vector in a non-empty convex parameter space $\mathbf{\Theta} \subseteq \SR^k$ (e.g., the positive quadrant). Additional dependence of $p(\cdot \mid \vtheta_t)$ on a static shape parameter $\vpsi$ and exogenous variables is allowed but suppressed for brevity.

\subsection{Filtering algorithm}

Score-driven filters are gradient-based methods whose filtering objective is based on the postulated log density of the observations \citep{lange2024robust}. Like \citeauthor{kalman1960new}'s (\citeyear{kalman1960new}) filter, SD filters can be viewed as  alternating between update and prediction steps. The updated and predicted states represent the researcher's estimates of the time-varying parameter based on the contemporaneous information set $\calF_t$ and the lagged information set $\calF_{t-1}$, respectively. For $t=1,\ldots,T$, the implicit SD (ISD) and explicit SD (ESD) updates are given by
\allowdisplaybreaks
\begin{align}
\text{ISD update: } \qquad \vtheta_{t \mid t}^\textnormal{im} & \;=\;\vtheta_{t \mid t-1}^\textnormal{im}\,+\,\mH_{t} \, \nabla\ell(\vy_t \mid \vtheta_{t \mid t}^\textnormal{im}),\label{Implicit parameter update step}\\
\text{ESD update: } \qquad \vtheta_{t \mid t}^\textnormal{ex} & \;=\;\vtheta_{t \mid t-1}^\textnormal{ex}\,+\,\mH_{t} \, \nabla\ell(\vy_t \mid \vtheta_{t \mid t-1}^\textnormal{ex}),\label{Explicit parameter update step}
\end{align}
where $\mH_t\in \SR^{k \times k}$ is an $\calF_{t-1}$-measurable positive-definite learning-rate matrix (i.e.,\ $\mH_t \succ \mZeros_k$, meaning $\mH_t$ is symmetric and $\text{\textbf{\textit{v}}}^{\top} \mH_t \text{\textbf{\textit{v}}} > 0$ for all $\text{\textbf{\textit{v}}} \neq \vzeros_k$), $\nabla:=\dd/\dd\vtheta$ is the gradient with respect to the second argument of $\ell(\vy \mid \vtheta)$, and $\ell(\vy \mid \vtheta):=\log p(\vy \mid \vtheta)$. Thus, $\nabla \ell(\vy \mid \vtheta)$ is the score. Although the learning-rate matrix may differ across methods, i.e., we may allow $\mH_t^j$ with $j\in\{\textnormal{im},\textnormal{ex}\}$, we suppress the superscript for simplicity. In \eqref{Implicit parameter update step}, the score is evaluated at the updated state, which makes the method implicit. By contrast, \eqref{Explicit parameter update step} is directly computable because the score is evaluated at the predicted state. As we show in Section~\ref{subsec:optimization}, the implicit update~\eqref{Implicit parameter update step} is the first-order condition for an optimization problem, whereas the explicit update~\eqref{Explicit parameter update step} solves a linearized version of the same problem. Interestingly, \citeauthor{kalman1960new}'s (\citeyear{kalman1960new}) update can be written in either form, albeit with different learning rates (see Appendix~\ref{A: Kalman}).

Turning to the prediction step, both filters use the linear first-order specification
\begin{equation}\label{Parameter prediction step}
    \text{prediction step: } \qquad \vtheta_{t \mid t-1}^j\;=\;(\mI_k - \mathbf{\Phi})\;\vomega\,+\,\mathbf{\Phi}\, \vtheta_{t-1 \mid t-1}^j, \qquad j \in \{\textnormal{im}, \textnormal{ex}\},
\end{equation}
for $t=1,\ldots,T$. Here, $\vomega$ is a $k \times 1$ vector, $\mathbf{\Phi}$ a $k \times k$ matrix, and $\mI_k$ the $k \times k$ identity. Although $\vomega$ and $\mathbf{\Phi}$ may differ across filters, i.e., we may allow $\vomega^j$ and $\mathbf{\Phi}^j$ with $j\in \{\textnormal{im},\textnormal{ex}\}$, we suppress superscripts when convenient. 
The initialization at time zero is taken as given, and we assume throughout that the prediction step maps $\mathbf{\Theta}$ into itself. If $\mathbf{\Theta}\neq \SR^k$, this can often be ensured by restricting $\vomega$ and $\mathbf{\Phi}$; for example, if $\mathbf{\Theta}$ is the positive quadrant, it suffices to take $\vomega\in \mathbf{\Theta}$, $\mathbf{\Phi}$ diagonal, and $\mZeros_k \preceq \mathbf{\Phi} \preceq \mI_k$. Although \eqref{Parameter prediction step} could be generalized to allow nonlinear or higher-order dynamics, we do not pursue this for tractability.

\subsection{Reformulation as optimization-based filters}
\label{subsec:optimization}

We reformulate the ISD update~\eqref{Implicit parameter update step} as a multivariate optimization problem:
\begin{equation}
 \text{ISD update: } \qquad \vtheta_{t \mid t}^{\textnormal{im}}=\underset{\vtheta\in \mathbf{\Theta}}{\operatorname{argmax}} \; \left\{
    \ell \left(\vy_t \mid \vtheta \right) - \frac{1}{2}\left\|\vtheta-\vtheta^\textnormal{im}_{t \mid t-1}\right\|_{\mP_t}^2\right\},
    \label{eq: Implicit optimization problem}
\end{equation}
where $\mP_t\in \SR^{k \times k}\succ \mZeros_k$ is the penalty matrix and $\| \text{\textbf{\textit{v}}} \|^2_{\mP_{t}} = \text{\textbf{\textit{v}}}^{\top}\mP_t \text{\textbf{\textit{v}}} $ is a squared weighted norm. The first-order condition for~\eqref{eq: Implicit optimization problem} is $\vzeros_k=\nabla \ell(\vy_t|\vtheta_{t|t}^\textnormal{im})-\mP_t(\vtheta_{t|t}^\textnormal{im}-\vtheta_{t|t-1}^\textnormal{im})$. Rearranging gives $\vtheta_{t|t}^\textnormal{im}=\vtheta_{t|t-1}^\textnormal{im}+\mP_t^{-1}\nabla \ell(\vy_t|\vtheta_{t|t}^\textnormal{im}) $, which recovers~\eqref{Implicit parameter update step} when $\mP_t$ equals the inverse learning-rate matrix. Hence, we set $\mP_{t} := \mH_{t}^{-1}\succ \mZeros_k$ throughout. Thus,~\eqref{Implicit parameter update step} is shorthand for optimization~\eqref{eq: Implicit optimization problem}: the ISD update maximizes the most recent log-likelihood contribution subject to a weighted quadratic penalty centered at the one-step-ahead prediction. In Section~\ref{sec: Invertibility}, Assumption~\ref{ass1} will ensure a smooth unique interior maximum, so that~\eqref{Implicit parameter update step} and~\eqref{eq: Implicit optimization problem} are equivalent.

In optimization~\eqref{eq: Implicit optimization problem}, we may assume $\mathbf{\Theta}=\SR^k$ to rule out boundary solutions and linearly approximate $\ell(\vy_t|\vtheta)$ around the prediction to obtain the ESD update~\eqref{Explicit parameter update step}:
\begin{equation}\label{Explicit optimization problem}
    \vtheta_{t \mid t}^{\textnormal{ex}}=\underset{\vtheta\in \SR^k}{\operatorname{argmax}} \; \left\{
    \underbrace{\ell(\vy_t \mid \vtheta_{t \mid t-1}^\textnormal{ex})+(\vtheta - \vtheta_{t \mid t-1}^\textnormal{ex})^{\top}\,\nabla\ell(\vy_t \mid \vtheta_{t \mid t-1}^\textnormal{ex})}_{\textnormal{linear approximation of }  \ell \left(\vy_t \mid \vtheta \right) \text{ at } \vtheta=\vtheta_{t|t-1}^\textnormal{ex} } - \frac{1}{2}\left\|\vtheta-\vtheta^\textnormal{ex}_{t \mid t-1}\right\|_{\mP_t}^2 \right\}.
\end{equation}
The corresponding first-order condition, $\vzeros_k =\nabla \ell(\vy_t|\vtheta_{t|t-1}^\textnormal{ex})-\mP_t(\vtheta^\textnormal{ex}_{t|t}-\vtheta^\textnormal{ex}_{t|t-1})$, can be rearranged as $\vtheta_{t|t}^\textnormal{ex}=\vtheta_{t|t-1}^\textnormal{ex}+\mP_t^{-1}\nabla \ell(\vy_t|\vtheta_{t|t-1}^\textnormal{ex}) $, thereby recovering~\eqref{Explicit parameter update step} with $\mP_{t}^{-1} = \mH_{t}$.

The full optimization~\eqref{eq: Implicit optimization problem} has two advantages over its linearized counterpart~\eqref{Explicit optimization problem}. First, it guarantees
$
\ell(\vy_t \mid \vtheta^{\textnormal{im}}_{t|t})
\geq
\ell(\vy_t \mid \vtheta^{\textnormal{im}}_{t|t-1}),
$
so the implicit update cannot worsen the fit of the postulated log density. The explicit update~\eqref{Explicit parameter update step} provides no such guarantee; see Appendix~\ref{A: discussion overshooting}. Second, the linearization in~\eqref{Explicit optimization problem} makes it harder to impose constraints on the parameter space $\mathbf{\Theta}$. We therefore assume throughout that the ESD filter uses $\mathbf{\Theta}^{\textnormal{ex}}=\mathbb{R}^k$. For intrinsically bounded parameters, such as volatility or correlation, ESD filters often rely on link functions mapping $\vtheta \in \mathbb{R}^k$ to the admissible space. These links introduce additional nonlinearity and may complicate the filter's theoretical analysis. By contrast, our theory for the ISD filter allows a general state space $\mathbf{\Theta}^{\textnormal{im}} \subseteq \mathbb{R}^k$, so link functions are not required, though they may still be used.

\subsection{Static (hyper-)parameter estimation}
\label{subsec:mle}

The static (hyper-)parameters of both filters can be estimated by maximum likelihood (ML). Suppose we use a static learning-rate matrix (i.e., $\mH^j_t=\mH^j=(\mP^j)^{-1}\succ \mZeros_k$ for all $t$ and $j\in \{\textnormal{im},\textnormal{ex}\}$) and collect all static parameters into a column vector $\vupsilon^j:=(\text{vech}(\mH^j)^{\top},\vomega^{j^{\top}},\text{vec}(\mathbf{\Phi}^j)^{\top},\vpsi^{j^{\top}})^{\top}$ with $j\in \{\text{im},\text{ex}\}$, where vech($\cdot$) and vec($\cdot$) denote the half-vectorization and vectorization matrix operations, respectively. Following \citet{creal2013generalized}, the static parameters are estimated as
\begin{equation}
    \hat{\vupsilon}^j:=\underset{\vupsilon^j \in \mathbf{\Upsilon}}{\operatorname{argmax}} \sum_{t=1}^T \ell \left(\vy_t \mid \vtheta^j_{t \mid t-1}, \vpsi^j \right),\qquad j\in \{\textnormal{im},\textnormal{ex}\}.
    \label{Eq_MLE}
\end{equation}
Here, $\mathbf{\Upsilon}$ is the subset of the static-parameter space for which $\mH^j \succ \mZeros_k$ and $\vpsi^j$ lies in its permissible domain. While the logarithmic density $\ell (\cdot\mid\cdot):=\log p(\cdot\mid\cdot)$ in \eqref{Eq_MLE} typically matches that used in the filter~\eqref{Implicit parameter update step} or~\eqref{Explicit parameter update step}, we can also allow the estimation criterion to differ from the logarithmic density used in the update. 
Following \citet{blasques2023quasi}, we refer to this as a quasi score-driven specification. Since our theory depends on the density used in the filtering recursion rather than on the static-parameter estimation criterion, our stability results and error bounds for the explicit class cover quasi score-driven filters as well.

\section{Stability of score-driven filters}\label{sec: Invertibility}

We study the exponential stability (e.g., \citealp{guo1995exponential}) and the closely related concept of invertibility (e.g., \citealp{straumann2006quasi}) of score-driven filters. These properties are important in their own right and also underpin maximum-likelihood estimation of static parameters (e.g., \citealp{blasques2022maximum}). Exponential stability means that, given identical data, filtered paths originating from different starting points converge exponentially fast over time. We focus on a particularly strong form of exponential stability that holds uniformly over all initializations as well as data sequences, which is especially relevant under possible filter misspecification.

\begin{definition}[Exponential stability] 
\label{def:stability}
Consider two starting points, $\vtheta_{0}$ and $\widetilde{\vtheta}_{0}$, and the corresponding paths, $\{\vtheta_t\}$ and $\{\widetilde{\vtheta}_t\}$, generated by the same data $\{\vy_t\}$ and the same filter (i.e., with identical static parameters). Here, $\vtheta_t$ may denote either the filtered state $\vtheta_{t|t}$ or the predicted state $\vtheta_{t+1|t}$. For a given set of static parameters $\vupsilon\in \mathbf{\Upsilon}$, the filter is said to be \textnormal{exponentially stable} if there exist a weight matrix $\mW\succ \mZeros_k$ and a contraction coefficient $\tau<1$ such that, uniformly over all $\vtheta_{0}, \widetilde{\vtheta}_{0}\in \mathbf{\Theta}$ and all data $\{\vy_t\}$,
\begin{equation}
 \| \vtheta_t-\widetilde{\vtheta}_t \|_{\mW}^2\;
 \leq\;
 \tau^t\, \| \vtheta_0-\widetilde{\vtheta}_0 \|_{\mW}^2,
 \qquad \forall t\geq 0.
\label{Eq_ExpFiltStab}
\end{equation}
\end{definition}

Exponential stability implies asymptotic forgetting of the initialization: for any given data sequence, the distance between any two paths converges to zero. Hence all paths are asymptotically equivalent, in the sense that they converge to the same data-driven path up to differences that vanish asymptotically. In many standard filters, moreover, the predicted and filtered paths are linked by one-step recursions, so that stability of either path typically implies that of the other. We introduce the concept of exponential stability to emphasize that this is a pathwise property of the \emph{filter} itself, rather than a property of the DGP: (i) a filter may be exponentially stable for some values of the static parameters, but not for others; (ii) under misspecification, the asymptotic filtered path need not coincide with the true parameter path; and (iii) if the data are not stationary and ergodic, the asymptotic path need not be stationary and ergodic either. 

Condition~\eqref{Eq_ExpFiltStab} is stronger than most conditions in the literature (e.g., \citealp{straumann2006quasi}; \citealp{blasques2018feasible}; \citealp{d2026empirical}), as it requires a ``sure'' contraction between any two paths, uniformly over all initializations and data sequences. Some flexibility remains, as $\mW$ is not fixed a priori; i.e., any positive definite matrix yielding the contraction in~\eqref{Eq_ExpFiltStab} is admissible. Later, we will choose $\mW$ to match the natural geometry of the filter.

Remark~\ref{rem:Examples of exponentially stable filters} shows that our new stability concept in Definition~\ref{def:stability} leads to classic stability conditions for affine filters. For such filters, standard Lyapunov theory (e.g., \citealp{stein1952some}) guarantees that an appropriate weight matrix $\mW\succ \mZeros_k$ exists if and only if the usual spectral-radius condition holds.

\begin{remark}[Exponential stability of affine filters]\label{rem:Examples of exponentially stable filters}
Many familiar filters, such as GARCH, are formulated in terms of predictions. Consider therefore the affine prediction filter
\begin{equation}\label{eq:affinefilter}
    \vtheta_{t+1|t} = (\mI_k-\mathbf{\Phi})\vomega + \mathbf{\Phi} \vtheta_{t|t-1} + \vf_t,
\end{equation}
which is affine in $\vtheta_{t|t-1}$ provided that $\vf_t$ is a ``forcing term'' that depends \emph{only} on the data (e.g., realized moments). Then, for two predicted paths $\{\vtheta_{t|t-1}\}$ and $\{\widetilde\vtheta_{t|t-1}\}$ driven by the same data sequence,
$$
\vtheta_{t+1|t}-\widetilde\vtheta_{t+1|t}
= \mathbf{\Phi}(\vtheta_{t|t-1}-\widetilde\vtheta_{t|t-1}).
$$
By Definition~\ref{def:stability}, the predicted path is exponentially stable if there exists a weight matrix $\mW\succ \mZeros_k$ such that $\|\mathbf{\Phi}\|_{\mW}^2<1$, where $\|\mathbf{\Phi}\|_{\mW}:=\sup_{\text{\textbf{\textit{v}}}\neq \vzeros_k}\|\mathbf{\Phi}\,\text{\textbf{\textit{v}}}\|_{\mW}/\|\text{\textbf{\textit{v}}}\|_{\mW}$ is the matrix norm induced by the weighted vector norm. By classic Lyapunov theory (e.g., \citealp{stein1952some}), this requirement is \emph{equivalent} to the usual spectral-radius condition $\varrho(\mathbf{\Phi})<1$. 
Thus, for well-known affine filters such as standard GARCH \citep{bollerslev1986generalized}, Poisson count filters \citep{davis2003observation}, integer GARCH \citep{ferland2006integer}, and ACD filters \citep{engle1998autoregressive}, Definition~\ref{def:stability} recovers the familiar spectral-radius stability condition; see also \citet{brandt1986stochastic} and \citet{bougerol1992stationarity}. Score-driven filters, however, are generally not affine in this sense, because the forcing term $\vf_t$ in~\eqref{eq:affinefilter} then relates to the \emph{score}, which depends not only on the data but also on the state itself. Definition~\ref{def:stability} therefore extends the familiar stability concept to a broader class of possibly non-affine filters.
\end{remark}

To derive tractable sufficient conditions for the exponential stability of SD filters, we now impose regularity conditions on the postulated log density. 
Assumption~\ref{ass1}(a) denotes the Hessian matrix of the postulated log density with respect to $\vtheta$ as $\cH(\vy,{\vtheta}): = \nabla^2 \ell (\vy \mid \vtheta)$. We assume this Hessian to be well defined, that is, the log density is twice differentiable, although not necessarily continuously so. Parts~(b)--(c) are needed to analyze the full optimization problem~\eqref{eq: Implicit optimization problem}, but not its linearized version~\eqref{Explicit optimization problem}. The reason is that the linearized problem is already strictly concave as $\mP\succ \mZeros_k$, and always admits interior solutions as $\mathbf{\Theta}^\text{ex}=\mathbb{R}^k$.

\begin{assumption}[Regularity conditions] \label{ass1} 
Let the penalty matrix $\mP=\mH^{-1}\succ\mZeros_k$ be static. \begin{enumerate}
    \item[a.] 
    The Hessian $\cH(\vy,{\vtheta})$ is well defined and satisfies $\alpha \mI_k \preceq -\cH(\vy,{\vtheta}) \preceq \beta \mI_k$ for all $\vtheta \in \mathbf{\Theta},\vy\in \mathbb{R}^n$ and some $\alpha \in (-\infty, \beta]$, while $\beta\in (0,\infty]$ may be unbounded; and
    \item[b.] Optimization~\eqref{eq: Implicit optimization problem} is strictly concave, implying $\lambda_{\min}(\mP^\textnormal{im})>\alpha^-:=\max\{0,-\alpha\}$; and
    \item[c.] Optimization~\eqref{eq: Implicit optimization problem} admits an interior solution $\vtheta_{t|t}^\textnormal{im}\in \mathbf{\Theta}^\textnormal{im}$ for all $t$.
\end{enumerate}
\end{assumption}

Part~(a) imposes a Hessian bound, as is standard in the analysis of stochastic gradient methods (e.g., \citealp[Ass.~3.1]{chen2020statistical}). We use a one-sided version of this condition (allowing $\beta=\infty$), applied uniformly over all $\vy \in \mathbb{R}^n$ rather than in expectation, since the DGP is unspecified. Under concavity of $\ell(\vy|\vtheta)$, the negative Hessian is positive semi-definite, so $\alpha\geq 0$. If $\alpha>0$, then $\alpha$ is the strong-concavity parameter. If the postulated log density is (locally) non-concave, then $\alpha<0$. Writing $\alpha=\alpha^+-\alpha^-$, where $\alpha^+:=\max\{0,\alpha\}$ and $\alpha^-:=\max\{0,-\alpha\}$, the quantity $\alpha^-$ measures the maximal violation of concavity and is assumed to be bounded in part~(a), i.e., $\alpha^-<\infty$. If the gradient is Lipschitz continuous, the negative Hessian is bounded above by $\beta<\infty$, so the Lipschitz constant $L:=\max\{\alpha^-,\beta\}$ is finite.

Part~(b) restricts the penalty matrix $\mP^\textnormal{im}$ by requiring its smallest eigenvalue to exceed $\alpha^-$, which ensures that the objective in optimization~\eqref{eq: Implicit optimization problem} is strictly concave. Since $\mP^\textnormal{im}=(\mH^\textnormal{im})^{-1}$, this is equivalent to $\lambda_{\max}(\mH^\textnormal{im})< 1/\alpha^-$, which under concavity reduces to the trivial requirement $\lambda_{\max}(\mH^\textnormal{im})<\infty$. Part~(c) is needed to analyze optimization~\eqref{eq: Implicit optimization problem} through its first-order condition~\eqref{Implicit parameter update step}. 

Our first main result, Theorem~\ref{thrm:invertibility}, involves the induced matrix norm
$\|\mA\|_{\mW} = \|\mW^{1/2}\mA \mW^{-1/2}\|$ for any positive-definite matrix $\mW$ and square real matrix $\mA$ of the same dimension (e.g., \citealp[p.\ 39]{jungers2009joint}). Here $\|\mA\| = \sqrt{\lambda_{\max}(\mA^{\top}\mA)}$ is the spectral norm of $\mA$, and $\mW^{1/2}$ is the unique positive-definite square root of $\mW$.

\begin{theorem}[Exponential stability of misspecified SD filters] \label{thrm:invertibility} Let Assumption~\ref{ass1} hold. Then the filters $\{\vtheta^j_{t|t}\}$ 
are exponentially stable (see Definition~\ref{def:stability}) with weight matrix $\mW=\mP=\mH^{-1}$ if the static-parameter vector $\vupsilon\in \mathbf{\Upsilon}$ is such that $\tau^j<1$ for $j\in \{\textnormal{im},\textnormal{ex}\}$, where
\begin{align}
\label{eq:invertibilityISD}
  \tau^{\textnormal{im}}&:= \|\mathbf{\Phi}\|^2_{\mP} \;\left( 1 - \frac{\alpha^+}{\lambda_{\max}(\mP) + \alpha^+} + \frac{\alpha^{-}}{\lambda_{\min}(\mP) - \alpha^{-}}\right)^2,
\\
\label{eq:invertibilityESD}
\tau^{\textnormal{ex}}&:= \|\mathbf{\Phi}\|^2_{\mP} \;
\left(1 - \min\left\{\frac{\alpha^+}{\lambda_{\max}(\mP)} -\frac{\alpha^-}{\lambda_{\min}(\mP)}
 , \, 2-\frac{\beta}{\lambda_{\min}(\mP)}\right\}
 \right)^2  .
\end{align}

\end{theorem}

\begin{remark}[Interpretable conditions] \label{rem2}
If the intuitive conditions (i) $\alpha\geq 0$ and (ii) $\|\mathbf{\Phi}\|_{\mP}\leq 1$ hold, with at least one holding strictly, then $\tau^{\textnormal{im}}<1$ and the ISD filter is exponentially stable. For the ESD filter, the same conclusion additionally requires $\lambda_{\max}(\mH)\leq 2/\beta$.
\end{remark}

In Appendices~\ref{sec: appendix prelim}--\ref{sec: proof of Theorem 1}, we present several intermediate results leading to the proof of Theorem~\ref{thrm:invertibility}. Intuitively, the conditions $\tau^{\textnormal{im}}<1$ and $\tau^{\textnormal{ex}}<1$ define subsets of the static-parameter space on which the ISD and ESD filters are exponentially stable, with $\tau^{\textnormal{ex}}<1$ potentially more restrictive, as discussed below. 
These stability results are uniform over admissible subsets of the static-parameter space, rather than tied to a single parameter value. Thus, much like the spectral-radius condition $\varrho(\mathbf{\Phi})<1$, the conditions $\tau^{\textnormal{im}}<1$ and $\tau^{\textnormal{ex}}<1$ identify nonempty (and typically sizeable) regions of the static-parameter space where the corresponding filter is exponentially stable. We exploit the freedom in Definition~\ref{def:stability} to choose the weight matrix by taking $\mW=\mP=\mH^{-1}$, which is consistent with the natural geometry of the filter and yields the new contraction rates~\eqref{eq:invertibilityISD}--\eqref{eq:invertibilityESD}.

In sum, Theorem~\ref{thrm:invertibility} provides a new multivariate stability result for both explicit and implicit filters that is easily verifiable and agnostic about the DGP. Indeed, the conditions $\tau^{\textnormal{im}}<1$ and $\tau^{\textnormal{ex}}<1$ depend only on filter parameters, namely $\alpha,\beta,\mP$, and $\mathbf{\Phi}$, and impose no restrictions on the DGP. Under these conditions, Theorem~\ref{thrm:invertibility} guarantees that, for any data sequence, all filtered paths converge toward one another, so any limiting path is unique. Relative to \citet[Thm.~1]{lange2024robust}, the ESD stability result is new, while the ISD result is strengthened by allowing possibly non-concave log densities and providing the sharper contraction rate~\eqref{eq:invertibilityISD}.

Generally, our filters are score driven and the DGP is unknown; we now highlight a consequence of Theorem~\ref{thrm:invertibility} for the special case of correct specification.

\begin{remark}[Correct specification]\label{remark:CS}
If the contraction conditions $\tau^{\textnormal{im}},\tau^{\textnormal{ex}}<1$ in Theorem~\ref{thrm:invertibility} hold and the data sequence $\{\vy_t\}$ is stationary and ergodic (SE), then the limiting filtered path is also SE, provided its dependence on past data remains measurable (for details, see \citealp[Prop.\ 4.3]{krengel2011ergodic}; \citealp[Thm.\ 1]{brandt1986stochastic}; \citealp[Thm.\ 3.1]{bougerol1993kalman}). Moreover, if both the filter and the DGP are score driven and share the same observation density and the same static parameters $\vupsilon\in \mathbf{\Upsilon}$, then $\| \vtheta^j_{t|t-1}  - \vvartheta_t \| 
\to 0$ exponentially fast \mbox{(as $t\to \infty$)} uniformly over all data $\{\vy_t\}$ and all initializations at time zero. Thus, the true path is perfectly recovered in the limit. This convergence is pathwise, rather than merely almost sure, because Definition~\ref{def:stability} is formulated uniformly over all data sequences and initializations; this is stronger than the almost-sure convergence commonly reported in the literature (e.g., \citealp[Rem.\ 3.1]{blasques2018feasible}).
\end{remark}

We now discuss the contraction conditions $\tau^{\textnormal{im}},\tau^{\textnormal{ex}}<1$ in more detail. As noted in Remark~\ref{rem2}, if the postulated log density $\ell(\vy|\vtheta)$ is (strictly) concave in $\vtheta$ (i.e., $\alpha\geq 0$), then simple restrictions on $\mathbf{\Phi}$ and $\mP$ suffice to ensure exponential stability of the ISD filter. More generally, $\alpha<0$ or $\|\mathbf{\Phi}\|_{\mP}> 1$ may still be allowed provided $\tau^{\textnormal{im}}<1$. The case $\beta=\infty$ is not problematic, since $\tau^{\textnormal{im}}$ does not depend on $\beta$.
For the ESD filter, by contrast, large values of $\beta$ can be problematic. The second argument of $\min\{\cdot,\cdot\}$ in equation~\eqref{eq:invertibilityESD} is $2-\beta/\lambda_{\min}(\mP)$, which must be nonnegative for stability. Using $1/\lambda_{\min}(\mP)=\lambda_{\max}(\mH)$, this requirement becomes $\lambda_{\max}(\mH)\leq 2/\beta$. Similar conditions are standard for explicit gradient methods; see, for example, \citet{boyd2004convex} (around Eq.~9.17), \citet{nesterov2018lectures} (around Eqns.~1.2.18--22), and \citet{wu2023implicit}. Thus, for ESD filters, the surprise is not that a learning-rate restriction is needed, but rather that such a simple and interpretable condition guarantees stability even in our dynamic and stochastic setting.

Another notable implication of $\lambda_{\max}(\mH)\leq 2/\beta$ is that the learning rate must shrink to zero as $\beta\to \infty$. In Section~\ref{sec:Simulation studies}, we find that misspecified ESD filters with $\beta=\infty$ and a positive learning rate can be made to diverge by increasing the volatility of the true state process. This suggests that $\lambda_{\max}(\mH)\leq 2/\beta$ is close to being necessary for exponential stability of the ESD filter in the sense of Definition~\ref{def:stability}. Although weaker notions of stability may still be available for some ESD filters with $\beta=\infty$, establishing them seems to require some knowledge of the DGP, which we do not assume. Instead, we seek a strong form of exponential filter stability that holds uniformly over \emph{any} DGP. Theorem~\ref{thrm:invertibility} is robust in this sense: its conclusion cannot fail even if the data $\{\vy_t\}$ are generated with the express purpose of making the filters diverge. The resulting conditions are intuitive, readily verifiable, and not overly stringent when the DGP is unknown.

\section{(Non-)asymptotic MSE bounds for score-driven filters}\label{sec:Error bounds for score-driven filters}

This section provides formal theoretical guarantees for the tracking accuracy of score-driven (SD) filters, even when misspecified. To this end, we must reintroduce some consideration of the true process. To measure performance, we consider the weighted mean squared filtering error $\textnormal{MSE}^{\mW}_{t\mid t}:=\mathbb{E}[\|\vtheta_{t \mid t}-\vtheta_t^{\star}\|_{\mW}^2]$ and the weighted mean squared prediction error $\textnormal{MSE}^{\mW}_{t\mid t-1}:=\mathbb{E}[\|\vtheta_{t \mid t-1}-\vtheta_{t}^{\star}\|_{\mW}^2]$, where $\mW \succ \mZeros_k$ is a positive-definite weight matrix and $\vtheta_t^{\star}$ denotes the pseudo-true parameter defined below.
The introduction of a \emph{weighted} MSE allows for some flexibility even if we are ultimately interested in the usual case $\mW=\mI_k$. For example, the usual MSE can be bounded as follows: $\textnormal{MSE}^{\mI_k}_{t|t} \leq 1/(\lambda_{\min}(\mP)) \textnormal{MSE}^{\mP}_{t|t}$.

\begin{definition}[Pseudo-true parameter]\label{Pseudo-truth definition} Consider a true distribution $p^{0}(\vy_t|\vvartheta_{t})$ modeled by some postulated distribution $p(\vy_t|\vtheta_t)$.\ Then $\vtheta^{\star}_{t} :=\mathrm{arg max}_{\vtheta \in \mathbf{\Theta}} \int p^{0}(\vy|\vvartheta_{t}) \ell(\vy|\vtheta) \rd \vy $ is the pseudo-true parameter, provided a unique solution exists. If $p(\vy_t|\cdot)=p^0(\vy_t|\cdot)$, then $\vtheta^{\star}_{t}=\vvartheta_t$.
\end{definition}

To track the pseudo-true parameter path $\{\vtheta_t^{\star}\}$, we require three moment conditions. Although these conditions are almost entirely nonrestrictive, they will, along with Assumption~\ref{ass1}, be sufficient to derive MSE bounds for misspecified SD filters.

\begin{assumption}[Moment conditions involving pseudo-truth] Consider a true distribution $p^{0}(\vy_t|\vvartheta_{t})$ modeled by a postulated distribution $p(\vy_t|\vtheta_t)$. \mbox{Assume for $t=1,\ldots,T$ that:}
    \begin{enumerate}
        \item[a.] The pseudo-truth $\vtheta_t^{\star}$ exists and its increments have finite second (cross) moments. That is,
        $\cov(\vtheta^{\star}_t-\vtheta^{\star}_{t-1})\preceq \mQ,$
        where $\mZeros_k \preceq \mQ \in \SR^{k \times k}$ with $q^2:= \tr(\mQ) < \infty$.
        \item[b.] The first moment of the postulated score, evaluated at the pseudo-truth, is zero; that is, ${\mathbb{E}}_{\vy_t}[\nabla\ell(\vy_t|\vtheta_t^{\star})] = \int p^{0}(\vy|\vvartheta_{t})\nabla\ell(\vy|\vtheta_t^{\star}) \rd\vy = \mathbf{0}_k$. Moreover, its second unconditional moment is bounded, which is ensured by $\mathbb{E}[\|\nabla\ell(\vy_{t} |\vtheta_t^{\star}) \|^2] \; \leq\;  \sigma^2\; <\; \infty.$
        \item[c.] At least one of the following 
        two 
        conditions holds: (i) the unconditional second moment of $\vtheta_t^{\star}$ is bounded, implying $s^2_{\omega} := \underset{t}{\sup} \mathbb{E}[\|\vtheta_t^{\star}-\vomega\|^2] < \infty$, $\forall \vomega \in \mathbb{R}^k$, and/or (ii) $\mathbf{\Phi} = \mI_k$. 
    \end{enumerate}\label{Assumption: Bounded moments}
\end{assumption}

Assumption~\ref{Assumption: Bounded moments}(a) is weaker than assumptions in the literature, which typically require pseudo-true parameter increments to be \emph{uniformly} bounded (e.g.,~\citealp{wilson2018adaptive}; \citealp{simonetto2020time}; \citealp{lanconelli2024maximum}). Assumption~\ref{Assumption: Bounded moments}(b) follows \citet[Ass.~$2.1$]{toulis2017asymptotic} in bounding the expected squared score at $\vtheta^\star_t$, which is less stringent than versions that bound the same quantity uniformly for all $\vtheta\in\mathbf{\Theta}$ (e.g.,~\citealp{lehmann1998theory}). Assumption~\ref{Assumption: Bounded moments}(c) implies that tracking a unit-root process is feasible only if we set $\mathbf{\Phi}=\mI_k$ in the prediction step~\eqref{Parameter prediction step}. 

Based on Assumption~\ref{Assumption: Bounded moments}, Theorem~\ref{Th: (Non-)asymptotic MSE bounds} below gives specific values for each filter such  that the $\mP$-weighted MSE at each time step $t$ can be bounded as follows:
\begin{alignat}{4}
 \text{updating-error bound :}\qquad    &\textnormal{MSE}^{\mP}_{t \mid t} \quad &&\leq& \quad a \quad &\textnormal{MSE}^{\mP}_{t \mid t-1} &\;\;+\;\; b;
     \label{eq: one-step updated MSE bound}\\
    \text{prediction-error bound:}\qquad  &\textnormal{MSE}^{\mP}_{t+1 \mid t} \quad &&\leq  & \quad c \quad &\textnormal{MSE}^{\mP}_{t \mid t} &\;\;+\;\; d,\label{eq: one-step predicted MSE bound}
\end{alignat}
for all $t\geq 1$. We refer to $a$ and $c$ as the contraction rates, while $b$ and $d$ are additive constants. In the first inequality, the values of $a$ and $b$ depend on which update step is used (ISD or ESD): the contraction rate $a$ depends on the shape of the log likelihood via $\alpha$ and $\beta$ as per Assumption~\ref{ass1}, while
$b$ relates to the noisiness $\sigma^2$ of scores as per Assumption~\ref{Assumption: Bounded moments}(b).  In the second inequality, the values of $c$ and $d$ relate to the prediction step: $c$ reflects the contraction rate via the autoregressive matrix $\mathbf{\Phi}$, while $d$ relates to the drifting of states via $q^2$ and $s_\omega^2$ as per Assumptions~\ref{Assumption: Bounded moments}(a) and (c).

Using a standard geometric-series result, the pair~\eqref{eq: one-step updated MSE bound}--\eqref{eq: one-step predicted MSE bound} yields the following \emph{finite-sample error bound} for all $t\geq 1$ provided that $a \, c \neq 1$:
\begin{equation}\label{eq: finite-sample error bound}
  \textnormal{MSE}^{\mP}_{t \mid t}\ \leq   \underbrace{a^{t}c^{t-1} \;\textnormal{MSE}^{\mP}_{1 \mid 0}}_{\text{initialization}} \; +\underbrace{\frac{1-a^{t} c^{t}}{1-a\,c}b}_{\text{noisy scores, $\sigma^2$}} +\underbrace{\frac{1-a^{t-1}c^{t-1}}{1-a\,c}a\,d}_{\text{drifting states, $q^2$ and $s_\omega^2$}},
\end{equation}
where three contributions to the weighted MSE are indicated. Assuming $a\, c < 1$ and letting $t\to \infty$, we obtain the following \emph{asymptotic error bounds}:
\begin{equation}
\label{eq: asymptotic error bound}
    \limsup _{t \rightarrow\infty} \;\textnormal{MSE}^{\mP}_{t\mid t}  \leq \frac{b\,+\,a\,d}{1 - a\,c},\qquad \qquad  \limsup _{t \rightarrow\infty} \;\textnormal{MSE}^{\mP}_{t+1\mid t}  \leq \frac{b\,c+d}{1- a\, c },
\end{equation}
where the effect of the initialization has disappeared. 
Both bounds increase with the product $a\, c$, which can be interpreted as a joint contraction rate analogous to $\tau^{\textnormal{im}}$ or $\tau^{\textnormal{ex}}$ in Theorem~\ref{thrm:invertibility}. Indeed, if $\tau^{\textnormal{im}},\tau^{\textnormal{ex}}<1$, then we can find values of $a$ and $c$ such that $a\, c<1$. If, moreover, the additive constants $b$ and $d$ are bounded, then (finite) MSE bounds exist.

\begin{theorem}[MSE bounds for misspecified SD filters]\label{Th: (Non-)asymptotic MSE bounds} Let Assumptions~\ref{ass1}--\ref{Assumption: Bounded moments} hold. Let the contraction conditions $\tau^{\textnormal{im}}<1$ and  $\tau^{\textnormal{ex}}<1$ in Theorem~\ref{thrm:invertibility} hold. Let $\cH(\vy_t,{\vtheta})$ be Riemann integrable in $\vtheta$ with probability one in $\vy_t$. Finite MSE bounds then exist; specifically, the \mbox{(non-)}asymptotic bounds~\eqref{eq: finite-sample error bound}--\eqref{eq: asymptotic error bound} hold with $a,b,c$, and $d$ as stated in Table~\ref{tab: overview}. In these bounds, the free parameters $\epsilon>0$ and $\chi>0$ can (and should) be chosen to ensure $a\, c<1$.
\end{theorem}

\begin{table}[h]
\centering
\caption{\label{tab: overview} Values of $a,b,c$, and $d$ in the \mbox{(non-)}asymptotic error bounds~\eqref{eq: finite-sample error bound}--\eqref{eq: asymptotic error bound}} 
\small
\begin{threeparttable}
\begin{tabular}{@{}l@{}l@{}c@{}c@{}}
\toprule
\textbf{Step}       & \textbf{Filter} &  \textbf{Contraction rate} & \textbf{Additive constant} 
\\
\midrule
\multirow{2}{*}{Update} & ISD & $a = \left( 1 - \frac{\alpha^+}{\lambda_{\max}(\mP) + \alpha^+} + \frac{\alpha^{-}}{\lambda_{\min}(\mP) - \alpha^{-}}\right)^2$   & $ 
 b=a\times\frac{\sigma^2}{\lambda_{\min}(\mP)}$ 
\\
 & ESD & $  a =  \left(1 - \min\left\{\frac{\alpha^+}{\lambda_{\max}(\mP)} -\frac{\alpha^-}{\lambda_{\min}(\mP)}
 ,\frac{ 2\lambda_{\min}(\mP) - \beta}{\lambda_{\min}(\mP)} \right\} \right)^2 \!  + \frac{\chi^2L^2}{\lambda_{\min}(\mP)^2}$  & 
 $ b=\frac{(1+1/\chi^2)\sigma^2}{\lambda_{\min}(\mP)}$
 \\
 \midrule
  Prediction & & $c = (1+\epsilon^2) \|\mathbf{\Phi}\|^2_{\mP}$ & $d = \lambda_{\max}(\mP)(1+\frac{1}{\epsilon^2})(\|\mI_k-\mathbf{\Phi}\| s_{\omega} + q)^2$

 \\
\bottomrule
\end{tabular}
\end{threeparttable}
\begin{tablenotes}
\item \emph{Note}: $\vomega$, $\mathbf{\Phi}$, and $\mP$ are the parameters in the filter. For the definition of $\alpha$ and $\beta$, see Assumption~\ref{ass1}(a), while $L:=\max\{\alpha^-,\beta\}$. For $q^2$, $\sigma^2$, and $s_\omega^2$, see Assumptions~\ref{Assumption: Bounded moments}(a), (b), and (c), respectively. Parameters $\epsilon>0$ and $\chi>0$ can be freely chosen as long as $a\, c<1$.
\end{tablenotes}
\end{table}

The proof is presented in Appendix~\ref{Proof of Theorem 2}. Under the conditions of Theorem~\ref{Th: (Non-)asymptotic MSE bounds}, Table~\ref{tab: overview} expresses the coefficients $a,b,c$, and $d$ appearing in the MSE bounds~\eqref{eq: finite-sample error bound}--\eqref{eq: asymptotic error bound} in terms of the filter's static parameters $\vomega,\mathbf{\Phi}$, and $\mP$ (as set by the researcher), 
$\alpha$ and $\beta$  (as defined in Assumption~\ref{ass1}), and $q$, $\sigma^2$ and $s_\omega^2$ (as defined in Assumption~\ref{Assumption: Bounded moments}). For the update step of the ESD filter, $a$ and $b$ also contain the free parameter $\chi>0$. For the prediction step, $c$ and $d$ similarly contain the free parameter $\epsilon>0$. These free parameters arise from our application of Young's inequality, which, for two vectors $\vu$ and $\text{\textbf{\textit{v}}}$, and a compatible matrix $\mW\succ\mZeros$, reads $ \| \vu + \text{\textbf{\textit{v}}} \|_{\mW}^2 \leq (1+\epsilon^2)\|\vu\|_{\mW}^2+ (1+1/\epsilon^2)\|\text{\textbf{\textit{v}}}\|_{\mW}^2$ for all $\epsilon>0$ (see Appendix~\ref{Proof of Theorem 2}). Theorem~\ref{Th: (Non-)asymptotic MSE bounds} therefore defines a \emph{family} of admissible bounds, indexed by the two Young's parameters $\chi>0$ and $\epsilon>0$, which influence both the numerator and denominator of the bounds.  Naturally, we can select the free parameters to yield the best possible bound. Indeed, Appendix~\ref{sec:Proof of Corollary 2.4} derives the smallest bound for the ISD filter in closed form.

Theorem~\ref{Th: (Non-)asymptotic MSE bounds} is, to the best of our knowledge, the first result to establish finite-sample and asymptotic MSE bounds for the \emph{entire} filtered and predicted paths of misspecified score-driven filters. Existing results are local in time: \citet{gorgi2023optimality} and \citet{creal2024moment} show that score-driven updates can improve one-step losses, but they do not control the propagation of errors along the filtered path. Similarly, \citet[Thm.~2]{lange2024robust} provide a one-step MSE improvement result for the ISD filter, but only for concave log densities and again without controlling the full filtered path. In contrast, Theorem~\ref{Th: (Non-)asymptotic MSE bounds} bounds the filtering and prediction errors relative to the pseudo-true path over \emph{multiple} time steps, including in the asymptotic limit as $t\to\infty$. Finally, unlike \citet{beutner2023consistency}, who establish convergence to the pseudo-true path under infill asymptotics, our bounds apply at a fixed observation frequency, where exact recovery is generally impossible and the relevant object is a non-vanishing tracking-error bound.

As our MSE bounds exist under the same conditions for which exponential stability in Theorem~\ref{thrm:invertibility} holds, much of the discussion there remains relevant. For the ISD filter, log concavity of the postulated density (i.e., $\alpha\geq 0$) along with restrictions on $\mathbf{\Phi}$ is sufficient to yield $a\, c<1$. For the ESD filter, in contrast, if we want to ensure $a<1$ we additionally need $\beta<\infty$ and $\lambda_{\max}(\mH)<2/\beta$. As we will see in Section~\ref{sec:Simulation studies}, these additional restrictions are not merely artifacts: it is easy to show that misspecified ESD filters may be divergent if $\beta=\infty$.

For our MSE bounds to be finite, we require $d<\infty$. We observe from Table~\ref{tab: overview} that $d$ contains the term $\|\mI_k-\mathbf{\Phi}\|\, \times \, s_{\omega} $, where $s^2_\omega:=\sup_{t} \mathbb{E}[\|\vtheta_t^{\star}-\vomega\|^2]$. The boundedness of this term is ensured by Assumption~\ref{Assumption: Bounded moments}(c), which imposes that (i) $s^2_\omega<\infty$ and/or (ii) $\mathbf{\Phi}=\mI_k$. If $s^2_\omega<\infty$, the resulting bound is minimized if $s_\omega^2$ is minimized, which occurs if $\vomega =  \mathbb{E}[\vtheta_t^{\star}]$; that is, ideally $\vomega$ should be the unconditional mean of $\vtheta_t^{\star}$. If $s_\omega^2=\infty$, on the other hand, $d<\infty$ is ensured by $\mathbf{\Phi}=\mI_k$.

While choosing $\mathbf{\Phi}=\mI_k$ ensures $d<\infty$, the prediction step is no longer contractive as $c=1+\epsilon^2>1$, although $\epsilon$ may be arbitrarily small. To obtain a finite MSE bound, we then require $a<1$ to ensure $a\, c<1$. For the ISD filter, this can be achieved if $\alpha>0$, which essentially means that each observation must carry a minimum (non-zero) amount of information. Thus, ISD filters can successfully track unit-root \mbox{(pseudo-)}true states (i.e., achieving an asymptotically bounded MSE) if the postulated density is strongly log concave. For ESD filters, as before, we additionally require $\beta<\infty$ and $\lambda_{\max}(\mH)<2/\beta$.

\subsection{Special case: Known observation density and linear state dynamics}

If the DGP is known, our error bounds can be improved in a manner analogous to the derivation of the classic Kalman filter. As in the classic setting, we assume linear state dynamics, but we deviate by allowing non-Gaussian state innovations and a general observation density.

\begin{assumption}[Known state-space model with linear state dynamics] For all $t=1,\ldots,T$, let the true density $p^0(\cdot|\cdot)$ coincide with the postulated density $p(\cdot|\cdot)$, while the true state evolves as 
        $
        \boldsymbol{{\vartheta}}_t=(\mI_k - \mathbf{\Phi}_{0}) \,\vomega_{0}+\mathbf{\Phi}_{0} \vvartheta_{t-1}+\vxi_t
        $,
        where $\vxi_t \sim \textnormal{i.i.d.}\,(\vzeros_k, \mathbf{\Sigma}_{\xi})$        with finite covariance matrix $\mZeros_k \preceq \mathbf{\Sigma}_{\xi} \in \SR^{k \times k}$ and $\sigma_{\xi}^2 := \tr(\mathbf{\Sigma}_{\xi}) < \infty$. The intercept parameter vector $\vomega_{0} \in \SR^k$ and autoregressive matrix \mbox{$\mathbf{\Phi}_{0} \in \SR^{k \times k}$} are known. 
    \label{Assumption: Correct specification}
\end{assumption}

When Assumption~\ref{Assumption: Correct specification} holds, our filters employ the true observation density $p(\vy_t|\cdot) = p^{0}(\vy_t|\cdot)$, while also using the true state-transition parameters,
 $\vomega = \vomega_{0}$ and $\mathbf{\Phi}=\mathbf{\Phi}_{0}$.
In this case, the pseudo-true and true parameters coincide; that is, $\vtheta_t^{\star} = \vvartheta_t$ for $t=1,\ldots,T$. However, as the DGP is a state-space model, our SD filters remain misspecified.

\begin{proposition}
\label{prop1} Let Assumptions~\ref{ass1}, \ref{Assumption: Bounded moments}(a,b), and~\ref{Assumption: Correct specification} hold. Let contraction conditions $\tau^{\textnormal{im}},\tau^{\textnormal{ex}}<1$ of Theorem~\ref{thrm:invertibility} and the Riemann integrability condition of Theorem~\ref{Th: (Non-)asymptotic MSE bounds} hold.
Then the MSE bounds~\eqref{eq: finite-sample error bound}--\eqref{eq: asymptotic error bound} hold with $a$ and $b$ as in Table~\ref{tab: overview} and $c=\|\mathbf{\Phi}_0\|^2_{\mP}$ and $d=\lambda_{\max}(\mP)\sigma^2_{\xi}$.
\end{proposition}

The proof is contained in Appendix~\ref{Proof of prop1}. Under Assumption~\ref{Assumption: Correct specification}, the values of $c=\|\mathbf{\Phi}_0\|^2_{\mP}$ and $d=\lambda_{\max}(\mP)\sigma^2_{\xi}$ are automatically bounded and no longer contain any free parameters. Although the MSE bound of the ESD filter still contains one free parameter (i.e., $\chi>0$), the MSE bound of the ISD filter now exclusively depends on $\alpha$, $\mP$, $\mathbf{\Phi}_0$, $\sigma^2$, and $\sigma_\xi^2$. Some freedom remains in selecting the penalty matrix or, equivalently, the learning-rate matrix. In Appendix~\ref{sec: proof of optimal learning rate}, we take this matrix to be a scalar multiple of the identity, enabling us to analytically derive the scalar learning rate that minimizes the asymptotic (Euclidean) MSE bound. This closed-form solution can be used to ``tune'' the learning rate if $\sigma^2$ and $\sigma_\xi^2$ are known or can be approximated. Moreover, in a special case related to the Kalman filter, Appendix~\ref{Proof Example 4} shows that this minimized MSE bound is tight (i.e., cannot be improved).

\section{Simulation studies}
\label{sec:Simulation studies}

We present three simulation studies comparing SD filters in terms of stability and performance. Specifically, we investigate a linear setting with high-dimensional states as in \cite{cutler2023stochastic} (Section~\ref{subsec:Least-squares recovery.}), nine nonlinear DGPs with univariate states as in \cite{koopman2016predicting} (Section~\ref{subsec:Koopman DGPs Monte Carlo}), and the popular dynamic Poisson count model (Section~\ref{subsec:Dynamic Poisson distribution with misspecified link function.}).

\subsection{Least-squares recovery as in \cite{cutler2023stochastic}}
\label{subsec:Least-squares recovery.}

\textbf{Observation equation.} Following \citet[pp.\ 453--4]{madden2021bounds} and \citet[pp.\ 41--2]{cutler2023stochastic}, we investigate a high-dimensional linear model. Conditional on the true latent state $\vvartheta_t\in \SR^k,$
each observation $\vy_t \in \mathbb{R}^n$ is independently drawn from a Gaussian distribution $ \mathrm{N}(\mA \vvartheta_{t}, \mathbf{\Sigma})$. The (static) matrix $\mA \in \mathbb{R}^{n \times k}$ is generated using
Haar-distributed orthogonal matrices to ensure that it has rank $k\leq n$, with
singular values equally spaced in the interval $[\sqrt{\alpha},\sqrt{\beta} \,]$, where $0<\alpha\leq \beta<\infty$. As we will see, this corresponds with the notation in Assumption~\ref{ass1}. The covariance matrix $\mathbf{\Sigma}$ is set as $\sigma^2/(n\beta)\mI_n$ with $0<\sigma^2
<\infty$, ensuring that $\sigma^2<\infty$ matches the notation in Assumption~\ref{Assumption: Bounded moments}(b).

\textbf{State transition.} The latent state $\vvartheta_{t} \in \mathbb{R}^k$ evolves according to a random walk $\vvartheta_{t+1} = \vvartheta_{t} + \vxi_t$,
where $\vxi_t$ is i.i.d.\ and non-Gaussian, drawn uniformly from the surface of a $k$-dimensional sphere with radius $\sigma_{\xi} > 0$, aligning with the notation in Assumption~\ref{Assumption: Correct specification}. The initial state $\vvartheta_{0}$ is drawn from the surface of a sphere with radius $10$.

\textbf{Postulated density.} Following \cite{cutler2023stochastic}, we are interested in least-squares recovery, meaning that all directions are equally important. Therefore, we postulate a Gaussian density with the identity as the covariance matrix:
\begin{equation}
    \ell(\vy_t|\vtheta) = - \frac{1}{2}\|\mA \vtheta - \vy_t\|^2 -\frac{k}{2}\log(2\pi).
    \label{Eq_MVND}
\end{equation}
The Hessian with respect to $\vtheta$ is constant at $\cH =- \mA^{\top}\mA$. Due to the construction of $\mA$, the eigenvalues of $-\cH$ lie in $[\alpha,\beta]$ with $0<\alpha\leq \beta<\infty$, consistent with the notation in Assumption~\ref{ass1}(a).
Although the logarithmic observation density~\eqref{Eq_MVND} is technically misspecified, the pseudo-true parameter still equals the true parameter (i.e., $\vvartheta_t=\vtheta_t^\star,\forall t$). Moreover, taking the postulated covariance matrix to be a \emph{scalar multiple} of the identity would lead to identical updates, differing only in the learning rate. Parameter $\sigma^2$ matches the notation in Assumption~\ref{Assumption: Bounded moments}(b), as  $ \mathbb{E} \|\nabla\ell(\vy_t|\vvartheta_t)\|^2= \mathbb{E} \|\mA^{\top}(\vy_t - \mA \vvartheta_{t})\|^2 =\tr(\mA^{\top} \mathbf{\Sigma}\mA) \leq
\beta \tr(\mathbf{\Sigma}) = \sigma^2 < \infty$.

\textbf{Filter specification.} Our filters are initialized at the origin and use the identity mapping for their prediction steps; that is, $\vtheta^j_{0|0}=\vzeros_k$ and $\vtheta^j_{t+1|t} = \vtheta^j_{t|t}$ with $j \in \{\text{ex, im}\}$. Following \cite{cutler2023stochastic}, we consider learning-rate matrices that are scalar multiples of the identity.
The ESD update with  $\mH_t^{\textnormal{ex}} = \eta^{\textnormal{ex}} \mI_k$ and $\eta^{\textnormal{ex}}>0$ then equals
\begin{equation*}
\text{ESD update:}\qquad    \vtheta_{t|t}^{\textnormal{ex}} \;=\;
    \vtheta_{t|t-1}^{\textnormal{ex}} + \eta^{\textnormal{ex}} \mA^{\top}(\vy_t-\mA \vtheta_{t|t-1}^{\textnormal{ex}}),
\end{equation*}
where $\nabla\ell(\vy_t|\vtheta^{\textnormal{ex}}_{t|t-1}) = \mA^{\top}(\vy_t-\mA \vtheta_{t|t-1}^{\textnormal{ex}})$ is the (explicit) score. For the ISD update, the first-order condition $\vtheta_{t|t}^{\textnormal{im}} =\vtheta_{t|t-1}^{\textnormal{im}} +\eta^{\textnormal{im}} \mA^{\top}(\vy_t-\mA \vtheta_{t|t}^{\textnormal{im}})$ can be solved for $\vtheta_{t|t}^{\textnormal{im}}$ to yield
\begin{align*}
\text{ISD update:}\qquad      \vtheta_{t|t}^{\textnormal{im}} &= \left(\mI_k + \eta^{\textnormal{im}} \mA^{\top} \mA\right)^{-1} \left(\vtheta_{t|t-1}^{\textnormal{im}} + \eta^{\textnormal{im}}\mA^{\top} \vy_t\right),\\
    &= \vtheta^\textnormal{im}_{t|t-1} + \eta^{\textnormal{im}} \left(\mI_k + \eta^{\textnormal{im}} \mA^{\top} \mA\right)^{-1} \mA^{\top}(\vy_t-\mA \vtheta_{t|t-1}^{\textnormal{im}})   ,
\end{align*}
which is a shrunken version of the ESD update with shrinkage coefficient $\left(\mI_k + \eta^{\textnormal{im}} \mA^{\top} \mA\right)^{-1}$. This ensures that the ISD update remains bounded as the learning rate $\eta^\textnormal{im}$ increases.

\textbf{Error bounds.}
For both filters in this setup,
Theorem~\ref{Th: (Non-)asymptotic MSE bounds} holds with $c^{\text{im}}=c^{\text{ex}}=1$ and
\begin{align*}
    a^{\textnormal{ex}} &= (1 - \min\{\alpha \eta^{\textnormal{ex}}, 2-\beta \eta^{\textnormal{ex}}\})^2 + (\chi\,\beta\, \eta^{\textnormal{ex} })^2,
    &b^{\textnormal{ex}} &= (1 + 1/\chi^{ 2}) \sigma^2 \eta^{\textnormal{ex}},&d^\textnormal{ex} & = \sigma_{\xi}^2/\eta^\textnormal{ex}.
    \\
    a^{\textnormal{im}} &= (1 + \alpha \eta^{\textnormal{im}})^{-2},
    &b^{\textnormal{im}} &= a^{\textnormal{im}} \sigma^2 \eta^{\textnormal{im}},& d^\textnormal{im} & = \sigma_{\xi}^2/\eta^\textnormal{im}.
\end{align*}
These values can be substituted into equations~\eqref{eq: finite-sample error bound}--\eqref{eq: asymptotic error bound} to obtain finite-sample and asymptotic error bounds in the $\mP^j$-norm, respectively.
As we are interested in least squares, we multiply these bounds by $1/\lambda_{\min}(\mP^j)=\eta^j$ to obtain (Euclidean) MSE bounds. We then minimize these MSE bounds by selecting the learning rates $\eta^j$ for $j\in \{\textnormal{im},\textnormal{ex}\}$. For the ISD filter, we use the analytic minimizer $\eta^{\textnormal{im}}_{\star}=1/\rho^\textnormal{im}_\star$ given in Appendix~\ref{sec: proof of optimal learning rate}. For the ESD filter, we numerically minimize the MSE bound with respect to both the learning rate $\eta^{\textnormal{ex}}$ and the free parameter $\chi$. 

\begin{figure}[tb]
    \centering
    \begin{subfigure}{0.475\textwidth}
        \centering
        \includegraphics[width=\textwidth]{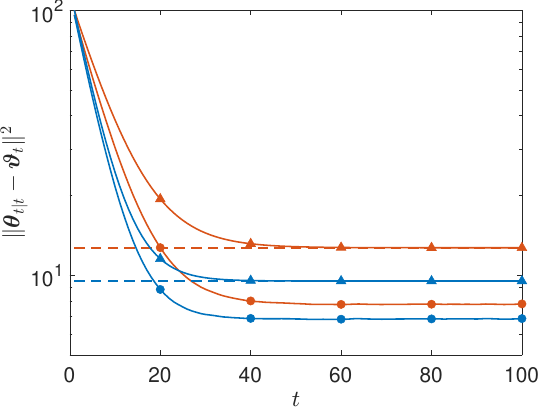}
        \label{fig:SD_least_squares_t_beta1}
    \end{subfigure}
    \hfill
    \begin{subfigure}{0.475\textwidth}
        \centering
        \includegraphics[width=\textwidth]{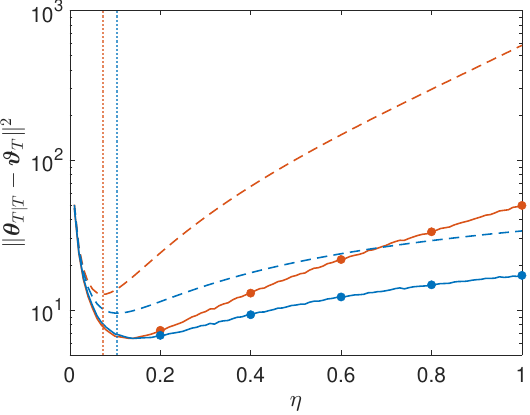}
        \label{fig:SD_least_squares_eta_beta1}
    \end{subfigure}
    \vspace{-4mm} 

    \begin{subfigure}{0.475\textwidth}
        \centering
        \includegraphics[width=\textwidth]{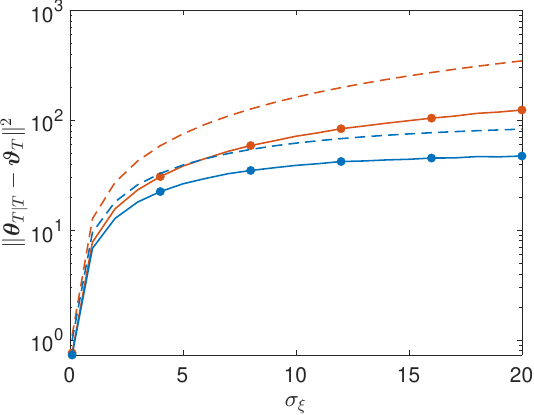}
        \label{fig:SD_least_squares_delta_beta1}
    \end{subfigure}
    \hfill
    \begin{subfigure}{0.475\textwidth}
        \centering
        \includegraphics[width=\textwidth]{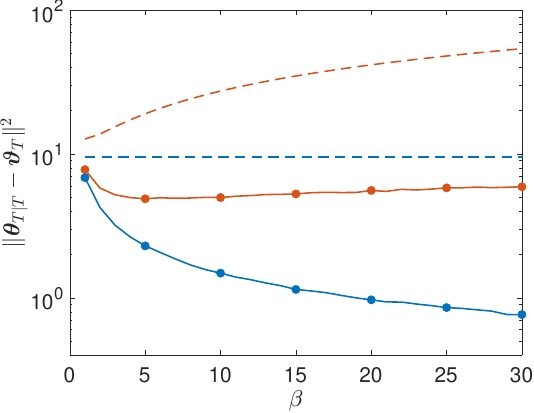}
        \label{fig:SD_least_squares_beta50}
    \end{subfigure}
        \vspace{-3mm}

    \begin{subfigure}{0.8\textwidth}
        \centering
        \adjustbox{trim=0 0 0 0,clip}{\includegraphics[width=\textwidth]{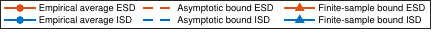}}
        \label{fig:legend}
    \end{subfigure}

    \vspace{-7mm}
    \caption{Semilog plots of empirical MSEs and MSE bounds (dotted) for least-squares recovery with respect to the time step $t$, learning rate $\eta$, state volatility $\sigma_{\xi}$, and Lipschitz gradient constant $\beta$, with average errors computed at horizon $T=500$ for the latter three plots. Empirical averages are computed over $1{,}000$ replications. Unless stated otherwise, parameters are $k=50$, $n=100$, $\alpha = \beta = 1$, $\sigma = 10$, and $\sigma_{\xi} = 1$, while learning rates $\eta^j$ for $j\in \{\textnormal{im},\textnormal{ex}\}$ are found by minimizing the asymptotic MSE bounds.}
    \label{fig:Least-squares recovery}
\end{figure}

\textbf{Simulation setup.} As in \cite{cutler2023stochastic}, our default parameters are $k=50$, $n=100$, $\alpha = \beta = 1$, $\sigma = 10$, and $\sigma_{\xi} = 1$.
We simulate $1{,}000$ paths of length $T=500$. For each replication, we compute the final squared error $\|\vtheta^{j}_{T|T} - \vvartheta_{T}\|^2$ for $j\in\{\textnormal{im},\textnormal{ex}\}$. We compute MSEs by averaging over all replications.

\textbf{Findings for SD filters.} Figure~\ref{fig:Least-squares recovery} shows the empirical MSEs of both SD filters for various values of $t$, $\eta$, $\sigma_{\xi}$, and $\beta$. The ISD filter achieves lower MSEs than the ESD filter across the entire range of settings considered. The top-left plot shows that the ISD filter has the lowest empirical MSE and MSE bound at all time steps. The initial MSE for both filters is $100$, as they are initialized at the origin while the true process starts on a sphere of radius $10$.
The top-right plot shows the performance of both filters for different learning rates, illustrating that the advantage of the ISD filter grows as the learning rate increases.
Vertical lines indicate the learning rates that minimize the MSE bounds, which in both cases lie slightly to the left of the learning rates that minimize the empirical MSEs.

The bottom-left plot shows that the advantage of ISD filters increases when the true state is more volatile (i.e., for larger values of $\sigma_\xi$).
Similarly, the bottom-right plot shows that this advantage also grows for larger values of $\beta$. Intuitively, while the ISD filter can take advantage of the increased curvature in the postulated density (i.e., when observations are more informative), the ESD filter cannot, as its learning rate must be capped at $2/\beta$.

\textbf{Three alternative tracking algorithms.}
Next, we consider three recent tracking algorithms: (i) the online version of Nesterov’s cornerstone modern optimization method \citep[ONM;][]{nesterov1983method} as discussed in~\cite{nesterov2018lectures}, (ii) the online gradient descent algorithm from \citet{madden2021bounds}, and (iii) the stochastic gradient method from \citet{cutler2023stochastic}. For ONM, we follow the implementation provided in \citet[sec.\ 6.1]{madden2021bounds}, where the method performed well empirically, although no performance guarantees (in finite-dimensional spaces) are known.
The algorithms by \citet{madden2021bounds} and \citet{cutler2023stochastic} both use explicit gradient methods and, other than using different learning rates, are equivalent to our ESD filter. These authors also derive error bounds and select their learning rates to minimize these bounds; for details, see \citet[Thm.\ 15]{cutler2023stochastic}, and \citet[Thm.\ 3.1]{madden2021bounds}.
Interestingly, \cite{cutler2023stochastic} cap the learning rate at $1/(2\beta)$, while our learning-rate cap for the ESD filter  is four times higher at $2/\beta$, which still guarantees stability by way of Theorem~\ref{thrm:invertibility}.
When $\alpha=\beta$, ONM is identical to online gradient descent as in~\citet{madden2021bounds}.
The learning rate in \citet{madden2021bounds} is $2/(\alpha+\beta)$, which differs from ours and \citeauthor{cutler2023stochastic}'s in that it is independent of the gradient noise $\sigma^2$. The bounds in \cite{madden2021bounds} use a different norm and are thus not shown in our graphs. 

\begin{figure}[tb]
    \centering
    \begin{subfigure}{0.476\textwidth}
        \centering
        \includegraphics[width=\textwidth]{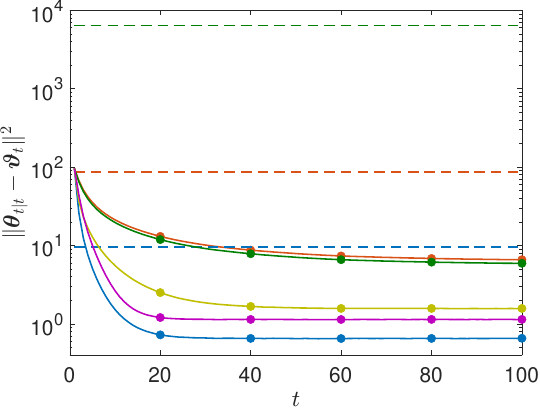}
        \label{fig:sub1}
    \end{subfigure}
    \hfill
    \begin{subfigure}{0.475\textwidth}
        \centering
        \includegraphics[width=\textwidth]{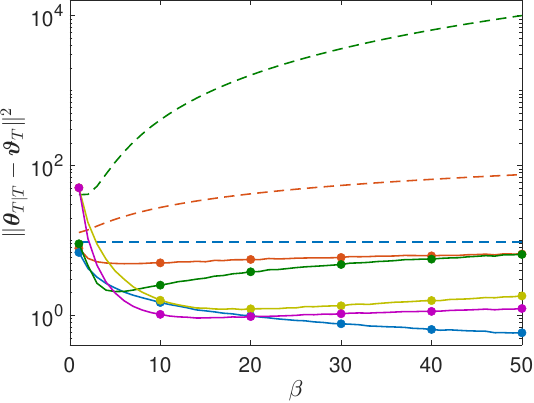}
        \label{fig:sub2}
    \end{subfigure}
    \vskip\baselineskip
    \vspace{-8mm}
    \begin{subfigure}{0.475\textwidth}
        \centering
        \includegraphics[width=\textwidth]{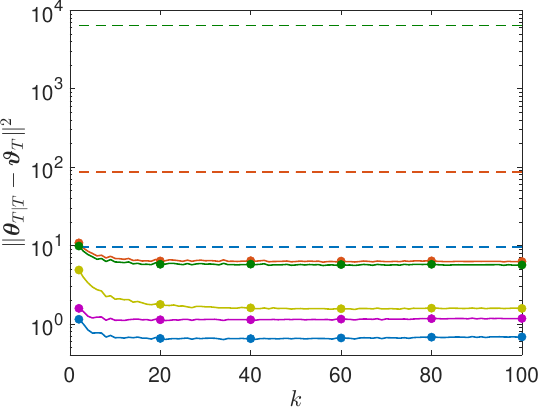}
        \label{fig:sub3}
    \end{subfigure}
    \hfill
    \begin{subfigure}{0.475\textwidth}
        \centering
        \includegraphics[width=\textwidth]{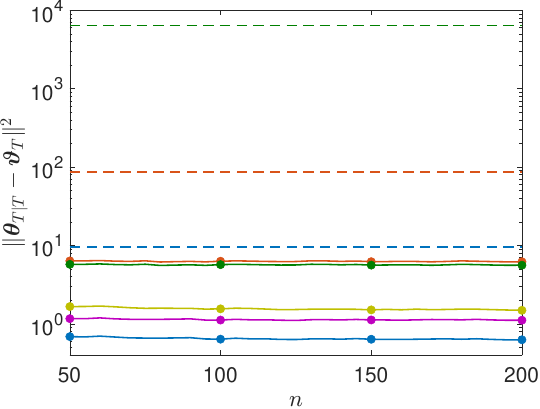}
        \label{fig:sub4}
    \end{subfigure}
    \vspace{-8mm}
    \begin{subfigure}{1\textwidth}
        \centering
        \includegraphics[width=\textwidth]{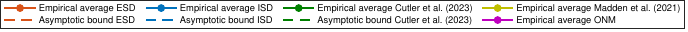}
        \label{fig:legend2}
    \end{subfigure}
    \caption{Semilog plots of guaranteed bounds and empirical tracking errors for least-squares recovery with respect to iteration $t$, Lipschitz gradient constant $\beta$, state dimension $k$, and observation dimension $n$, with average errors computed at horizon $T=500$ for the latter three plots. Empirical averages are computed over $1{,}000$ trials. Default parameter values: $\alpha=1, \beta=40, \sigma=10, \sigma_{\xi}=1, \eta=\eta_{\star}$, $k=50$, $n=100$.}
    \label{fig:Least-squares recovery comparison}
\end{figure}

\textbf{Comparison of tracking algorithms.} To investigate the differences between all five tracking algorithms, we consider a setting with a large disparity between $\alpha$ and $\beta$, taking $\alpha=1$ and $\beta=40$.
For completeness, results for our default parameters (for which $\alpha=\beta=1$ as in \citealp{cutler2023stochastic}) are given in Appendix~\ref{subsec:Simulation study 1: Tracking under default settings}. Figure~\ref{fig:Least-squares recovery comparison} shows the MSEs of all five algorithms
and, when available, the asymptotic MSE bounds.

We find that the ISD filter (i) outperforms all other methods across all time steps $t$, (ii) is the only algorithm that yields consistently lower MSEs when the Lipschitz constant $\beta$ is increased, (iii) has the lowest MSE for all investigated state dimensions $k$ and observation dimensions $n$, and (iv) provides the strongest \mbox{(non-)}asymptotic performance guarantees (i.e., the lowest MSE bounds). Of the alternatives, ONM performs best overall, echoing the findings in~\cite{madden2021bounds}, although there are no known guarantees for this algorithm.

\subsection{Performance guarantees for nine DGPs in \cite{koopman2016predicting}}\label{subsec:Koopman DGPs Monte Carlo}

\textbf{DGPs.} We take nine distributions from \cite{koopman2016predicting}, listed in Table~\ref{tab: MSE} with exact specifications in Appendix~\ref{app: DGPs}, and combine them with linear state dynamics $\vartheta_t = \phi_0 \vartheta_{t-1} + \xi_t$, initialized at $\vartheta_0 = 0$, where the state innovations $\xi_t \sim \text{i.i.d.}(0,\sigma_\xi^2)$ are drawn from a Student's $t$ distribution with six degrees of freedom. As our theoretical guarantees only require two moments, we allow fat-tailed state increments. For completeness, we also consider the Gaussian case (in Appendix~\ref{app: DGPs}). In both cases, the state $\vartheta_t$ takes values in $\mathbb{R}$, such that all distributions contain link functions (e.g., mapping $\vartheta_t$ to $\mathbb{R}_{>0}$ for volatility).

The static parameters are $\phi_{0}=0.97$ and $\sigma_\xi \in \{0.15,0.3,0.6\}$ for low-, medium-, and high-volatility settings. \cite{koopman2016predicting} investigated only the low-volatility case ($\sigma_\xi=0.15$) with Gaussian state innovations, finding that (explicit) SD filters perform relatively accurately. We take no stance on which volatility setting is more realistic; we are merely interested in investigating the implications for SD filter stability and accuracy. Are there, for example, values of $\sigma_{\xi}$ for which SD filters lose track of the underlying state or diverge?

\textbf{Filters.} We follow \cite{koopman2016predicting} in assuming that the observation densities in Table~\ref{tab: MSE} are correctly specified, meaning that our postulated density $p(\cdot|\theta_t)$ matches (the functional form of) the true density $p^0(\cdot|\vartheta_t)$. This also implies that the ISD and ESD filters are implemented using the same (correctly specified) link functions. Because the DGP is a state-space model, however, both SD filters remain misspecified.
Some densities include additional shape parameters, which are treated as unknown and estimated via maximum likelihood~\eqref{Eq_MLE}. While ESD updates~\eqref{Explicit parameter update step} are given in closed form, ISD updates~\eqref{eq: Implicit optimization problem} can be computed using standard numerical methods (for details, see Appendix~\ref{subsec:Computing the implicit-gradient update}). 

\textbf{MSE bounds.} For each density, Table~\ref{tab: MSE} reports the corresponding values of $\alpha$ and $\beta$ from Assumption~\ref{ass1}(a), where $\alpha\geq 0$ indicates log concavity and $\beta<\infty$ signifies a bounded Hessian. For the two non-concave cases, we impose the condition $\mP=\rho>\alpha^-$ to ensure that Assumption~\ref{ass1}(b) holds; in our simulations, this restriction was never binding. Assumption~\ref{ass1}(c) is automatically satisfied for all DGPs, as the state space is unbounded (i.e., $\Theta=\SR$). Given the linear state equation, Assumption~\ref{Assumption: Bounded moments} also holds. Using the specified values of $\alpha$ and $\beta$, Table~\ref{tab: MSE} indicates with check marks whether MSE bounds for the ISD and ESD filters exist (i.e., are finite). This is the case for nine and two DGPs, respectively.

\textbf{Simulation setting.} For each DGP, we simulate $1{,}000$ time series of length $10{,}000$. The ``in-sample'' period of the first $1{,}000$ observations is used for the estimation of static parameters (i.e., $\omega,\phi,\rho$ and shape parameters), while the remaining ``out-of-sample'' period is used to compute MSEs of predictions $\{\theta^{j}_{t|t-1}\}$ for $j\in \{\textnormal{im},\textnormal{ex}\}$ relative to true states $\{\vartheta_t\}$. 

{
\begin{table}[tbh]
\caption{\label{tab: MSE} Out-of-sample MSE of $\{\theta^j_{t|t-1}\}$ for $j\in\{\textnormal{im},\textnormal{ex}\}$ relative to true states $\{\vartheta_t\}$.}
\centering 
\begin{footnotesize}
\begin{threeparttable}
\begin{tabular}{l@{\hspace{0.3cm}}l@{\hspace{0.1cm}}c@{\hspace{0.1cm}}c@{\hspace{0.1cm}}c@{\hspace{0.1cm}}c@{\hspace{0.1cm}}c@{\hspace{0.1cm}}c@{\hspace{0.1cm}}c@{\hspace{0.1cm}}c@{\hspace{0.1cm}}c@{\hspace{0.1cm}}c@{\hspace{0.1cm}}c@{\hspace{0.1cm}}c}
\toprule
\multirow{2}{*}{\bf DGP} &
&   \multicolumn{2}{c}{\multirow{2}{*}{\bf Assumption~\ref{ass1}}}
& \multicolumn{2}{c}{\multirow{1}{*}{\bf MSE}}
&  \multicolumn{2}{c}{\bf Low volatility}
& \multicolumn{2}{c}{\bf Medium volatility}
& \multicolumn{2}{c}{\bf High volatility}
\\
  & & & &	\multicolumn{2}{c}{\multirow{1}{*}{\bf bound?}}   &   \multicolumn{2}{c}{($\sigma_\xi=0.15$)} & \multicolumn{2}{c}{($\sigma_\xi=0.30$)} & \multicolumn{2}{c}{($\sigma_\xi=0.60$)}
\\

\cmidrule(r{5pt}l{5pt}){1-2} \cmidrule(r{5pt}l{5pt}){3-4} \cmidrule(r{5pt}l{5pt}){5-6} \cmidrule(r{5pt}l{5pt}){7-8} \cmidrule(r{5pt}l{5pt}){9-10} \cmidrule(r{5pt}l{5pt}){11-12}

Type &	Distribution &
\multicolumn{1}{c}{$\alpha$} & \multicolumn{1}{c}{$\beta$} & \multicolumn{1}{c}{ISD} & \multicolumn{1}{c}{ESD} & \multicolumn{1}{c}{ISD} & \multicolumn{1}{c}{ESD} & \multicolumn{1}{c}{ISD} & \multicolumn{1}{c}{ESD} & \multicolumn{1}{c}{ISD} & \multicolumn{1}{c}{ESD}	\\

\cmidrule(r{5pt}l{5pt}){1-2} \cmidrule(r{5pt}l{5pt}){3-4} \cmidrule(r{5pt}l{5pt}){5-6} \cmidrule(r{5pt}l{5pt}){7-8} \cmidrule(r{5pt}l{5pt}){9-10} \cmidrule(r{5pt}l{5pt}){11-12}

Count & 	Poisson
&  \multicolumn{1}{c}{0}
& \multicolumn{1}{c}{$\infty$}
& \cmark
& \xmark
& \multicolumn{1}{c}{$0.146$}
& \multicolumn{1}{c}{$0.149\phantom{\dagger}$}
& \multicolumn{1}{c}{$0.408$}
& \multicolumn{1}{c}{$\infty$}
&  \multicolumn{1}{c}{$1.74$ }
& \multicolumn{1}{c}{$\infty$}

\\
Count &	Neg. Binomial
& 	\multicolumn{1}{c}{0}
& \multicolumn{1}{c}{$\infty$}
& \cmark
& \xmark
& \multicolumn{1}{c}{$0.159$}
& \multicolumn{1}{c}{$0.160\phantom{\dagger}$}
& \multicolumn{1}{c}{$0.430$}
& \multicolumn{1}{c}{$\infty$}
& \multicolumn{1}{c}{$1.68$ }
& \multicolumn{1}{c}{$\infty$ }
\\

Intensity  &	Exponential
& \multicolumn{1}{c}{0}
& \multicolumn{1}{c}{$\infty$}
& \cmark
& \xmark
& \multicolumn{1}{c}{$0.146$}
& \multicolumn{1}{c}{$0.150\phantom{\dagger}$}
& \multicolumn{1}{c}{$0.370$}
& \multicolumn{1}{c}{$0.899$}
&\multicolumn{1}{c}{$1.06$}
& \multicolumn{1}{c}{${\sim} 10^3$}
\\

Duration  &	Gamma
& \multicolumn{1}{c}{0}
& \multicolumn{1}{c}{$\infty$}
& \cmark
& \xmark
& \multicolumn{1}{c}{$0.157$}
& \multicolumn{1}{c}{$0.162\phantom{\dagger}$}
& \multicolumn{1}{c}{$0.481$}
& \multicolumn{1}{c}{$0.577$}
& \multicolumn{1}{c}{$1.32$}
& \multicolumn{1}{c}{${\sim} 10^6$}

\\

Duration  &	Weibull
& \multicolumn{1}{c}{0}
& \multicolumn{1}{c}{$\infty$}
& \cmark
& \xmark
& \multicolumn{1}{c}{$0.125$}
& \multicolumn{1}{c}{$0.128\phantom{\dagger}$}
& \multicolumn{1}{c}{$0.307$}
& \multicolumn{1}{c}{$0.345$}
& \multicolumn{1}{c}{$0.80$}
& \multicolumn{1}{c}{${\sim} 10^3$}

\\

Volatility  &	Gaussian
& \multicolumn{1}{c}{0}
& \multicolumn{1}{c}{$\infty$}
& \cmark
& \xmark
& \multicolumn{1}{c}{$0.193$}
& \multicolumn{1}{c}{$0.199\phantom{\dagger}$}
& \multicolumn{1}{c}{$0.506$}
& \multicolumn{1}{c}{$0.647$}
& \multicolumn{1}{c}{$1.51$}
& \multicolumn{1}{c}{${\sim} 10^7$}

\\

Volatility  &	Student's \emph{t}
& \multicolumn{1}{c}{$0$}
& \multicolumn{1}{c}{$\frac{\nu+1}{8}$}
& \cmark
& \cmark
& \multicolumn{1}{c}{$0.226$}
& \multicolumn{1}{c}{$0.226\phantom{\dagger}$}
& \multicolumn{1}{c}{$0.608$}
& \multicolumn{1}{c}{$0.615$}
& \multicolumn{1}{c}{$1.56$}
& \multicolumn{1}{c}{$1.61$}
\\

Dependence  &	Gaussian
& \multicolumn{1}{c}{$-\frac{1}{4}$}
& \multicolumn{1}{c}{$\infty$}
& \cmark
& \xmark
& \multicolumn{1}{c}{$0.237$}
& \multicolumn{1}{c}{$0.239^\dagger$}
& \multicolumn{1}{c}{$0.593$}
& \multicolumn{1}{c}{$\infty$}
& \multicolumn{1}{c}{$1.58$}
&  \multicolumn{1}{c}{$\infty$}
\\

Dependence  &	Student's \emph{t}
& \multicolumn{1}{c}{$-\frac{1}{4}$}
& \multicolumn{1}{c}{$\frac{\nu+1}{4}$}
& \cmark
& \cmark
& \multicolumn{1}{c}{$0.251$}
& \multicolumn{1}{c}{$0.251\phantom{\dagger}$}
& \multicolumn{1}{c}{$0.619$}
& \multicolumn{1}{c}{$0.624$}
& \multicolumn{1}{c}{$1.52$}
& \multicolumn{1}{c}{$1.55$}

\\
\bottomrule
 \end{tabular}
\begin{tablenotes}
\item \emph{Note}: MSE = mean squared error. ISD = implicit score driven. ESD = explicit score driven.
{$\dagger$} For the Gaussian dependence model in the low-volatility setting, the ESD filter diverged in the out-of-sample period for a single replication; for simplicity, we ignore this path and report a finite MSE.
\end{tablenotes}
\end{threeparttable}
\end{footnotesize}
\end{table}
}

\textbf{Findings.} Table~\ref{tab: MSE} shows that the ISD filter outperforms the ESD filter across all DGPs and volatility settings. In the low-volatility setting (as in \citealp{koopman2016predicting}), the differences in performance are marginal. Although the ESD filter diverged during the out-of-sample period for one replication of the Gaussian dependence model, for simplicity we ignore this path and report a finite MSE. This divergence may explain why \cite{koopman2016predicting} further reduced the state volatility for both dependence DGPs to $\sigma_\xi=0.10$.

In the medium-volatility setting, the performance gap becomes more pronounced. For three DGPs, the ESD filter diverges during the out-of-sample period, with a substantial proportion of paths affected (e.g., ${\sim} 10\%$ of paths for the Gaussian dependence model). We report the corresponding MSEs as infinite.
For two DGPs with $\beta<\infty$, the difference between the ISD and ESD filters remains marginal.

In the high-volatility setting, the potential instability of the ESD filter becomes evident across all models except those with $\beta<\infty$. Even when it does not strictly diverge (i.e., to infinity), the resulting MSEs can still be extremely large (e.g., ${\sim}10^7$). In contrast, the MSE of the ISD filter never exceeds $1.74$ (for the Poisson distribution).

In sum, ESD filters remain competitive across all volatility scenarios only when an MSE bound is available; otherwise, they may eventually diverge, whether in low-, \mbox{medium-,} or high-volatility settings. For the Gaussian dependence model, even the low-volatility setting is problematic, although the probability of divergence appears to be small. The value of $\sigma_\xi$ at which a substantial proportion of paths diverge is model dependent. This possibility of divergence for misspecified explicit SD filters was not identified by \cite{koopman2016predicting}, likely because their analysis focused on empirical performance, without theoretical guarantees, under low-variance Gaussian state increments.

Although our focus has been on ESD filters with constant learning rates, these stability issues do not disappear, and may even worsen, when using time-varying learning rates. In the next subsection, we demonstrate that such instability persists for the Poisson count model across all standard learning-rate approaches.

\subsection{Poisson count model with various link and scaling functions}
\label{subsec:Dynamic Poisson distribution with misspecified link function.}

\textbf{DGP.} We consider count observations $y_t\in \SN$ generated by a dynamic Poisson distribution $ p^{0}(y_t|\mu_{t}) = {\mu_t^{y_t} \exp(-\mu_{t})}/{y_t!}$ with mean $\mu_t=\exp(\vartheta_t)$, where $\mu_t>0$ is strictly positive due to the exponential link. The latent process $\{\vartheta_t\}$ follows linear dynamics $\vartheta_{t} = 0.98 \vartheta_{t-1} + \xi_t$ with Gaussian increments $\xi_t \sim \text{i.i.d. } \mathrm{N}(0, \sigma_{\xi}^2)$.
This model, along with slight variations, has been extensively studied in the literature, with notable applications including the modeling of {U.K.} van-driver deaths \citep{harvey1989time, durbin1997monte, durbin2000time}.

\textbf{ESD filter.} Explicit SD filters for the Poisson count model
have been considered in  \citet[sec.~2.3]{davis2003observation}, \citet[sec.~5.2]{gorgi2018integer} and \citet[sec.~4.3]{gorgi2023optimality}. In line with this literature, we assume that the Poisson distribution and the exponential link $\mu_t=\exp(\vartheta_t)$ are known to the researcher, such that they can be used to calculate the score, $y_t-\exp(\theta)$, and Fisher's information quantity, $\exp(\theta)$. We follow the literature in considering several \emph{scaling} options by multiplying the score by the inverse Fisher information to the power $\zeta \in \{0,1/2,1\}$. However, since the resulting scaled score $\exp(-\zeta\theta)\, [y_t-\exp(\theta)]$ fails to be Lipschitz continuous in $\theta \in \SR$ for any scaling $\zeta\in \{0,1/2,1\}$, neither stability nor performance guarantees can be established for the ESD filter.

\textbf{ISD filter.} For the ISD filter, we consider only a static learning rate. While the Poisson distribution is again assumed to be known, we explore scenarios where the relationship between $\vartheta_t$ and $\mu_t$ is unknown. Specifically, we consider two cases in which the researcher postulates either (i) an exponential link $\mu^{\textnormal{im}}_{t|t}=\exp(\theta^{\textnormal{im}}_{t|t})$, which is correct, or (ii) a quadratic link $\mu^{\textnormal{im}}_{t|t}={(\theta^{\textnormal{im}}_{t|t})}^2$, which is misspecified.

The parameter space $\Theta$ is $\mathbb{R}$ under the exponential link and $\mathbb{R}_{\geq 0}$ under the quadratic link (this ensures monotonicity). For the quadratic link, simple parameter constraints (i.e.,\ positivity of $\omega$ and $\phi$) ensure that the prediction step~\eqref{Parameter prediction step} maps $\SR_{\geq 0}$ to itself. This capacity to handle bounded parameter spaces is a distinctive feature of the ISD filter.

Under the two link functions, the logarithmic Poisson distribution is either (i) concave in $\theta\in \SR$ (i.e., $\alpha=0$), or (ii) strongly concave in $\theta\in \SR_{\geq 0}$ (with $\alpha=2$). In both cases, Assumptions~\ref{ass1}(a,b) are satisfied.
For the exponential link, the ISD update can be computed using Newton's optimization method (see Appendix~\ref{subsec:Computing the implicit-gradient update}). For the quadratic link, a closed-form solution is available (see Appendix~\ref{subsec:Estimated learning rates}). Notably, if the prediction lies in the interior of $\Theta=\SR_{\geq 0}$, then so does the update. Even when both the prediction and the update lie on the boundary (which may occur when $y_t=0$), the first-order condition~\eqref{Implicit parameter update step} is still satisfied. Thus, while Assumption~\ref{ass1}(c) is not strictly met, it can be effectively circumvented.
Finally, Assumption~\ref{Assumption: Bounded moments} holds in both cases (see Appendix~\ref{subsec:Estimated learning rates}).

In sum, performance guarantees are available for both versions of the ISD filter. When the correct (exponential) link function is used, these bounds apply to the tracking error relative to the true states $\{\vartheta_t\}$. In contrast, when the incorrect (quadratic) link is employed, the bounds pertain to the tracking error relative to the \emph{pseudo}-true states $\{\theta^\star_t\}=\{ \exp(\vartheta_t/2) \}$. In the latter case, the bound includes a free parameter, for which we substitute an analytic expression that minimizes the bound (see Appendix~\ref{sec:Proof of Corollary 2.4}). The resulting minimized MSE bound further depends on $q^2=\sup_{t} \mathbb{E}[\|\theta_t^{\star}-\theta_{t-1}^{\star}\|^2] <\infty$ and $s_{\omega}^2=\sup_{t} \mathbb{E} [\|\theta_t^{\star}-\omega\|^2]<\infty$. For the purpose of computing the bound, both quantities are assumed to be known; this is standard practice in the stochastic-optimization literature (e.g.,~\citealp[p.\ 8]{nesterov2018lectures}).

\textbf{Simulation setup.} We vary the state variation $\sigma_{\xi}$ over the range $(0,0.5]$. For each value of $\sigma_{\xi}$, we simulate $500$ time series of length $10{,}000$. The ``in-sample'' period, consisting of the first $1{,}000$ observations, is used to estimate the static parameters, while the remaining ``out-of-sample'' period is used to assess the filtering performance.

\textbf{Parameter estimation.} Filters with exponential link functions are implemented with Assumption~\ref{Assumption: Correct specification}, meaning the true parameters $\omega_{0}=0$ and $\phi_{0}=0.98$ are used in the prediction step~\eqref{Parameter prediction step} (i.e., we set $\omega^j=\omega_0$ and $\phi^j=\phi_0$ for $j\in \{\textnormal{im},\textnormal{ex}\}$) and the learning rates $\eta^j$ for $j \in \{\textnormal{im}, \textnormal{ex}\}$ are estimated using~\eqref{Eq_MLE}. For the ISD filter with the quadratic link, all static parameters ($\omega^\textnormal{im}$, $\phi^\textnormal{im}$, $\eta^\textnormal{im}$) are estimated using~\eqref{Eq_MLE}.

\begin{figure}[tb]
    \centering
    \begin{subfigure}{.47\textwidth}
        \centering
        \includegraphics[width=\linewidth]{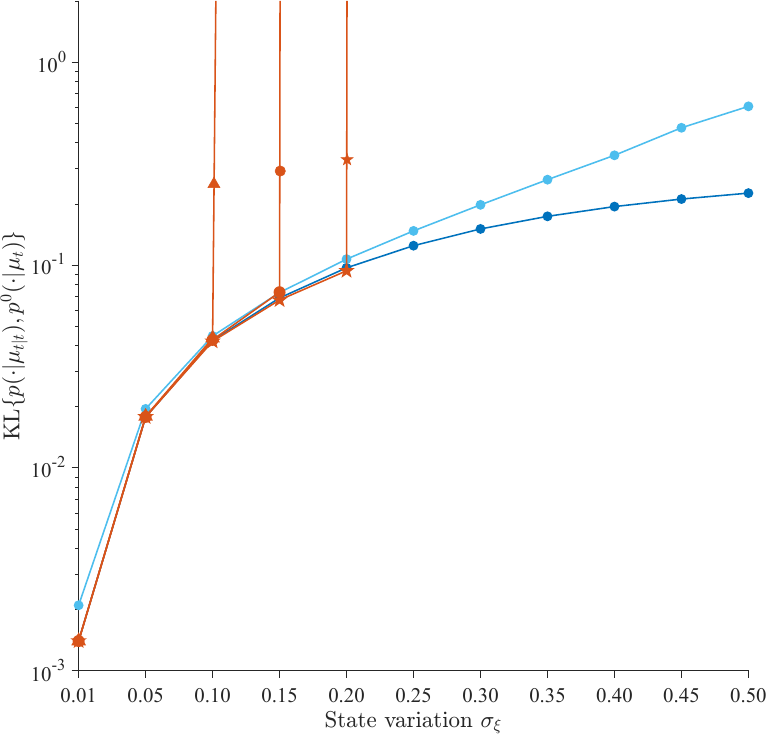}
        \label{fig:KL divergencev1}
    \end{subfigure}
    \hfill
    \begin{subfigure}{.47\textwidth}
        \centering
        \includegraphics[width=\linewidth]{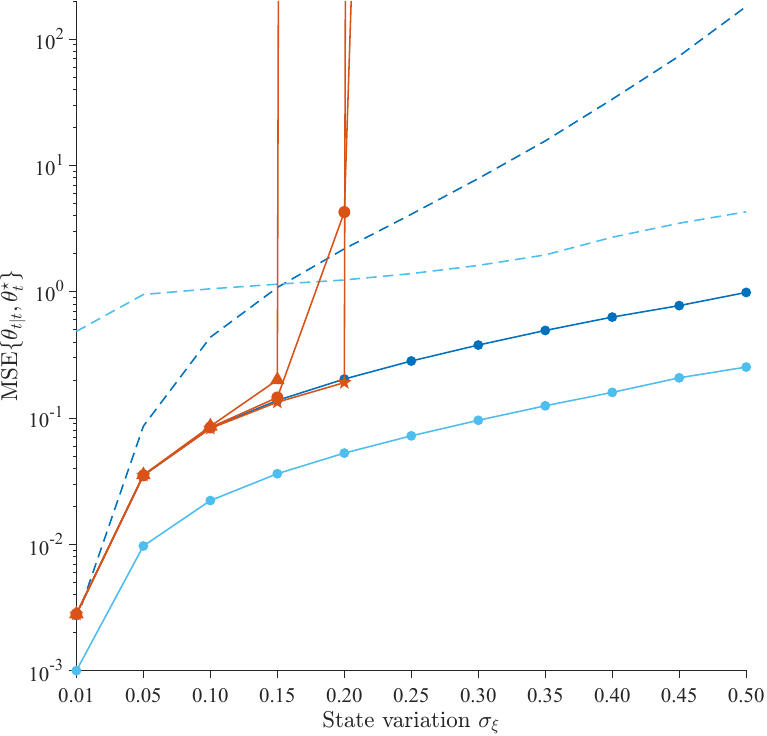}
        \label{fig:Out-of-sample MSE, misspecificationv1.}
    \end{subfigure}

    \vspace{-2mm}

    \begin{subfigure}{1\textwidth}
        \centering
        \includegraphics[width=\textwidth]{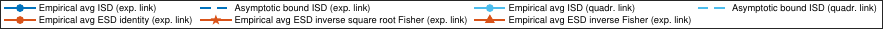}
        \label{fig:legend_Poissonv1}
    \end{subfigure}

    \vspace{-8mm}

    \caption{Plots of the out-of-sample empirical errors and guaranteed bounds for tracking (the logarithm of the rate of) a dynamic Poisson distribution (i.e., the true state  $\{\vartheta_{t}\}$) with respect to its variation $\sigma_{\xi}$. The left-hand plot shows the Kullback-Leibler (KL) divergence between the postulated and true densities $p(\cdot|\mu_{t|t})$ and $p^{0}(\cdot|\mu_{t})$, where $\mu_{t|t}$ and $\mu_{t}$ denote the filtered and true rates, respectively. The right-hand plot shows the MSE between the filtered state $\theta_{t|t}$ and the pseudo-true state $\theta_t^{\star}$ (= $\vartheta_{t}$ if Assumption~\ref{Assumption: Correct specification} holds). Empirical averages are computed over $500$ trials.} 
    \label{fig:Poisson figuresv1}
\end{figure}

\textbf{Findings.} The left-hand panel of Fig.~\ref{fig:Poisson figuresv1} shows the out-of-sample Kullback-Leibler (KL) divergence between the filtered distribution $p(\cdot|\mu_{t|t})$ and the true distribution $p^{0}(\cdot|\mu_{t})$, given by $\text{KL}\{p(\cdot|\mu_{t|t}),p^{0}(\cdot|\mu_{t})\}=\mu_{t}\log(\mu_{t}/\mu_{t|t}) + \mu_{t|t} - \mu_{t}$. Because each filter assumes a Poisson distribution, the KL divergence depends solely on the discrepancy between the filtered mean $\mu^{j}_{t|t}$ for $j\in \{\textnormal{im},\textnormal{ex}\}$ and the true mean $\mu_t$. At low levels of state volatility, all filters achieve comparable KL divergence values. However, as state variability increases, all three ESD filters become unstable. Both ISD filters remain stable across all state variations, even as their estimated learning rates are considerably higher (see Appendix~\ref{subsec:Estimated learning rates}).

The right-hand panel of Fig.~\ref{fig:Poisson figuresv1} shows the out-of-sample MSEs of the filtered states $\{\theta^j_{t|t}\}$, evaluated against the true states $\{\vartheta_t\}$ when using the exponential link, or against the pseudo-true states $\{\theta^\star_t=\exp(\vartheta_t/2)\}$ when using the quadratic link.
Meaningful comparisons can only be made between filters that use the same link function.
As the state variability increases, all three ESD filters become unstable. In contrast, both ISD filters remain stable across all state variations, consistent with our theoretical results.

\vspace{-0.2cm}
\section{Conclusion}
\label{sec:conc}

This paper established theoretical guarantees for the stability and filtering performance of multivariate explicit and implicit score-driven filters under possible misspecification. We derived practical sufficient conditions for exponential stability of the filter, uniformly over all data sequences, and combined these conditions with mild assumptions on the DGP to obtain finite-sample and asymptotic MSE bounds between filtered and pseudo-true paths. These bounds can be minimized analytically, facilitating the tuning of static \mbox{(hyper-)}parameters such as the learning rate. Monte Carlo studies validated the theory and highlighted the advantages of implicit filters. In a high-dimensional linear model, the proposed bounds sharpened existing results by up to three orders of magnitude. Consistent with the theory, misspecified explicit filters may diverge when finite MSE bounds are unavailable, whereas implicit filters remain stable and accurately track the pseudo-true path.


{\section*{Acknowledgments}\label{Acknowledgments}

We thank Mariia Artemova, Janneke van Brummelen, Timo Dimitriadis, Andrew Harvey, Bernd Heidergott, Frank Kleibergen, Stan Koobs, Yicong Lin, André Lucas, Sam van Meer, Ramon de Punder, Alberto Quaini, Neil Shephard, Bernhard van der Sluis, Pierluigi Vallarino, and Phyllis Wan for their comments and suggestions. We also thank the participants of the 12th Econometric Internal PhD Conference Rotterdam 2024, Tinbergen Institute PhD seminar 2024, Oxford Dynamic Econometrics Conference 2024, New York Camp Econometrics 2024, Netherlands Econometric Study Group 2024, ISEO Summer School 2024, International Association for Applied Econometrics 2024, European Economic Association--Econometric Society European Meeting 2024, Brown Bag seminar VU Amsterdam 2025, Society for Financial Econometrics 2025, National Bureau of Economic Research--National Science Foundation Time Series Conference 2025, Harvard Statistics PhD seminar 2025, and Janeway Institute Workshop on Recent Developments in Time Series Modelling, Cambridge, 2026.}

\putbib[paper-ref]
\end{bibunit}

\clearpage
\appendix

\begin{center}
\Large \textbf{ Online supplement to:  \\
[6pt]
``Gradient-based filtering under misspecification: 
\\
Stability and error bounds''}\\[3pt]
\normalsize
S. W. Donker van Heel, R.-J. Lange, B. van Os, and D. van Dijk
\\
\end{center}

\pagenumbering{arabic}
\renewcommand*{\thepage}{S\arabic{page}}
\setcounter{page}{1}

\setcounter{equation}{0}
\renewcommand{\theequation}{\Alph{section}.\arabic{equation}}
\setcounter{table}{0}
\renewcommand{\thetable}{\Alph{section}.\arabic{table}}
\setcounter{figure}{0}
\renewcommand{\thefigure}{\Alph{section}.\arabic{figure}}
\renewcommand{\thesubsection}{\Alph{section}.\arabic{subsection}}

\phantomsection
\addcontentsline{toc}{chapter}{Online Supplement}

\vspace{-1cm}
\etocsetnexttocdepth{2}   
\localtableofcontents     


\begin{bibunit}[chicago]

\section{Proofs of main results}
\label{Appendix A. Proofs}

\subsection{Preliminaries}
\label{sec: appendix prelim}

Throughout our proofs, we make extensive use of the squared weighted norm $\|\vx\|^2_{\mW} := \vx^{\top}\mW\vx$ for any $\vx \in \mathbb{R}^k$ and any positive-definite matrix $\mW \in \mathbb{R}^{k\times k}$. With slight abuse of notation, we also use the shorthand $\|\vx\|^2_{\mW}$ for a matrix  $\mW$ that is not positive definite; in this case, the expression remains well defined but strictly speaking does not constitute a norm.
We also use the $\mW$-weighted (induced) matrix norm, defined as
$$
\|\mA\|_{\mW}^2:= \underset{\vy \in \mathbb{R}^{k} \setminus \{\vzeros_k\}}{\sup} \frac{\|\mA\vy\|_{\mW}^2}{\|\vy\|_{\mW}^2} \qquad \geq\quad    \frac{\|\mA\vx\|_{\mW}^2}{\|\vx\|_{\mW}^2}, \quad \forall \vx \in \mathbb{R}^k \setminus \{\vzeros_k\}.
$$
The inequality demonstrates that the usual submultiplicative property holds, i.e.,
\begin{equation}
\|\mA\vx\|_{\mW}^2 \;\leq\; \|\mA\|_{\mW}^2 \, \|\vx\|_{\mW}^2,\qquad \forall \vx \in \mathbb{R}^k,\label{eq: submultiplicativity weighted matrrix norm}
\end{equation}
for any matrix $\mA \in \mathbb{R}^{k \times k}$ and any positive-definite matrix $\mW \in \mathbb{R}^{k \times k}$.

The notation $\mA^{1/2}$ for positive-semidefinite $\mA \in \mathbb{R}^{k\times k}$ refers to the symmetric matrix square root that can be obtained via eigendecomposition. Specifically, let $\mV\in \mathbb{R}^{k\times k}$ denote the matrix of eigenvectors of $\mA$ and let $\mD\in \mathbb{R}^{k\times k}$ denote the diagonal matrix of eigenvalues in corresponding order such that $\mA = \mV \mD \mV^{\top}$; then, $\mA^{1/2}:=  \mV \tilde{\mD} \mV^{\top}$, where $\tilde{\mD}\in \mathbb{R}^{k\times k}$ is a diagonal matrix containing the principal square root eigenvalues such that $\tilde{\mD}\tilde{\mD} =
\mD$. From this definition and the orthogonality of $\mV$ it is immediately apparent that $\mA^{1/2}$ is symmetric and that $\mA^{1/2}\mA^{1/2} = \mV \tilde{\mD} \mV^{\top}\mV \tilde{\mD} \mV^{\top} = \mV \tilde{\mD} \mI_k \tilde{\mD} \mV^{\top}=  \mV \mD \mV^{\top} = \mA$. It can also be established that $\mA^{1/2}$ is the unique symmetric matrix such that $\mA^{1/2}\mA^{1/2} = \mA$; see \citet[Thm{.} 7{.}2{.}6a]{horn2012matrix}.

\subsection{Lemma~\ref{lem3}}
\label{A: lem3}
The next useful eigenvalue property\footnote{We are grateful to Phyllis Wan for suggesting to clarify the proof using this lemma.} is used  in the proof of Lemma~\ref{lem1}, which in turn is used in the proof of Theorem~\ref{thrm:invertibility}.

\begin{lemma}\label{lem3} Let $\mA,\mB\in \SR^{k \times k}$ be symmetric, while $\mB$ is positive definite. Then
\begin{equation}
\min \left\{\frac{\lambda_{\min }(\mA)}{\lambda_{\min }(\mB)},\frac{\lambda_{\min }(\mA)}{\lambda_{\max }(\mB)}\right\} \; \mI_k \; \preceq\;  \mB^{-\frac{1}{2}} \mA \mB^{-\frac{1}{2}} \; \preceq\;
\max \left\{\frac{\lambda_{\max }(\mA)}{\lambda_{\min }(\mB)},\frac{\lambda_{\max }(\mA)}{\lambda_{\max }(\mB)}\right\} \; \mI_k.
\end{equation}
\end{lemma}

\begin{proof} First, we prove the inequality relating to the minimum eigenvalue, i.e.,
\begin{equation*}
\min \left\{\frac{\lambda_{\min }(\mA)}{\lambda_{\min }(\mB)},\frac{\lambda_{\min }(\mA)}{\lambda_{\max }(\mB)}\right\} \;\leq\; \lambda_{\min }\left(\mB^{-\frac{1}{2}} \mA \mB^{-\frac{1}{2}}\right) .
\end{equation*}
To see this, we write the smallest eigenvalue as
\begin{equation*}
\lambda_{\min }\left(\mB^{-\frac{1}{2}} \mA \mB^{-\frac{1}{2}}\right)=\min_{\vx
\neq \vzeros_k} \frac{\vx^{\top} \mB^{-\frac{1}{2}} \mA \mB^{-\frac{1}{2}} \vx}{\vx^{\top}  \vx}=\min_{\vy\neq \vzeros_k} \frac{\vy^{\top} \mA \vy}{\vy^{\top} \mB \vy},
\end{equation*}
where we used $\vy=\mB^{-1/2}\vx$. Depending on the sign of $\lambda_{\min }(\mA)$, there are two cases:
\begin{align}
\lambda_{\min }(\mA)\geq 0\quad \Rightarrow \quad \min_{\vy\neq \vzeros_k} \frac{\vy^{\top} \mA \vy}{\vy^{\top} \mB \vy} &\geq \frac{\lambda_{\min }(\mA)}{\lambda_{\max }(\mB)} ,
\notag
\\
\lambda_{\min }(\mA)<0\quad \Rightarrow \quad \min_{\vy\neq \vzeros_k} \frac{\vy^{\top} \mA \vy}{\vy^{\top} \mB \vy} &= \min _{\vy: \vy^{\top} \mA \vy<0} \frac{\vy^{\top} \mA \vy}{\vy^{\top} \mB \vy} \geq \min_{\vy\neq \vzeros_k} \frac{\vy^{\top} \mA \vy}{\lambda_{\min }(\mB) \vy^{\top} \vy } =\frac{\lambda_{\min }(\mA)}{\lambda_{\min }(\mB)}.
\notag
\end{align}
In both cases, the numerator is the smallest eigenvalue of $\mA$. When the numerator is positive (negative), the denominator is the largest (smallest) eigenvalue of $\mB$. Hence the minimum of both possible fractions is the relevant one.

Second, we prove the inequality relating to the maximum eigenvalue, i.e.,
\begin{equation*}
 \lambda_{\max }\left(\mB^{-\frac{1}{2}} \mA \mB^{-\frac{1}{2}}\right)\; \leq \; \max \left\{\frac{\lambda_{\max }(\mA)}{\lambda_{\min }(\mB)},\frac{\lambda_{\max }(\mA)}{\lambda_{\max }(\mB)}\right\}.
\end{equation*}
To see this, we write the largest eigenvalue as
\begin{equation*}
\lambda_{\max }\left(\mB^{-\frac{1}{2}} \mA \mB^{-\frac{1}{2}}\right)=\max_{\vx
\neq \vzeros_k} \frac{\vx^{\top} \mB^{-\frac{1}{2}} \mA \mB^{-\frac{1}{2}} \vx}{\vx^{\top}  \vx}=\max_{\vy\neq \vzeros_k} \frac{\vy^{\top} \mA \vy}{\vy^{\top} \mB \vy},
\end{equation*}
where we used $\vy=\mB^{-1/2}\vx$.  Depending on the sign of $\lambda_{\max }(\mA)$, there are two cases:
\begin{align}
\lambda_{\max }(\mA)\geq 0\quad \Rightarrow \quad \max_{\vy\neq \vzeros_k} \frac{\vy^{\top} \mA \vy}{\vy^{\top} \mB \vy} &\leq \frac{\lambda_{\max }(\mA)}{\lambda_{\min }(\mB)} ,
\notag
\\
\lambda_{\max }(\mA)<0\quad \Rightarrow \quad \max_{\vy \neq \vzeros_k} \frac{\vy^{\top} \mA \vy}{\vy^{\top} \mB \vy} &= \max_{\vy: \vy^{\top} \mA \vy<0} \frac{\vy^{\top} \mA \vy}{\vy^{\top} \mB \vy} \leq \max_{\vy\neq \vzeros_k} \frac{\vy^{\top} \mA \vy}{\lambda_{\max }(\mB) \vy^{\top} \vy } =\frac{\lambda_{\max }(\mA)}{\lambda_{\max }(\mB)}.
\notag
\end{align}
In both cases, the numerator is the largest eigenvalue of $\mA$. When the numerator is positive (negative), the denominator is the smallest (largest) eigenvalue of $\mB$. Hence the maximum of both possible fractions is the relevant one.
\end{proof}

\subsection{Lemma~\ref{lem1}}
\label{ap: Proof of Lemma 1}

Lemma~\ref{lem1} relies on Lemma~\ref{lem3} in the previous section. Lemma~\ref{lem1} will be used in the proofs of Theorems~\ref{thrm:invertibility} and~\ref{Th: (Non-)asymptotic MSE bounds}.

\begin{lemma}[Update stability]
\label{lem1} Let Assumption~\ref{ass1} hold. Fix $t\geq 1$. Consider two predictions $\vtheta^j_{t|t-1}\in \mathbf{\Theta}$ with $j\in \{\textnormal{im},\textnormal{ex}\}$ and associated ISD and ESD updates~\eqref{Implicit parameter update step} and~\eqref{Explicit parameter update step}, yielding $\vtheta^\textnormal{im}_{t \mid t}$ and $\vtheta^\textnormal{ex}_{t \mid t}$, respectively. Then, uniformly in $\vtheta^{\textnormal{im}}_{t|t-1},\vtheta^{\textnormal{ex}}_{t|t-1}\in \mathbf{\Theta}$
and for all $\vy_t$,
\allowdisplaybreaks
\begin{align}
       \left\|\frac{\dd \vtheta^\textnormal{im}_{t \mid t}}{\dd {\vtheta^{\textnormal{im}}_{t \mid t-1}}^{\!\!\!\!\!\!\top}}\right\|_{\mP} &\leq 1 - \frac{\alpha^+}{\lambda_{\max}(\mP) + \alpha^+} + \frac{\alpha^{-}}{\lambda_{\min}(\mP) - \alpha^{-}},
       \label{eq:update stability ISD}
       \\
        \left\|\frac{\dd \vtheta^\textnormal{ex}_{t \mid t}}{\dd {\vtheta^\textnormal{ex}_{t \mid t-1}}^{\!\!\!\!\!\!\top}}\right\|_{\mP} & \leq 1 - \min\left\{ \frac{\alpha^+}{ \lambda_{\max}(\mP)}- \frac{\alpha^-}{ \lambda_{\min}(\mP)},2-\frac{\beta}{\lambda_{\min}(\mP)}  \right\}.
           \label{eq:update stability ESD}
\end{align}

\end{lemma}

\textbf{Discussion of Lemma~\ref{lem1}}. The lemma considers the sensitivity of the updated parameter with respect to the predicted parameter, as measured by the Jacobian matrix, in the $\mP$-weighted matrix norm. We call the updating step stable (in the $\mP$-norm) if the weighted norm does not exceed unity. By equation~\eqref{eq:update stability ISD}, for the implicit update to be stable for all observations $\vy_t$ and all predictions, it is both necessary and sufficient to have $\alpha\geq 0$, in which case the right-hand side does not exceed unity. In particular, the update is always stable if $\alpha>0$, while it may be expansive if $\alpha<0$. The largest possible expansion is bounded due to Assumption~\ref{ass1}(b) (i.e., $\lambda_{\min}(\mP)>\alpha^-$). 

As inequality~\eqref{eq:update stability ESD} shows, ESD update stability additionally requires $\beta<\infty$; otherwise the right-hand side would be unbounded. While $\alpha\geq 0$ remains necessary, it is no longer sufficient: if $\beta$ is large, such that the second argument of $\min\{\cdot,\cdot\}$ dominates, we also need $2-\beta/\lambda_{\min}(\mP)$ to be positive. Using $1/\lambda_{\min}(\mP)=\lambda_{\max}(\mH)$, this additional requirement can be concisely written as $\lambda_{\max}(\mH)\leq 2/\beta$.


\begin{proof} \textbf{Update stability for ISD filters.} For clarity of exposition, we omit the superscript on $\vtheta_{t|t-1}$. Differentiating the ISD update step~\eqref{Implicit parameter update step}, the Jacobian of the implicit update with respect to the prediction $\vtheta_{t|t-1}$ is
\begin{equation*}
\frac{\dd \vtheta^{\textnormal{im}}_{t \mid t}}{\dd \vtheta_{t \mid t-1}^{\top}} = \mI_k + \mP^{-1} \cH^{\textnormal{im}}_{t} \frac{\dd \vtheta^{\textnormal{im}}_{t \mid t}}{\dd \vtheta_{t \mid t-1}^{\top}},
\end{equation*}
where $\cH^{\textnormal{im}}_{t}:=\nabla^2\ell(\vy_t \mid \vtheta^{\textnormal{im}}_{t \mid t})$ is the Hessian matrix evaluated at the ISD update. Pre-multiply by the penalty matrix $\mP$ to get
\begin{equation*}
\mP \frac{\dd \vtheta^{\textnormal{im}}_{t \mid t}}{\dd \vtheta_{t \mid t-1}^{\top}} = \mP + \cH^{\textnormal{im}}_{t} \frac{\dd \vtheta^{\textnormal{im}}_{t \mid t}}{\dd \vtheta_{t \mid t-1}^{ \top}},
\end{equation*}
and  solve for the Jacobian to obtain
\begin{equation}
\frac{\dd \vtheta^{\textnormal{im}}_{t \mid t}}{\dd \vtheta_{t \mid t-1}^{\top}}  = (\mP -  \cH^{\textnormal{im}}_{t})^{-1}\mP = (\mI_k - \mP^{-1}\cH^{\textnormal{im}}_{t})^{-1}.
\label{eq: Jac}
\end{equation}
Here, the inverse of $\mP-\cH^{\textnormal{im}}_{t}$ exists as $\mP-\cH^{\textnormal{im}}_{t}\succ \mZeros_k $ by Assumption~\ref{ass1}(b), i.e., the penalty exceeds any possible non-concavity of the log density such that the regularized objective in optimization~\eqref{eq: Implicit optimization problem} is strictly concave. The second equality follows from $(\mP-\cH^{\textnormal{im}}_{t})^{-1}\mP= [\mP^{-1}(\mP-\cH^{\textnormal{im}}_{t})]^{-1}=
(\mI_k-\mP^{-1}\cH^{\textnormal{im}}_{t})^{-1}$. Next, we investigate
\allowdisplaybreaks
\begin{align}
    \smash{\left\|\frac{\dd \vtheta^{\textnormal{im}}_{t \mid t}}{\dd \vtheta_{t \mid t-1}^\top}\right\|_{\mP}} &= \left\|(\mI_k - \mP^{-1}\cH^{\textnormal{im}}_{t})^{-1} \right\|_{\mP} \quad \text{\footnotesize by~\eqref{eq: Jac}}
    \notag
    \\
    &= \left\|\mP^{1/2} \, (\mI_k - \mP^{-1}\cH^{\textnormal{im}}_{t})^{-1}\,\mP^{-1/2}\right\| \quad \text{\footnotesize as $\|\mA\|_{\mB} = \|\mB^{1/2}\mA \mB^{-1/2}\|$}
    \notag \\
    &= \left\|(\mI_k +  \mP^{-1/2}(-\cH^{\textnormal{im}}_t)\mP^{-1/2})^{-1}\right\|
    \quad \text{\footnotesize as $\mB^{1/2} \mA^{-1} \mB^{-1/2} =(\mB^{1/2} \mA \mB^{-1/2})^{-1}$}
    \notag\\
    &= \lambda_{\max }\left(\,[\mI_k +  \mP^{-1/2}(-\cH^{\textnormal{im}}_t)\mP^{-1/2}]^{-1}\,   \right)
      \quad \text{\footnotesize as $\|\mB\|=\lambda_{\max}(\mB)$ if $\mB$ is p.d.}
      \notag\\
     &=1/ \lambda_{\min }(\mI_k +  \mP^{-1/2}(-\cH^{\textnormal{im}}_t)\mP^{-1/2} )
     \notag\\
     &= [1+ \lambda_{\min }(\mP^{-1/2}(-\cH^{\textnormal{im}}_t)\mP^{-1/2}   ) ]^{-1}.
   \label{eq: intermediate}
     \end{align}
The fourth line uses that $\mI_k -  \mP^{-1/2}\cH^{\textnormal{im}}_t\mP^{-1/2}$ is (symmetric and) positive definite, which follows from the symmetry of $\mP$ along with $\mP - \cH_t^{\textnormal{im}} \succ \mZeros_k$ as implied by Assumption~\ref{ass1}(b). The last two lines use $\lambda_{\max}(\mA^{-1}) = 1/\lambda_{\min}(\mA)$ and $\lambda_{\min}(\mI_k + \mA) = 1 + \lambda_{\min}(\mA)$ for an arbitrary matrix $\mA \in \mathbb{R}^{k \times k}$. To upper bound the last quantity, we must lower bound $\lambda_{\min }(\mP^{-1/2}(-\cH^{\textnormal{im}}_t)\mP^{-1/2}   )>-1$, as $1/(1+x)$ is decreasing in $x$ for $x>-1$. To this end we use Lemma~\ref{lem3} in Appendix~\ref{A: lem3} to yield
     \begin{align}
 \smash{\left\|\frac{\dd \vtheta^{\textnormal{im}}_{t \mid t}}{\dd \vtheta_{t \mid t-1}^\top}\right\|_{\mP}}     &\leq \left[1+\min\left\{\frac{\lambda_{\min }(-\cH^{\textnormal{im}}_t)}{\lambda_{\min}(\mP)},\frac{\lambda_{\min }(-\cH^{\textnormal{im}}_t)}{\lambda_{\max}(\mP)}\right\}\right]^{-1}
     \quad \text{\footnotesize by \eqref{eq: intermediate} and Lemma~\ref{lem3}}
     \notag
          \\
    &   \leq \left[1+\min\left\{\frac{\alpha}{\lambda_{\min}(\mP)},\frac{\alpha}{\lambda_{\max}(\mP)}\right\}\right]^{-1}
    \quad \text{\footnotesize as $\alpha\leq \lambda_{\min }(-\cH^{\textnormal{im}}_t)$ by Assumption~\ref{ass1}(a)}
     \notag
     \\
     &= \left[1+\frac{\alpha^+}{\lambda_{\max}(\mP)}-\frac{\alpha^-}{\lambda_{\min}(\mP)}\right]^{-1}\quad \text{\footnotesize as $\alpha=\alpha^+ - \alpha^-$ and $\lambda_{\min}(\mP)>0$}
     \notag
     \\
     &=1-\frac{\alpha^{+}}{\lambda_{\max }(\mP)+\alpha^+}+\frac{\alpha^{-}}{\lambda_{\min }(\mP)-\alpha^-}, \quad \text{\footnotesize by algebra}\label{eq: ISD Jacobian equation}
\end{align}
where $\alpha^+:=\max\{0,\alpha\}$ and $\alpha^-:=\max\{0,-\alpha\}$. In sum, in the concave case (i.e., $\alpha\geq 0$), we obtain the standard contraction coefficient $1-\alpha^+/(\lambda_{\max}(\mP)+\alpha^+)=\lambda_{\max}(\mP)/(\lambda_{\max}(\mP)+\alpha^+)\leq 1$ (see also~\citealp[p.\ 9]{lange2024robust}). In the non-concave case  (i.e., $\alpha< 0$), we obtain the maximal expansion coefficient $1+\alpha^-/(\lambda_{\min}(\mP)-\alpha^-)=\lambda_{\min}(\mP)/(\lambda_{\min}(\mP)-\alpha^-)$.  The strict concavity of the regularized objective (i.e., Assumption~\ref{ass1}(b)) implies $\lambda_{\max}(\mP) \geq \lambda_{\min}(\mP)>\alpha^-$, such that the denominator is strictly positive in either case. 

\textbf{Update stability for ESD filters.} For the clarity of exposition, we omit the superscript on $\vtheta_{t|t-1}$. Differentiating the ESD update step~\eqref{Explicit parameter update step}, the Jacobian of the update with respect to the prediction is
\begin{equation*}
\frac{\dd \vtheta^{\textnormal{ex}}_{t \mid t}}{\dd \vtheta_{t \mid t-1}^{\top}} = \mI_k + \mP^{-1} \cH^{\textnormal{ex}}_{t} ,
\end{equation*}
where $\cH^{\textnormal{ex}}_{t}:=\nabla^2\ell(\vy_t \mid \vtheta^{\textnormal{ex}}_{t \mid t-1})$ is the Hessian matrix evaluated at the ESD prediction. Then
\allowdisplaybreaks
\begin{align}
   \left\| \frac{\dd \vtheta^{\textnormal{ex}}_{t \mid t}}{\dd \vtheta_{t \mid t-1}^{\top}}\right\|_{\mP} &= \left\|\mI_k + \mP^{-1} \cH^{\textnormal{ex}}_t\right\|_{\mP} = \left\|\mP^{1/2}(\mI_k + \mP^{-1} \cH^{\textnormal{ex}}_t)\mP^{-1/2}\right\| \notag\\
    &=\left\|\mI_k + \mP^{-1/2} \cH^{\textnormal{ex}}_t\mP^{-1/2}\right\|\notag\\
    &=\max \left\{ \lambda_{\max}(\mI_k + \mP^{-1/2} \cH^{\textnormal{ex}}_t\mP^{-1/2}),-\lambda_{\min}(\mI_k + \mP^{-1/2} \cH^{\textnormal{ex}}_t\mP^{-1/2})\right\},
    \label{eq: eigenvalue bounds}
\end{align}
where the first line uses the definition of the induced matrix norm $\|\mA\|_{\mW} = \|\mW^{1/2}\mA \mW^{-1/2}\|$ for any symmetric positive-definite matrix $\mW$ and square matrix $\mA$ of equal dimension, and the second line uses $\|\mA\| = \sqrt{\lambda_{\max}(\mA^2)} = \max\{\lambda_{\max}(\mA),-\lambda_{\min}(\mA)\}$ for any
symmetric matrix $\mA$ with real eigenvalues. Because we work exclusively with real-valued matrices, symmetry is sufficient to guarantee that all eigenvalues are real.

First, focusing on the maximal eigenvalue of $\mI_k + \mP^{-1/2} \cH^{\textnormal{ex}}_t\mP^{-1/2}$, we can bound it by using Lemma~\ref{lem3} in Appendix~\ref{A: lem3} as follows:
\begin{align}
\lambda_{\max}(\mI_k + \mP^{-1/2} \cH^{\textnormal{ex}}_t\mP^{-1/2}) &=1+ \lambda_{\max}( \mP^{-1/2} \cH^{\textnormal{ex}}_t\mP^{-1/2})
\notag
\\
&=
1 -\lambda_{\min}(\mP^{-1/2} (-\cH^{\textnormal{ex}}_t)\mP^{-1/2})
\notag
\\
&\leq 1-\min\left\{\frac{\lambda_{\min} (-\cH^{\textnormal{ex}}_t)}{\lambda_{\min}(\mP)},\frac{\lambda_{\min} (-\cH^{\textnormal{ex}}_t)}{\lambda_{\max}(\mP)} \right\} \quad \text{\footnotesize by Lemma~\ref{lem3}}
\notag
\\
&\leq 1-\min\left\{\frac{\alpha}{\lambda_{\min}(\mP)},\frac{\alpha}{\lambda_{\max}(\mP)} \right\}\quad \text{\footnotesize as $\alpha\leq -\cH^{\textnormal{ex}}_t$ by Assumption~\ref{ass1}(a)}
\notag
\\
&= 1-\frac{\alpha^+}{\lambda_{\max}(\mP)}+\frac{\alpha^-}{\lambda_{\min}(\mP)}.
\label{new result1}
\end{align}
The first and second lines use $\lambda_{\max}(\mI_k + \mA) = 1 + \lambda_{\max}(\mA)$ and $\lambda_{\max}(\mA)=-\lambda_{\min}(-\mA)$, respectively, for an arbitrary matrix $\mA \in \mathbb{R}^{k \times k}$. The inequality in the third line holds because $1-x$ is decreasing in $x$ and we can lower bound $x=\lambda_{\min}(\mP^{-1/2} (-\cH^{\textnormal{ex}}_t)\mP^{-1/2})$ by Lemma~\ref{lem3} in Appendix~\ref{A: lem3} with $\mA=-\cH_t^\textnormal{ex}$ and $\mB=\mP$. The inequality in the fourth line holds by Assumption~\ref{ass1}(a). The final equality follows by $\alpha=\alpha^+-\alpha^-$ (where $\alpha^+:=\max\{0,\alpha\}$ and $\alpha^-:=\max\{0,-\alpha\}$) and $\lambda_{\min}(\mP)>0$.

Second, focusing on (the negative of) the smallest eigenvalue of $\mI_k + \mP^{-1/2} \cH^{\textnormal{ex}}_t\mP^{-1/2}$, we again use Lemma~\ref{lem3} as follows:
\begin{align}
-\lambda_{\min}(\mI_k +\mP^{-1/2}\cH^{\textnormal{ex}}_t\mP^{-1/2})
&= \lambda_{\max}(-\mI_k +\mP^{-1/2}(-\cH^{\textnormal{ex}}_t)\mP^{-1/2})
\notag
\\
&= -1 + \lambda_{\max}(\mP^{-1/2}(-\cH^{\textnormal{ex}}_t)\mP^{-1/2})\notag
\\
&\leq -1 + \max\left\{\frac{\lambda_{\max}(-\cH^{\textnormal{ex}}_t)}{\lambda_{\min}(\mP)},\frac{\lambda_{\max}(-\cH^{\textnormal{ex}}_t)}{\lambda_{\max}(\mP)}\right \}\quad \text{\footnotesize by Lemma~\ref{lem3}}
\notag\\
&\leq -1 + \frac{\beta}{\lambda_{\min}(\mP)}\quad  \text{\footnotesize by Assumption~\ref{ass1}(a)}
\notag
\\
&= 1 - \left(2-\frac{\beta}{\lambda_{\min}(\mP)} \right).
\label{new result2}
\end{align}
The first and second lines use $-\lambda_{\min}(\mA)=\lambda_{\max}(-\mA)$ and $\lambda_{\max}(-\mI_k + \mA) = -1 + \lambda_{\max}(\mA)$, respectively, for arbitrary matrix $\mA \in \mathbb{R}^{k \times k}$. The inequality in the third line uses Lemma~\ref{lem3} in Appendix~\ref{A: lem3} with $\mA=-\cH_t^\textnormal{ex}$ and $\mB=\mP$. The inequality in the fourth line holds by Assumption~\ref{ass1}(a), which says $\lambda_{\max}(-\cH_t^\textnormal{ex})\leq \beta$. The final line, which holds trivially, is used below. 

Third, combining \eqref{eq: eigenvalue bounds} with bounds \eqref{new result1} and \eqref{new result2}, we obtain
\begin{align}
    \left\|\frac{\dd \vtheta^{\textnormal{ex}}_{t \mid t}}{\dd \vtheta_{t \mid t-1}^{\top}}\right\|_{\mP} &=
    \max \left\{ \lambda_{\max}(\mI_k + \mP^{-1/2} \cH^{\textnormal{ex}}_t\mP^{-1/2}),-\lambda_{\min}(\mI_k + \mP^{-1/2} \cH^{\textnormal{ex}}_t\mP^{-1/2})\right\} \text{\footnotesize by \eqref{eq: eigenvalue bounds}}\notag
\\
   & \leq 1 - \min \left\{ \frac{\alpha^{+}}{\lambda_{\max}(\mP) } -  \frac{\alpha^{-}}{\lambda_{\min}(\mP) }, 2-\frac{\beta}{\lambda_{\min}(\mP)}  \right\}
   \text{\footnotesize by \eqref{new result1} and \eqref{new result2}}
   \label{eq: Jacobian bound ESD update 2}
   \end{align}
This concludes the proof.
\end{proof}

\subsection{Proof of Theorem~\ref{thrm:invertibility}}
\label{sec: proof of Theorem 1}

The proof of Theorem~\ref{thrm:invertibility} makes use of Lemma~\ref{lem1} in the previous section.

\begin{proof} Let $f^{j}_t : \mathbf{\Theta} \rightarrow \mathbf{\Theta}$ denote the update function at time $t$ for $j\in\{\textnormal{im},\textnormal{ex}\}$. For example, for the ISD update we have $\vtheta^{\textnormal{im}}_{t|t} = f^{\textnormal{im}}_t(\vtheta_{t|t-1}) = f^{\textnormal{im}}(\vtheta_{t|t-1}|\vy_t,\mP)=\underset{\vtheta\in \mathbf{\Theta}}{\operatorname{argmax}} \; \{
\ell (\vy_t \mid \vtheta ) - \frac{1}{2}\|\vtheta-\vtheta^\textnormal{im}_{t \mid t-1}\|_{\mP}^2\}$. Because the proof structure is not contingent on whether the update is explicit or implicit, we suppress the filter-type superscript in the proof below and use $f_t$ and $\vtheta_{t|t}$.

Next, let $g_t^{\va}: [0,1] \rightarrow \mathbb{R}$ for some $\va \in \mathbb{R}^{k}$, where $g_t^{\va}(u) := \langle \va, f_t(u \vtheta_{t|t-1} + (1-u)\widetilde{\vtheta}_{t|t-1}) \rangle$, $u \in [0,1]$ and $\vtheta_{t|t-1}, \widetilde{\vtheta}_{t|t-1} \in \mathbf{\Theta}$ are two predictions. By convexity of the parameter space $\mathbf{\Theta}$, we have $u \vtheta_{t|t-1} + (1-u)\widetilde{\vtheta}_{t|t-1} \in \mathbf{\Theta}, \forall u\in[0,1]$, such that $g_t^{\va}$ is well defined.

For both the ISD and ESD update mappings, we have that the Jacobian of the update mapping $f_t$ is well-defined if the Hessian of the log-likelihood exists, see again Lemma~\ref{lem1}. In addition, if $f_t(\vtheta)$ is differentiable in $\vtheta$ everywhere, then $g_t^{\va}(u)$ is differentiable in $u$ everywhere. In sum, Assumption~\ref{ass1} implies $g_t^{\va}(u)$ is differentiable and hence continuous everywhere. Therefore, by the mean-value theorem, we have $\forall \va \in \mathbb{R}^{k}$ that $\exists \, u^{\star} \in [0,1]$ such that
\begin{align}
\langle \va , \vtheta_{t|t}  - \widetilde{\vtheta}_{t|t}  \rangle &=  \langle \va , f_t(\vtheta_{t|t-1})  - f_t(\widetilde{\vtheta}_{t|t-1})  \rangle \notag \\
&= g_t^{\va}(1) - g_t^{\va}(0) \notag\\
&= \frac{\dd g^{\va}_t(u)}{\dd u} \Big \rvert_{u =u^{\star}} (1-0)\notag\\
&= \langle \va, \mathcal{J}_{f_t}(\vtheta_t^{\star}) ( \vtheta_{t|t-1}  - \widetilde{\vtheta}_{t|t-1})\rangle,
\label{MVT result}
\end{align}
where $\mathcal{J}_{f_t}(\vtheta_t^{\star}) := \frac{\dd f_t}{\dd \vtheta^{\top}}\Big \rvert_{\vtheta = \vtheta_t^{\star}}$ is the $k \times k $ Jacobian of the update function $f_t$ evaluated at the midpoint $\vtheta_t^{\star} := u^{\star}\vtheta_{t|t-1} + (1-u^{\star})\widetilde{\vtheta}_{t|t-1}$.

Next, we use our result with $\va = \mP(\vtheta_{t|t}  - \widetilde{\vtheta}_{t|t}) / \| \vtheta_{t|t}  - \widetilde{\vtheta}_{t|t} \|_{\mP} \in \mathbb{R}^k$. This yields
\allowdisplaybreaks
\begin{align}
    \| \vtheta_{t|t}  - \widetilde{\vtheta}_{t|t} \|_{\mP} &= \langle \mP(\vtheta_{t|t}  - \widetilde{\vtheta}_{t|t}) / \| \vtheta_{t|t}  - \widetilde{\vtheta}_{t|t} \|_{\mP}, \vtheta_{t|t}  - \widetilde{\vtheta}_{t|t} \rangle\notag \\
    &= \langle \mP(\vtheta_{t|t}  - \widetilde{\vtheta}_{t|t}) / \| \vtheta_{t|t}  - \widetilde{\vtheta}_{t|t} \|_{\mP}, \mathcal{J}_{f_t}(\vtheta_t^{\star}) ( \vtheta_{t|t-1}  - \widetilde{\vtheta}_{t|t-1})\rangle\notag \\
    &= \| \vtheta_{t|t}  - \widetilde{\vtheta}_{t|t} \|^{-1}_{\mP} \, \langle \mP^{1/2}(\vtheta_{t|t}  - \widetilde{\vtheta}_{t|t}), \mP^{1/2}\mathcal{J}_{f_t}(\vtheta_t^{\star}) ( \vtheta_{t|t-1}  - \widetilde{\vtheta}_{t|t-1})\rangle\notag\\
    &\leq \| \vtheta_{t|t}  - \widetilde{\vtheta}_{t|t} \|^{-1}_{\mP} \| \vtheta_{t|t}  - \widetilde{\vtheta}_{t|t} \|_{\mP} \|\mathcal{J}_{f_t}(\vtheta_t^{\star}) ( \vtheta_{t|t-1}  - \widetilde{\vtheta}_{t|t-1})\|_{\mP}\notag\\
    &\leq \| \mathcal{J}_{f_t}(\vtheta_t^{\star})\|_{\mP} \|\vtheta_{t|t-1}  - \widetilde{\vtheta}_{t|t-1}\|_{\mP},
\end{align}
where the second line uses the mean-value theorem~\eqref{MVT result} above, the fourth uses the Cauchy-Schwarz inequality and $\|\mP^{1/2}\vx\| = \|\vx\|_{\mP}, \forall \vx \in \mathbb{R}^{k}$ and the final line uses the submultiplicative property~\eqref{eq: submultiplicativity weighted matrrix norm} of the $\mP$-weighted matrix norm.

Using the prediction step~\eqref{Parameter prediction step}, we obtain
\begin{align}
    \| \vtheta_{t|t}  - \widetilde{\vtheta}_{t|t} \|_{\mP} &\leq \| \mathcal{J}_{f_t}(\vtheta_t^{\star})\|_{\mP} \|\vtheta_{t|t-1}  - \widetilde{\vtheta}_{t|t-1}\|_{\mP}\notag \\
    &= \| \mathcal{J}_{f_t}(\vtheta_t^{\star})\|_{\mP} \|\mathbf{\Phi}(\vtheta_{t-1|t-1}  - \widetilde{\vtheta}_{t-1|t-1})\|_{\mP}\notag\\
    &\leq \| \mathcal{J}_{f_t}(\vtheta_t^{\star})\|_{\mP} \|\mathbf{\Phi}\|_{\mP} \|\vtheta_{t-1|t-1}  - \widetilde{\vtheta}_{t-1|t-1}\|_{\mP},\label{eq: One-step different starting point bounds}
\end{align}
using again the submultiplicative property~\eqref{eq: submultiplicativity weighted matrrix norm} of the $\mP$-weighted norm. The result indicates that the update-to-update mapping at time $t$ is contractive in the norm $\|\cdot\|_{\mP}$ if $\| \mathcal{J}_{f_t}(\vtheta_t^{\star})\|_{\mP} \|\mathbf{\Phi}\|_{\mP}<1$.

In Lemma~\ref{lem1}, we derived bounds for the ISD updates~\eqref{eq: ISD Jacobian equation} and ESD updates~\eqref{eq: Jacobian bound ESD update 2}. With slight abuse of notation, we have $\mathcal{J}_{f^{j}_t}(\vtheta_t^{\star}) := \frac{\dd f^{j}_t}{\dd \vtheta^{\top}}\Big \rvert_{\vtheta = \vtheta_t^{\star}} = \frac{\dd \vtheta^{j}_{t \mid t}}{\dd \vtheta_{t \mid t-1}^{\top}}\Big \rvert_{\vtheta_{t|t-1} = \vtheta_t^{\star}}$ for $j\in\{\textnormal{im},\textnormal{ex}\}$. Substituting these bounds into equation~\eqref{eq: One-step different starting point bounds}, we have
\begin{align}
  \| \vtheta^{j}_{t|t}  - \widetilde{\vtheta}^{j}_{t|t} \|_{\mP} &\leq \sqrt{\tau^j}\,
        \| \vtheta^{j}_{t-1|t-1}  - \widetilde{\vtheta}^{j}_{t-1|t-1} \|_{\mP},\quad \text{where}
        \label{Eq_App_UptoUp}
        \\[1.5ex]
\label{Eq_App_tauIM}
       \sqrt{\tau^{\textnormal{im}}} &:=\|\mathbf{\Phi}\|_{\mP} \left(1 - \frac{\alpha^+}{\lambda_{\max}(\mP) + \alpha^+} + \frac{\alpha^{-}}{\lambda_{\min}(\mP) - \alpha^{-}}\right),
\\
\label{Eq_App_tauEX}
        \sqrt{\tau^{\textnormal{ex}}} &:= \|\mathbf{\Phi}\|_{\mP} \left( 1 - \min\left\{\frac{\alpha^{+}}{\lambda_{\max}(\mP)}   - \frac{\alpha^{-}}{\lambda_{\min}(\mP)}    , 2-\frac{\beta}{\lambda_{\min}(\mP)}  \right\} \right),
\end{align}
such that the ISD and ESD update-to-update mappings at time $t$ are contractive if $\sqrt{\tau^{\textnormal{im}}}<1$ and $\sqrt{\tau^{\textnormal{ex}}}<1$, respectively. Because $\sqrt{\tau^{\textnormal{im}}}, \sqrt{\tau^{\textnormal{ex}}}\geq 0$ (see again the proof of Lemma~\ref{lem1}), these conditions are equivalent to $\tau^{\textnormal{im}}<1$ and $\tau^{\textnormal{ex}}<1$, exactly the sufficient conditions stated in Theorem~\ref{thrm:invertibility}.

Since $\sqrt{\tau^{\textnormal{im}}}, \sqrt{\tau^{\textnormal{ex}}},\|\cdot\|\geq 0$, squaring both sides of \eqref{Eq_App_UptoUp} and repeatedly applying it yields
\begin{equation*}
       \| \vtheta_{t|t}^j  - \widetilde{\vtheta}^{j}_{t|t} \|^2_{\mP} \leq (\tau^j)^{t}
        \| \vtheta^j_{0|0}  - \widetilde{\vtheta}^j_{0|0} \|^2_{\mP},
\end{equation*}
where $\vtheta^{j}_{0|0}, \widetilde{\vtheta}^{j}_{0|0} \in \mathbf{\Theta}$ are two starting points. Hence for each filter, Definition~\ref{def:stability} is satisfied with $\mW=\mP$. Thus, under the (sufficient) condition $\tau^j < 1$, it follows that
\begin{equation*}
        \lim_{t\to \infty}\| \vtheta^j_{t|t}  - \widetilde{\vtheta}^j_{t|t} \|_{\mP}^2  =0,
\end{equation*}
for any starting points $\vtheta^{j}_{0|0}$ and $\widetilde{\vtheta}^{j}_{0|0}$ and any data sequence $\{\vy_t\}$, and this convergence to zero is exponentially fast. Finally, for any two positive-definite matrices $\mW, \tilde{\mW} \in \mathbb{R}^{k \times k}$, and for any $ \vx \in \mathbb{R}^{k}: \|\vx\|^2_{\mW} = \vx^{\top}\mW\vx=0 \Leftrightarrow \vx=\vzeros_k \Leftrightarrow \|\vx\|^2_{\tilde{\mW}} = 0$ (norm equivalence), which also implies exponential convergence of the filtered paths in the Euclidean norm $\| \cdot \|$.\end{proof}

\subsection{Proof of Theorem~\ref{Th: (Non-)asymptotic MSE bounds}}
\label{Proof of Theorem 2}

The proof of Theorem~\ref{Th: (Non-)asymptotic MSE bounds} makes use of Lemma~\ref{lem1} in Appendix~\ref{ap: Proof of Lemma 1}.

\begin{proof}
\noindent Here, we derive the values of $a, b, c, d$ given in Table~\ref{tab: overview} in terms of other quantities defined in Assumptions~\ref{ass1}--\ref{Assumption: Bounded moments}. Specifically, we derive $a$ and $b$ for the ISD and ESD updates, and $c$ and $d$ for the prediction step. 

\textbf{Preliminaries.} Throughout this proof, we make use of two different expectation operators. First, we use the unconditional expectation $\mathbb{E}[\cdot]$, which acts on $\{\vy_t\}$ using the true densities $\{p^0(\cdot \mid \vvartheta_t)\}$ and then on the true state path $\{\vvartheta_t\}$ using its joint density. Second, we use the conditional expectation operator that acts only on $\vy_t$ given a particular state $\vvartheta_t$. That is, $\underset{\vy_t}{\mathbb{E}}[\cdot] := \int \cdot \, p^0(\vy|\vvartheta_t) d\vy$. By the tower property, it follows that $\mathbb{E}[\underset{\vy_t}{\mathbb{E}}[\cdot]] = \mathbb{E}[\cdot]$.

\textbf{Young's inequality.} For both the ESD update and the prediction step, we make use of Young's inequality. Specifically, we use that for any $\vu, \text{\textbf{\textit{v}}}\in \mathbb{R}^k$ and any positive definite $\mW \in \mathbb{R}^{k \times k}$:
\allowdisplaybreaks
\begin{align}
    \| \vu + \text{\textbf{\textit{v}}} \|_{\mW}^2 &= \| \mW^{1/2}\vu + \mW^{1/2}\text{\textbf{\textit{v}}} \|^2\notag\\
    &= \|\mW^{1/2}\vu\|^2 + \|\mW^{1/2}\text{\textbf{\textit{v}}}\|^2 + 2 \langle \mW^{1/2}\vu , \mW^{1/2}\text{\textbf{\textit{v}}} \rangle\notag\\
    &\leq \|\vu\|_{\mW}^2 + \|\text{\textbf{\textit{v}}}\|_{\mW}^2 + 2 |\langle \mW^{1/2}\vu , \mW^{1/2}\text{\textbf{\textit{v}}} \rangle| \notag\\
    &\leq \|\vu\|_{\mW}^2 + \|\text{\textbf{\textit{v}}}\|_{\mW}^2 + 2\|\mW^{1/2}\vu\|\|\mW^{1/2}\text{\textbf{\textit{v}}}\|\notag\\
    &\leq \|\vu\|_{\mW}^2 + \|\text{\textbf{\textit{v}}}\|_{\mW}^2 + \epsilon^2\|\mW^{1/2}\vu\|^2 + (1/\epsilon^{2})\|\mW^{1/2}\text{\textbf{\textit{v}}}\|^2\notag\\
    &= (1+\epsilon^2)\|\vu\|_{\mW}^2+ (1+1/\epsilon^2)\|\text{\textbf{\textit{v}}}\|_{\mW}^2 ,
    \label{Eq_Young}
\end{align}
for any $\epsilon>0$, where in the fourth line we used the Cauchy-Schwarz inequality and in the fifth line we used Young's inequality for products (i.e., $2 x y\leq \epsilon^2 x^2 +1/\epsilon^2 y^2$ for all $x,y\geq 0,\epsilon>0$).

\textbf{Cauchy-Schwarz inequality.} For the prediction step, we also use the Cauchy-Schwarz inequality. For any $\vu, \text{\textbf{\textit{v}}}\in \mathbb{R}^k$, we have
\allowdisplaybreaks
\begin{align}
    \mathbb{E}[\| \vu + \text{\textbf{\textit{v}}} \|^2] &= \mathbb{E}[\|\vu\|^2 + \|\text{\textbf{\textit{v}}}\|^2 + 2 \langle \vu , \text{\textbf{\textit{v}}} \rangle]\notag\\
    &\leq \mathbb{E}[\|\vu\|^2] + \mathbb{E}[\|\text{\textbf{\textit{v}}}\|^2] + 2 \mathbb{E}[|\langle \vu , \text{\textbf{\textit{v}}} \rangle|] \notag\\
    &\leq \mathbb{E}[\|\vu\|^2] + \mathbb{E}[\|\text{\textbf{\textit{v}}}\|^2] + 2\mathbb{E}[\|\vu\|\|\text{\textbf{\textit{v}}}\|]\notag\\
    &\leq \mathbb{E}[\|\vu\|^2] + \mathbb{E}[\|\text{\textbf{\textit{v}}}\|^2] + 2\sqrt{\mathbb{E}[\|\vu\|^2] \mathbb{E}[\|\text{\textbf{\textit{v}}}\|^2]}
    \notag\\
    & = \left(\sqrt{\mathbb{E}[\|\vu\|^2]} + \sqrt{\mathbb{E}[\|\text{\textbf{\textit{v}}}\|^2]}\right)^2,
    \label{Eq_Expectation_CS}
\end{align}
where the third line uses the Cauchy-Schwarz inequality and the penultimate line uses the Cauchy-Schwarz inequality for random variables: $|\mathbb{E}[XY]| \leq \sqrt{\mathbb{E}[X^2]\mathbb{E}[Y^2]}$ for scalar-valued random variables $X,Y$, which we use with $X=\|\vu\|$ and $Y=\|\text{\textbf{\textit{v}}}\|$.

\textbf{Analysis of ISD update step.} The first-order condition of the ISD update for a static penalty matrix $\mP$ reads:
\begin{equation*}
\vtheta_{t \mid t} = \vtheta_{t \mid t-1} + \mP^{-1} \nabla \ell(\vy_t \mid \vtheta_{t \mid t}).
\end{equation*}
We move $\mP^{-1} \nabla \ell(\vy_t \mid \vtheta_{t \mid t})$ to the left-hand side, pre-multiply both sides by the symmetric square root of the penalty matrix, denoted $\mP^{\frac{1}{2}}$, and subtract $\mP^{\frac{1}{2}} \vtheta_t^{\star} - \mP^{-\frac{1}{2}} \nabla \ell(\vy_t \mid \vtheta_t^{\star})$ from both sides to obtain
\begin{equation}
\mP^{\frac{1}{2}}(\vtheta_{t \mid t} - \vtheta_t^{\star}) - \mP^{-\frac{1}{2}} (\nabla \ell(\vy_t \mid \vtheta_{t \mid t}) - \nabla \ell(\vy_t \mid \vtheta_{t}^{\star})) = \mP^{\frac{1}{2}}(\vtheta_{t \mid t-1} - \vtheta_t^{\star}) + \mP^{-\frac{1}{2}}\nabla \ell(\vy_t \mid \vtheta_{t}^{\star}).
\label{Eq_Proof_Opt_ISD}
\end{equation}
Using Riemann integrability of the Hessian of the log-likelihood function, we may write
\begin{equation*}
\nabla \ell(\vy_t \mid \vtheta_{t \mid t}) - \nabla \ell(\vy_t \mid \vtheta_{t}^{\star}) = \cH^{\star}_{t|t} ( \vtheta_{t \mid t} -  \vtheta_{t}^{\star}),
\end{equation*}
where $\cH^{\star}_{t|t} := \int_{0}^{1} \left.\frac{\partial^2  \ell(\vy_t|\vtheta)}{\partial\vtheta \partial\vtheta^{\top}}\right|_{\substack{\vtheta \,= \, u \,\vtheta_{t|t} \,+\, (1-u)\,\vtheta_t^{\star}}} \mathrm{d}u$ is the average Hessian between $\vtheta_{t|t}$ and $\vtheta_{t}^{\star}$. Substituting this result into equation~\eqref{Eq_Proof_Opt_ISD} produces
\begin{equation*}
(\mI_k - \mP^{-1/2}\cH^{\star}_{t|t}\mP^{-1/2})\mP^{\frac{1}{2}}(\vtheta_{t \mid t} - \vtheta_t^{\star}) = \mP^{\frac{1}{2}}(\vtheta_{t \mid t-1} - \vtheta_t^{\star}) + \mP^{-\frac{1}{2}}\nabla \ell(\vy_t \mid \vtheta_{t}^{\star}),
\end{equation*}
where by Assumption~\ref{ass1}(b), $\mP \succ \cH^{\star}_{t|t} \Rightarrow \mI_k \succ \mP^{-1/2}\cH^{\star}_{t|t}\mP^{-1/2}\Rightarrow \mI_k - \mP^{-1/2}\cH^{\star}_{t|t}\mP^{-1/2} \succ \mZeros_k$, such that taking the inner product on both sides gives
\begin{equation*}
\|\mP^{1/2}(\vtheta_{t \mid t} - \vtheta_t^{\star})\|^2_{(\mI_k - \mP^{-1/2}\cH^{\star}_{t|t}\mP^{-1/2})^2} 
= \|\vtheta_{t \mid t-1} - \vtheta_t^{\star}\|^2_{\mP} + \|\nabla \ell(\vy_t \mid \vtheta_t^{\star})\|^2_{\mP^{-1}} + 2\langle \vtheta_{t \mid t-1} - \vtheta_t^{\star}, \nabla \ell(\vy_t \mid \vtheta_t^{\star}) \rangle.
\end{equation*}
Next, we take the unconditional expectation on both sides and use that $\mathbb{E}[\|\nabla \ell(\vy_t \mid \vtheta_t^{\star})\|^2_{\mP^{-1}}]\leq \lambda_{\max}(\mP^{-1})\mathbb{E}[\|\nabla \ell(\vy_t \mid \vtheta_t^{\star})\|^2] \leq \sigma^2/\lambda_{\min}(\mP)$ by Assumption~\ref{Assumption: Bounded moments}(b) and that $\mathbb{E}[\langle \vtheta_{t \mid t-1} - \vtheta_t^{\star}, \nabla \ell(\vy_t \mid \vtheta_t^{\star}) \rangle] = \mathbb{E}[\underset{\vy_t}{\mathbb{E}}[\langle \vtheta_{t \mid t-1} - \vtheta_t^{\star}, \nabla \ell(\vy_t \mid \vtheta_t^{\star}) \rangle]] = \mathbb{E}[\langle \vtheta_{t \mid t-1} - \vtheta_t^{\star}, \underset{\vy_t}{\mathbb{E}}[\nabla \ell(\vy_t \mid \vtheta_t^{\star})]\rangle] = 0$ by the tower property and the fact that $\underset{\vy_t}{\mathbb{E}}[\nabla \ell(\vy_t \mid \vtheta_t^{\star})]=\vzeros_k$ because the pseudo-true parameter uniquely maximizes the expected log likelihood (see Definition~\ref{Pseudo-truth definition}). This produces
\begin{equation}
\mathbb{E}[\|\mP^{1/2}(\vtheta_{t \mid t} - \vtheta_t^{\star})\|^2_{(\mI_k - \mP^{-1/2}\cH^{\star}_{t|t}\mP^{-1/2})^2}] \leq \mathbb{E}[\|\vtheta_{t \mid t-1} - \vtheta_t^{\star}\|^2_{\mP}] + \sigma^2/\lambda_{\min}(\mP).
\label{Eq_ISD_Proof_Halfway}
\end{equation}
For the left-hand side, we may use the following lower bound
\begin{equation*}
\lambda_{\min}((\mI_k - \mP^{-1/2}\cH^{\star}_{t|t}\mP^{-1/2})^2)\|\vtheta_{t \mid t} - \vtheta_t^{\star}\|^2_{\mP} \leq \|\mP^{1/2}(\vtheta_{t \mid t} - \vtheta_t^{\star})\|^2_{(\mI_k - \mP^{-1/2}\cH^{\star}_{t|t}\mP^{-1/2})^2},
\end{equation*}
as $\mI_k - \mP^{-1/2}\cH^{\star}_{t|t}\mP^{-1/2} \succ \mZeros_k$, it follows that $\lambda_{\min}((\mI_k - \mP^{-1/2}\cH^{\star}_{t|t}\mP^{-1/2})^2) = \lambda_{\min}(\mI_k - \mP^{-1/2}\cH^{\star}_{t|t}\mP^{-1/2})^2 > 0$. Using the lower bound in \eqref{Eq_ISD_Proof_Halfway} and dividing both sides by $\lambda_{\min}(\mI_k - \mP^{-1/2}\cH^{\star}_{t|t}\mP^{-1/2})^2$ gives
\begin{equation*}
\mathbb{E}[\|\vtheta_{t \mid t} - \vtheta_t^{\star}\|^2_{\mP}] \leq \lambda_{\min}(\mI_k - \mP^{-1/2}\cH^{\star}_{t|t}\mP^{-1/2})^{-2}\left(\mathbb{E}[\|\vtheta_{t \mid t-1} - \vtheta_t^{\star}\|^2_{\mP}] + \sigma^2/\lambda_{\min}(\mP)\right).
\end{equation*}
Using the same methodology as in the proof of Lemma~\ref{lem1}, equation~\eqref{eq: ISD Jacobian equation}, we may obtain:
\begin{equation}
    \lambda_{\min}(\mI_k - \mP^{-1/2}\cH^{\star}_{t|t}\mP^{-1/2})^{-1} \leq  1 - \frac{\alpha^+}{\lambda_{\max}(\mP) + \alpha^+} + \frac{\alpha^{-}}{\lambda_{\min}(\mP) - \alpha^{-}}.
    \label{Eq_ISD_Proof_lmin}
\end{equation}
As both sides are nonnegative, we may square both sides. Using the definition of the $\mP$-weighted MSE produces the final result:
\begin{equation*}
\text{MSE}^{\mP}_{t|t} \leq a \text{MSE}^{\mP}_{t|t-1} + b,
\end{equation*}
where $a = \left(1 - \frac{\alpha^+}{\lambda_{\max}(\mP) + \alpha^+} + \frac{\alpha^{-}}{\lambda_{\min}(\mP) - \alpha^{-}}\right)^2$ and $b=a\sigma^2/\lambda_{\min}(\mP)$, which confirms the expressions for the ISD filter in Table~\ref{tab: overview}.

\textbf{Analysis of ESD update step.} The ESD update reads:
\begin{equation*}
\vtheta_{t \mid t} = \vtheta_{t \mid t-1} + \mP^{-1}\nabla \ell(\vy_t \mid \vtheta_{t \mid t-1}).
\end{equation*}
Pre-multiplying both sides by $\mP^{1/2}$ and subtracting $\mP^{1/2}\vtheta^{\star}_t$ on both sides yields
\begin{equation*}
\mP^{1/2}(\vtheta_{t \mid t}  - \vtheta^{\star}_t)= \mP^{1/2}(\vtheta_{t \mid t-1} - \vtheta^{\star}_t) + \mP^{-1/2}\nabla \ell(\vy_t \mid \vtheta_{t \mid t-1}).
\end{equation*}
Computing the squared norm on both sides and taking an unconditional expectation yields
\begin{align}
&\mathbb{E}[\|\vtheta_{t \mid t}  - \vtheta^{\star}_t\|^2_{\mP}]
 \notag \\ 
&=\mathbb{E}[\|\vtheta_{t \mid t-1}  - \vtheta^{\star}_t\|^2_{\mP}] + 2\mathbb{E}[\langle \nabla \ell(\vy_t \mid \vtheta_{t \mid t-1}), \vtheta_{t \mid t-1} - \vtheta^{\star}_t  \rangle] + \mathbb{E}[\|\nabla \ell(\vy_t \mid \vtheta_{t \mid t-1})\|^2_{\mP^{-1}}].
\label{Eq_Proof_Opt_ESD}
\end{align}
For the second term on the right-hand side of \eqref{Eq_Proof_Opt_ESD}, we may write
\begin{align}
2\mathbb{E}[\langle \nabla \ell(\vy_t \mid \vtheta_{t \mid t-1}), \vtheta_{t \mid t-1} - \vtheta^{\star}_t  \rangle]
&= 2\mathbb{E}[\langle \nabla \ell(\vy_t \mid \vtheta_{t \mid t-1}) - \nabla \ell(\vy_t \mid \vtheta^{\star}_t), \vtheta_{t \mid t-1} - \vtheta^{\star}_t  \rangle] \notag \\
&= 2\mathbb{E}[\langle \cH^{\star}_{t|t-1}( \vtheta_{t \mid t-1} - \vtheta_{t}^{\star}), \vtheta_{t \mid t-1} - \vtheta^{\star}_t  \rangle] \notag \\
&= \mathbb{E}[\|\vtheta_{t \mid t-1}  - \vtheta^{\star}_t\|^2_{2\cH^{\star}_{t|t-1}}],
\label{last}
\end{align}
where the first equality uses the tower property together with $\underset{\vy_t}{\mathbb{E}}[\nabla \ell(\vy_t \mid \vtheta_t^{\star})]=\vzeros_k$ from Assumption~\ref{Assumption: Bounded moments}(b), and that $\vtheta_{t \mid t-1} - \vtheta^{\star}_t$ is $\mathcal{F}_{t-1}$-measurable. The second equality uses Riemann integrability of the Hessian to write
\begin{equation*}
\nabla \ell(\vy_t \mid \vtheta_{t \mid t-1}) - \nabla \ell(\vy_t \mid \vtheta_{t}^{\star}) = \cH^{\star}_{t|t-1} ( \vtheta_{t \mid t-1} - \vtheta_{t}^{\star}),
\end{equation*}
where $\cH^{\star}_{t|t-1} := \int_{0}^{1} \left.\frac{\partial^2  \ell(\vy_t|\vtheta)}{\partial\vtheta \partial\vtheta^{\top}}\right|_{\substack{\vtheta \,= \, u \,\vtheta_{t|t-1} \,+\, (1-u)\,\vtheta_t^{\star}}} \mathrm{d}u$ is the average Hessian between $\vtheta_{t|t-1}$ and $\vtheta_{t}^{\star}$. Equation~\eqref{last} technically contains an abuse of notation as the Hessian matrix typically is not positive definite; below, however, this term will be combined with others, resulting in a proper norm.

For the final term on the right-hand side of \eqref{Eq_Proof_Opt_ESD}, we have that
\begin{align}
  &\mathbb{E}[\| \nabla \ell(\vy_t \mid \vtheta_{t \mid t-1})\|^2_{\mP^{-1}}] =\mathbb{E}\left[\|\nabla \ell(\vy_t \mid \vtheta_{t \mid t-1}) - \nabla \ell(\vy_t \mid \vtheta_{t}^{\star}) +\nabla \ell(\vy_t \mid \vtheta_{t}^{\star})\|_{\mP^{-1}}^2\right] \notag\\
  &\leq \mathbb{E}\left[(1 + \chi^2)\|\nabla \ell(\vy_t \mid \vtheta_{t \mid t-1})- \nabla \ell(\vy_t \mid \vtheta_{t}^{\star}) \|_{\mP^{-1}}^2  + (1+1/\chi^{2})\|\nabla \ell(\vy_t \mid \vtheta_{t}^{\star})\|_{\mP^{-1}}^2\right]\notag\\
  &= (1 + \chi^2)\mathbb{E}\left[\| \cH^{\star}_{t|t-1} ( \vtheta_{t \mid t-1} - \vtheta_{t}^{\star})  \|^2_{\mP^{-1}}\right] + (1+1/\chi^2)\mathbb{E}\left[\|\nabla \ell(\vy_t \mid \vtheta_{t}^{\star})\|^2_{\mP^{-1}}\right]\notag\\
  &\leq (1 + \chi^2)\mathbb{E}\left[\| \vtheta_{t \mid t-1} - \vtheta_{t}^{\star}\|^2_{\cH^{\star}_{t|t-1}\mP^{-1}\cH^{\star}_{t|t-1}}\right] + (1+1/\chi^2)\sigma^2 /\lambda_{\min}(\mP),
  \label{last2}
\end{align}
where the second line uses Young's
inequality as specified in \eqref{Eq_Young} with $\epsilon=\chi$, $\vu = \nabla \ell(\vy_t \mid \vtheta_{t \mid t-1}) - \nabla \ell(\vy_t \mid \vtheta_{t}^{\star})$, $\text{\textbf{\textit{v}}} = \nabla \ell(\vy_t \mid \vtheta_{t}^{\star})$ and $\mW = \mP^{-1} $, the third uses the definition of $\cH^{\star}_{t|t-1}$ and the fourth that $\mathbb{E}[\|\nabla \ell(\vy_t \mid \vtheta_t^{\star})\|^2_{\mP^{-1}}]\leq \sigma^2/\lambda_{\min}(\mP)$ by Assumption~\ref{Assumption: Bounded moments}(b).

Substituting~\eqref{last} and~\eqref{last2} into equation~\eqref{Eq_Proof_Opt_ESD}, we obtain
\begin{align}
    \mathbb{E}[\|\vtheta_{t \mid t}  - \vtheta^{\star}_t\|^2_{\mP}] 
    &\leq  \mathbb{E}[\|\vtheta_{t \mid t-1}  - \vtheta^{\star}_t\|^2_{\mP}] + \mathbb{E}[\|\vtheta_{t \mid t-1}  - \vtheta^{\star}_t\|^2_{2\cH^{\star}_{t|t-1}}] \notag \\
    &\quad + (1 + \chi^2)\mathbb{E}\left[\| \vtheta_{t \mid t-1} - \vtheta_{t}^{\star}\|^2_{\cH^{\star}_{t|t-1}\mP^{-1}\cH^{\star}_{t|t-1}}\right] + \frac{(1+1/\chi^2)\sigma^2}{\lambda_{\min}(\mP)}  \notag\\
    &= \mathbb{E}[\|\vtheta_{t \mid t-1}  - \vtheta^{\star}_t\|^2_{\mP + 2\cH^{\star}_{t|t-1} + (1+\chi^2)\cH^{\star}_{t|t-1}\mP^{-1}\cH^{\star}_{t|t-1}}] + \frac{(1+1/\chi^2)\sigma^2}{\lambda_{\min}(\mP)}\notag\\
    &= \mathbb{E}[\|\mP^{1/2} (\vtheta_{t \mid t-1}  - \vtheta^{\star}_t)\|^2_{(\mI_k + \mP^{-1/2}\cH^{\star}_{t|t-1}\mP^{-1/2})^2}] \notag\\
    &\quad + \chi^2\mathbb{E}[\|\mP^{1/2}( \vtheta_{t \mid t-1}  - \vtheta^{\star}_t)\|^2_{(\mP^{-1/2}\cH^{\star}_{t|t-1}\mP^{-1/2})^2}] + \frac{(1+1/\chi^2)\sigma^2}{\lambda_{\min}(\mP)}.
    \label{eq:intermediate weighted MSE bounds}
\end{align}
The two weighted norms on the right-hand side of~\eqref{eq:intermediate weighted MSE bounds} can be bounded by the largest eigenvalues of their weight matrices. For the first weight matrix, the bound on $\lambda_{\max}(\mI_k + \mP^{-1/2}\cH^{\star}_{t|t-1}\mP^{-1/2})$ from \eqref{eq: Jacobian bound ESD update 2} in Lemma~\ref{lem1} has nonnegative right-hand side under Assumption~\ref{ass1}, so squaring yields
\begin{equation*}
     \lambda_{\max}\left((\mI_k + \mP^{-1/2}\cH^{\star}_{t|t-1}\mP^{-1/2})^2\right) \leq \left(1 - \min \left\{ \frac{\alpha^{+}}{\lambda_{\max}(\mP) } -  \frac{\alpha^{-}}{\lambda_{\min}(\mP) }, 2-\frac{\beta}{\lambda_{\min}(\mP)}  \right\}\right)^2.
\end{equation*}
For the second weight matrix, using again inequality~\eqref{eq: Jacobian bound ESD update 2} in Lemma~\ref{lem1}, we also have
 \begin{align}
      \lambda_{\max}&(\mP^{-1/2}\cH^{\star}_{t|t-1}\mP^{-1}\cH^{\star}_{t|t-1}\mP^{-1/2}) = \lambda_{\max}((\mP^{-1/2}\cH^{\star}_{t|t-1}\mP^{-1/2})^2)\notag\\
      & = \max\{\lambda_{\max}(\mP^{-1/2}\cH^{\star}_{t|t-1}\mP^{-1/2}), - \lambda_{\min}(\mP^{-1/2}\cH^{\star}_{t|t-1}\mP^{-1/2})\}^2 \notag\\
      &= \max\{\lambda_{\max}(\mI_k + \mP^{-1/2}\cH^{\star}_{t|t-1}\mP^{-1/2}) - 1, 1 - \lambda_{\min}(\mI_k + \mP^{-1/2}\cH^{\star}_{t|t-1}\mP^{-1/2})\}^2 \notag\\
      &\leq \max \left\{ \max \left\{\frac{\lambda_{\max}(\mP)-\alpha}{\lambda_{\max}(\mP) },  \frac{\lambda_{\min}(\mP)-\alpha}{\lambda_{\min}(\mP) } \right\} -1, 1 + \frac{ \beta-\lambda_{\min}(\mP)}{\lambda_{\min}(\mP)} \right\}^2  \notag \\
     &\leq \max \left\{ \frac{-\alpha}{\lambda_{\max}(\mP) },  \frac{-\alpha}{\lambda_{\min}(\mP) }, \frac{\beta}{\lambda_{\min}(\mP)}\right\}^2 \notag  \\
     & \leq\max \left\{ \frac{\alpha^{-}}{\lambda_{\min}(\mP) }, \frac{\beta}{\lambda_{\min}(\mP)} \right\}^2 = \frac{L^2}{\lambda_{\min}(\mP)^2},\notag
 \end{align}
 using $\beta>0$, $\alpha^{-}/\lambda_{\min}(\mP) \geq \alpha^{-}/\lambda_{\max}(\mP)$, and $L:=\max\{\alpha^{-},\beta\}$. In the penultimate line, $-\alpha/\lambda_{\max}(\mP)$ cannot be the largest of the three numbers, so that it can be dropped. Combining these results, we obtain the final result:
 \begin{equation*}
 \text{MSE}^{\mP}_{t|t} \leq a \text{MSE}^{\mP}_{t|t-1} + b,
 \end{equation*}
 where $a = \left(1 - \min \left\{ \frac{\alpha^{+}}{\lambda_{\max}(\mP) } -  \frac{\alpha^{-}}{\lambda_{\min}(\mP) }, 2-\frac{\beta}{\lambda_{\min}(\mP)}  \right\}\right)^2 + \frac{\chi^2L^2}{\lambda_{\min}(\mP)^2}$ and $b=\frac{(1+1/\chi^2)\sigma^2}{\lambda_{\min}(\mP)}$, which confirms the expressions for the ESD filter in Table~\ref{tab: overview}.

\textbf{Analysis of the prediction step.} We start by subtracting the pseudo-true state $\vtheta_t^{\star}$ from the prediction step in equation~\eqref{Parameter prediction step}, multiplying by $\mP^{1/2}$ and taking the squared norm:
\begin{align}
\|\vtheta_{t+1 \mid t} &- \vtheta_{t+1}^{\star}\|_{\mP}^2 = \left\|(\mI_k - \mathbf{\Phi}) \vomega + \mathbf{\Phi} \vtheta_{t \mid t} - \vtheta_{t+1}^{\star}\right\|_{\mP}^2\notag\\
&= \left\| \mathbf{\Phi} (\vtheta_{t \mid t} - \vtheta_t^{\star}) + (\mathbf{\Phi} - \mI_k)(\vtheta_t^{\star} - \vomega) + (\vtheta_t^{\star} - \vtheta_{t+1}^{\star}) \right\|_{\mP}^2\notag\\
&\leq (1 + \epsilon^2) \|\mathbf{\Phi} (\vtheta_{t \mid t} - \vtheta_t^{\star})\|_{\mP}^2 
\notag
\\
&\qquad \qquad 
+ \left( 1 + \frac{1}{\epsilon^2} \right) \| (\mathbf{\Phi} - \mI_k)(\vtheta_t^{\star} - \vomega) + (\vtheta_t^{\star} - \vtheta_{t+1}^{\star}) \|_{\mP}^2,\label{eq: prediction with unknown DGP intermediate step}
\end{align}
where in the second line we added and subtracted $(\mathbf{\Phi} - \mI_k)\vtheta_t^{\star}$ and the third line uses Young's inequality as specified in~\eqref{Eq_Young} with $\vu = \mathbf{\Phi} (\vtheta_{t \mid t} - \vtheta_t^{\star})$, $\text{\textbf{\textit{v}}} = (\mathbf{\Phi} - \mI_k)(\vtheta_t^{\star} - \vomega) + (\vtheta_t^{\star} - \vtheta_{t+1}^{\star})$ and $\mW = \mP$.
Using the submultiplicative property~\eqref{eq: submultiplicativity weighted matrrix norm} of the $\mP$-weighted matrix norm
and taking the unconditional expectation of~\eqref{eq: prediction with unknown DGP intermediate step} yields
\begin{equation}
\text{MSE}_{t+1|t}^{\mP} \leq (1 + \epsilon^2) \|\mathbf{\Phi}\|_{\mP}^2 \text{MSE}_{t|t}^{\mP} + \left( 1 + \frac{1}{\epsilon^2} \right) \mathbb{E}[\| (\mathbf{\Phi} - \mI_k)(\vtheta_t^{\star} - \vomega) + (\vtheta_t^{\star} - \vtheta_{t+1}^{\star}) \|_{\mP}^2],
\label{Eq_Proof_MisspecPred}
\end{equation}
where $\|\mathbf{\Phi}\|_{\mP}^2$ is the squared matrix norm of $\mathbf{\Phi}$ induced by the vector norm $\|\tilde{\vu}\|_{\mP}^2$ for $\tilde{\vu}\in \mathbb{R}^k$.

Using preliminary result~\eqref{Eq_Expectation_CS} with $\vu = (\mathbf{\Phi} - \mI_k)(\vtheta_t^{\star} - \vomega)$ and $\text{\textbf{\textit{v}}} = \vtheta_t^{\star} - \vtheta_{t+1}^{\star}$, we obtain
\begin{align}
\mathbb{E}[\| (\mathbf{\Phi} &- \mI_k)(\vtheta_t^{\star} - \vomega) + (\vtheta_t^{\star} - \vtheta_{t+1}^{\star}) \|_{\mP}^2] \notag\\&\leq  \lambda_{\max}(\mP) \left( \sqrt{\mathbb{E}[\| (\mathbf{\Phi} - \mI_k)(\vtheta_t^{\star} - \vomega) \|^2]} + \sqrt{\mathbb{E}[\|\vtheta_t^{\star} - \vtheta_{t+1}^{\star}\|^2]} \right)^2\notag\\
&\leq   \lambda_{\max}(\mP)\left( \sqrt{\mathbb{E}[\|\mathbf{\Phi} - \mI_k\|^2\|\vtheta_t^{\star} - \vomega\|^2]} + \sqrt{\mathbb{E}[\|\vtheta_{t+1}^{\star} - \vtheta_{t}^{\star}\|^2]} \right)^2\notag\\
&=  \lambda_{\max}(\mP)\left( \|\mI_k - \mathbf{\Phi}\|\sqrt{\mathbb{E}[\|\vtheta_t^{\star} - \vomega\|^2]} + \sqrt{\mathbb{E}[\|\vtheta_{t+1}^{\star} - \vtheta_{t}^{\star}\|^2]} \right)^2\notag\\
&\leq  \lambda_{\max}(\mP)\left( \|\mI_k - \mathbf{\Phi}\|s_{\omega} + q \right)^2\label{eq: unknown DGP drift},
\end{align}
where the third line uses the definition of induced matrix norm and the final line uses that
$s_{\omega}^2:=\mathbb{E}[\|\vtheta_t^{\star} - \vomega\|^2]$ and $\mathbb{E}[\|\vtheta_{t+1}^{\star} - \vtheta_{t}^{\star}\|^2] = \mathbb{E}[\tr((\vtheta_{t+1}^{\star} - \vtheta_{t}^{\star})(\vtheta_{t+1}^{\star} - \vtheta_{t}^{\star})^{\top})] = \tr(\mathbb{E}[(\vtheta_{t+1}^{\star} - \vtheta_{t}^{\star})(\vtheta_{t+1}^{\star} - \vtheta_{t}^{\star})^{\top}]) \leq \tr(\mQ) =: q^2 < \infty$.

Combining inequalities~\eqref{eq: unknown DGP drift} and~\eqref{Eq_Proof_MisspecPred} yields the final result
\begin{equation*}
\text{MSE}_{t+1|t}^{\mP}\; \leq\; c\, \text{MSE}_{t|t}^{\mP} + d,
\end{equation*}
where $c = (1 + \epsilon^2) \|\mathbf{\Phi}\|_{\mP}^2$ and $d = \left( 1 + \frac{1}{\epsilon^2} \right)\lambda_{\max}(\mP)\left( \|\mI_k - \mathbf{\Phi}\|s_{\omega} + q \right)^2$, which confirms the expressions for the prediction step in Table~\ref{tab: overview}.\end{proof}

\subsection{Proof of Proposition~\ref{prop1}}
\label{Proof of prop1}

\begin{proof}
 If, in addition to Assumptions~\ref{ass1} and \ref{Assumption: Bounded moments}(a,b),  Assumption~\ref{Assumption: Correct specification} holds, then
\begin{align}
    \text{MSE}^{\mP}_{t+1|t} &= \mathbb{E}[\|\vtheta_{t+1 \mid t} - \vvartheta_{t+1}\|^2_{\mP}]\notag \\
    &= \mathbb{E}[\|(\mI_k - \mathbf{\Phi})\vomega + \mathbf{\Phi}\vtheta_{t \mid t} - (\mI_k - \mathbf{\Phi}_0)\vomega_{0}  - \mathbf{\Phi}_0 \vvartheta_{t}  - \vxi_{t+1} \|^2_{\mP}]\notag\\
    &= \mathbb{E}[\|\mathbf{\Phi}_{0} (\vtheta_{t \mid t} - \vvartheta_{t}) - \vxi_{t+1} \|^2_{\mP}]\notag\\
    &= \mathbb{E}[\|\mathbf{\Phi}_{0} (\vtheta_{t \mid t} - \vvartheta_{t})\|_{\mP}^2] + \mathbb{E}[\|\vxi_{t+1}\|_{\mP}^2] - 2 \mathbb{E}[\langle \mP\mathbf{\Phi}_{0} (\vtheta_{t \mid t} - \vvartheta_{t}), \vxi_{t+1} \rangle]\notag\\
    &= \mathbb{E}[\|\mathbf{\Phi}_{0} (\vtheta_{t \mid t} - \vvartheta_{t})\|_{\mP}^2] + \mathbb{E}[\|\vxi_{t+1}\|_{\mP}^2]\notag\\
    &\leq \|\mathbf{\Phi}_{0}\|_{\mP}^2 \mathbb{E}[\|\vtheta_{t \mid t} - \vvartheta_{t}\|_{\mP}^2] + \lambda_{\max}(\mP)\mathbb{E}[\|\vxi_{t+1}\|^2]\notag\\
    &\leq \|\mathbf{\Phi}_{0}\|_{\mP}^2 \text{MSE}^{\mP}_{t|t} + \lambda_{\max}(\mP)\sigma^2_{\xi},\notag
\end{align}
where the third line uses that $\vomega = \vomega_{0}$ and $\mathbf{\Phi} = \mathbf{\Phi}_0$, the fifth uses that $\vxi_{t+1}$ is independent of $\mathbf{\Phi}_{0} (\vtheta_{t \mid t} - \vvartheta_{t})$ and that $\mathbb{E}[\vxi_{t+1}] = \mathbf{0}_k$, while the sixth line uses the submultiplicative property~\eqref{eq: submultiplicativity weighted matrrix norm} of the $\mP$-weighted matrix norm. The last line uses that $\mathbb{E}[\|\vxi_{t+1}\|^2] = \mathbb{E}[\text{tr}(\vxi_{t+1}\vxi_{t+1}^{\top})] = \text{tr}(\mathbb{E}[\vxi_{t+1}\vxi_{t+1}^{\top}]) = \text{tr}(\mathbf{\Sigma}_{\xi}) = \sigma^2_{\xi} < \infty$.
We conclude that $c = \|\mathbf{\Phi}_{0}\|_{\mP}^2$  and $d = \lambda_{\max}(\mP)\sigma^2_{\xi}$. 
\end{proof}

\setcounter{equation}{0}
\section{Further theoretical results}
\label{Appendix B. Further theoretical results}

\subsection{Kalman filter as ISD and ESD update}
\label{A: Kalman}

Although the ISD and ESD filters generally produce different outcomes, they yield identical results (albeit with different learning rates) in the case of a Gaussian observation density where the mean is a linear transformation of the parameter being tracked.

\begin{example}[\citeauthor{kalman1960new}'s (\citeyear{kalman1960new}) level update as both ISD and ESD]
\label{ex1}Consider the observation $\vy_t =  \vd +\mZ\vvartheta_t + \vvarepsilon_t$, where $\vy_t,\vd\in \SR^n$,  $\mZ\in \SR^{n\times k}$, $\vvartheta_t\in\SR^k$, and  $\vvarepsilon_t\sim\textnormal{i.i.d.}\, \rN(\vzeros_n,\mathbf{\Sigma}_{\varepsilon})$ with $\mathbf{\Sigma}_{\varepsilon} \succ \mZeros_n$.
Consider Kalman's predicted and updated covariance matrices $\mP^\textnormal{\tiny KF}_{t|t-1},\mP^\textnormal{\tiny KF}_{t|t}\succ \mZeros_k$, which are both $\mathcal{F}_{t-1}$ measurable.
Assume the postulated density $p(\cdot|\cdot)$ matches the true density $p^0(\cdot|\cdot)$. Then,
Kalman's level update can be interpreted as (i) an ISD
update as in equation~\eqref{Implicit parameter update step}, with the learning-rate matrix $\mH^{\textnormal{im}}_t=\mP^{\textnormal{\tiny KF}}_{t|t-1}$; and (ii) an ESD update as in equation~\eqref{Explicit parameter update step}, with $\mH^{\textnormal{ex}}_t=\mP^{\textnormal{\tiny KF}}_{t|t}$.
The implicit learning rate exceeds the explicit one, in that\ $\mH^{\textnormal{im}}_t\succeq \mH^{\textnormal{ex}}_t$, since $\mP^{\textnormal{\tiny KF}}_{t|t-1} \succeq \mP^{\textnormal{\tiny KF}}_{t|t}$.
This dual interpretation suggests that generalizations of the Kalman filter may be constructed from either its ISD or ESD representation, as in \citet{lange2024bellman} and \citet{ollivier2018online}, respectively.
\end{example}

\begin{proof}

\textbf{Linear Gaussian observation equation.} Observations $\vy_t \in \SR^n$ are generated by
\begin{equation*}
 \vy_t =\vd\,+\, \mZ\,\vvartheta_t\,+\,\vvarepsilon_t,\qquad \vvarepsilon_t \sim \text{i.i.d.}\, \mathrm{N}(\vzeros_n,\mathbf{\Sigma}_{\varepsilon}),
\end{equation*}
where $\vd,\vvarepsilon_t\in \SR^n$, $\mZ\in \SR^{n\times k}$,
$\vvartheta_t\in \SR^k$ and $\mathbf{\Sigma}_{\varepsilon} \succ \mZeros_n$. If the postulated density $p(\cdot|\vtheta)$ matches the true density $p^0(\cdot|\vvartheta)$, then the log-likelihood function and score are
\allowdisplaybreaks
\begin{align}
\text{log density:} && \ell(\vy | \vtheta) & \propto -\smash{\frac{1}{2}} (\vy-\vd-\mZ\vtheta)^{\top}\mathbf{\Sigma}_{\varepsilon}^{-1}(\vy-\vd-\mZ\vtheta),
\notag
 \\
\text{score:}&& \nabla \ell(\vy|\vtheta) &= \mZ^{\top}\mathbf{\Sigma}_{\varepsilon}^{-1}(\vy-\vd-\mZ\vtheta).
\label{score}
\end{align}
Let the predicted parameter $\vtheta_{t|t-1}\in  \SR^k$ be given and fixed, which is used to compute both the ISD update~\eqref{Implicit parameter update step} and the ESD update~\eqref{Explicit parameter update step}; hence, we omit the superscript on $\vtheta_{t|t-1}$.

\textbf{Kalman's covariance update.} Let Kalman's predicted and updated covariance matrices $\mP^{\textnormal{\tiny KF}}_{t|t-1}\succ \mZeros_k$ and $\mP^{\textnormal{\tiny KF}}_{t|t}\succ \mZeros_k$ be given and fixed. They are related via
\begin{align}
\mP^{\textnormal{\tiny KF}}_{t|t}&=
\mP^{\textnormal{\tiny KF}}_{t|t-1}-\mP^{\textnormal{\tiny KF}}_{t|t-1}\mZ^{\top} (\mZ\mP^{\textnormal{\tiny KF}}_{t|t-1}\mZ^{\top}+\mathbf{\Sigma}_{\varepsilon})^{-1}\mZ \mP^{\textnormal{\tiny KF}}_{t|t-1},
\notag
\\
&=[(\mP^{\textnormal{\tiny KF}}_{t|t-1})^{-1}+\mZ^{\top}\mathbf{\Sigma}_{\varepsilon}^{-1}\mZ]^{-1}.
\label{Kalman filter covariance update}
\end{align}
The first line gives the standard covariance-matrix updating step; see \citet[p.\ 106]{harvey1990forecasting} or \citet[p.\ 86]{durbin2012time}. The second line is used below and can be found in several places (e.g.\ \citealp[p.\ 504]{fahrmeir1992posterior}, \citealp[eq.\ 36]{lambert2022recursive}); it follows from the Woodbury matrix-inversion lemma (see equation~\eqref{matrixlemma1} below).

\textbf{Kalman update as ISD update.} We take the ISD update~\eqref{Implicit parameter update step} with $\mH^\textnormal{im}_t=
\mP^{\textnormal{\tiny KF}}_{t|t-1}$ and substitute the score~\eqref{score} to yield
\begin{align}
    \vtheta^\textnormal{im}_{t|t}&=\vtheta_{t|t-1}+ \mP^{\textnormal{\tiny KF}}_{t|t-1} \nabla \ell(\vy_t| \vtheta^\textnormal{im}_{t|t}),
    \notag
    \\
    &=\vtheta_{t|t-1}+\mP^{\textnormal{\tiny KF}}_{t|t-1}\,\mZ^{\top}\mathbf{\Sigma}_{\varepsilon}^{-1}(\vy_t-\vd-\mZ\vtheta^\textnormal{im}_{t|t}).\notag
\end{align}
Pre-multiplying by $(\mP^{\textnormal{\tiny KF}}_{t|t-1})^{-1}$ and isolating $\vtheta^\textnormal{im}_{t|t}$ on the left-hand side, we obtain
\begin{align}
\small
    \vtheta^\textnormal{im}_{t|t} &\small = (\mZ^{\top}\mathbf{\Sigma}_{\varepsilon}^{-1}\mZ+(\mP^{\textnormal{\tiny KF}}_{t|t-1})^{-1})^{-1}[(\mP^{\textnormal{\tiny KF}}_{t|t-1})^{-1} \vtheta_{t|t-1}+\mZ^{\top}\mathbf{\Sigma}_{\varepsilon}^{-1}(\vy_t-\vd)],
    \notag
    \\
    &\small = (\mZ^{\top}\mathbf{\Sigma}_{\varepsilon}^{-1}\mZ+(\mP^{\textnormal{\tiny KF}}_{t|t-1})^{-1})^{-1}[\big\{(\mP^{\textnormal{\tiny KF}}_{t|t-1})^{-1}+{\mZ^{\top}\mathbf{\Sigma}_{\varepsilon}^{-1}\mZ}-{\mZ^{\top}\mathbf{\Sigma}_{\varepsilon}^{-1}\mZ}\big\} \vtheta_{t|t-1}+\mZ^{\top}\mathbf{\Sigma}_{\varepsilon}^{-1}(\vy_t-\vd)],
        \notag
        \\
        &\small= \vtheta_{t|t-1}+ (\mZ^{\top}\mathbf{\Sigma}_{\varepsilon}^{-1}\mZ+(\mP^{\textnormal{\tiny KF}}_{t|t-1})^{-1})^{-1} \mZ^{\top} \mathbf{\Sigma}_{\varepsilon}^{-1} (\vy_t-\vd-\mZ\vtheta_{t|t-1}),       \notag
              \\
        &\small= \vtheta_{t|t-1}+ \mP^{\textnormal{\tiny KF}}_{t|t-1}\mZ^{\top}(\mZ\mP^{\textnormal{\tiny KF}}_{t|t-1}\mZ^{\top}+\mathbf{\Sigma}_{\varepsilon})^{-1} (\vy_t-\vd-\mZ\vtheta_{t|t-1}),\notag
     \end{align}
where the last line, which follows from a standard matrix-inversion lemma (see equation~\eqref{matrixlemma2} below), is the standard Kalman-filter level update; see e.g.\ \citet[p.\ 106]{harvey1990forecasting} or \citet[p.\ 86]{durbin2012time}.

\textbf{Kalman update as ESD update.}  We take the ESD update~\eqref{Explicit parameter update step} with $\mH^\textnormal{ex}_t=
\mP^{\textnormal{\tiny KF}}_{t|t}$ and substitute the score~\eqref{score} to yield
\begin{eqnarray}
    \vtheta^\textnormal{ex}_{t|t}&=&\vtheta_{t|t-1}+ \mP^{\textnormal{\tiny KF}}_{t|t} \,\nabla \ell(\vy_t|\vtheta_{t|t-1}),
    \notag
  \\
 & =&\vtheta_{t|t-1}+ \mP^{\textnormal{\tiny KF}}_{t|t} \,\mZ^{\top}\mathbf{\Sigma}_{\varepsilon}^{-1}(\vy_t-\vd-\mZ\vtheta_{t|t-1}),
    \notag
    \\
      &\overset{\eqref{Kalman filter covariance update}}{=}& \vtheta_{t|t-1}+ ((\mP^{\textnormal{\tiny KF}}_{t|t-1})^{-1}+\mZ^{\top}\mathbf{\Sigma}_{\varepsilon}^{-1}\mZ)^{-1}\, \mZ^{\top} \mathbf{\Sigma}_{\varepsilon}^{-1} (\vy_t-\vd-\mZ\vtheta_{t|t-1}),       \notag
        \\
    &=& \vtheta_{t|t-1}+ \mP^{\textnormal{\tiny KF}}_{t|t-1}\mZ^{\top}(\mZ\mP^{\textnormal{\tiny KF}}_{t|t-1}\mZ^{\top}+\mathbf{\Sigma}_{\varepsilon})^{-1} (\vy_t-\vd-\mZ\vtheta_{t|t-1}), \notag
\end{eqnarray}
where the last line, which follows from the same matrix-inversion lemma (see equation~\eqref{matrixlemma2} below), is again the standard form of Kalman's level update (see references given above).

\textbf{Matrix-inversion lemmas.} The above derivation relied on two matrix-inversion lemmas for positive-definite matrices $\mA, \mB\succ \mZeros_k$ and an arbitrary (size-compatible) matrix~$\mC$:
\begin{align}
 (\mA+\mC^{\top}\mB^{-1}\mC)^{-1}\mC^{\top}\mB^{-1} &= \mA^{-1}\mC^{\top} (\mB+\mC\mA^{-1}\mC^{\top})^{-1},
    \label{matrixlemma2}
    \\
    (\mA+\mC^{\top}\mB^{-1}\mC)^{-1} &=\mA^{-1}-\mA^{-1}\mC^{\top}(\mB+\mC\mA^{-1}\mC^{\top})^{-1}\mC\mA^{-1}.
    \label{matrixlemma1}
\end{align}
While the second identity is a special case of the Woodbury matrix identity (e.g.\ \citealp{sherman1950adjustment} and \citealp{henderson1981deriving}), a simple proof for the first identity is hard to find. A short proof of both identities can be found in \cite{lange2024short}.
\end{proof}

\subsection{Minimizing MSE bound of ISD filter w.r.t.\ Young's parameter}
\label{sec:Proof of Corollary 2.4}

For the ISD filter when Assumption~\ref{Assumption: Bounded moments} holds but Assumption~\ref{Assumption: Correct specification} does not, the MSE bound contains a single free parameter, $\epsilon>0$, which can be analytically optimized.

\begin{corollary}[Optimal Young's parameter for the ISD filter]\label{Cor: Optimal ISD Young} Let  Assumptions~\ref{ass1}--\ref{Assumption: Bounded moments} hold and assume that sufficient condition~\eqref{eq:invertibilityISD} for stability in Theorem~\ref{thrm:invertibility} is met, implying $\tau^{\textnormal{im}} = ac/(1+\epsilon^2) < 1$ for some $\epsilon>0$. 
For $0<\tau^{\textnormal{im}} < 1$, the value of the Young's parameter $\epsilon^2 > 0$ that minimizes the asymptotic MSE bound~\eqref{eq: asymptotic error bound} is then given by
    \begin{equation*}
        \epsilon_{\star}^2 = \frac{1-\tau^{\textnormal{im}}}{\tau^{\textnormal{im}}+\sqrt{\tau^{\textnormal{im}}+ \tau^{\textnormal{im}} (1 - \tau^{\textnormal{im}})\frac{\sigma^2}{\lambda_{\max}(\mP)\lambda_{\min}(\mP)(\|\mI_k-\mathbf{\Phi}\|\, s_{\omega} + q)^2}}} < \infty.
    \end{equation*}
\end{corollary}

\begin{proof} \textbf{Computing the bound.} Consider the ISD filter when Assumption~\ref{Assumption: Bounded moments} holds, but Assumption~\ref{Assumption: Correct specification} does not. In addition, let $\tau^{\textnormal{im}} = ac/(1 +\epsilon^2) < 1$ (see equation~\eqref{eq:invertibilityISD}) and define $\tilde{d} := d/(1 + \frac{1}{\epsilon^2})$, where $d$ is as in Table~\ref{tab: overview}. Unlike $d$, $\tilde{d}$ is independent of $\epsilon^2$.
As we must ensure $a\,c<1$ to guarantee finite asymptotic MSE bounds, it is convenient to take $\epsilon^2:= (\kappa(1-\tau^{\textnormal{im}}))/\tau^{\textnormal{im}}$, such that $ac = \tau^{\textnormal{im}} (1 +\epsilon^2) = \tau^{\textnormal{im}} + \kappa(1- \tau^{\textnormal{im}}) < 1$, $\forall \kappa \in (0,1)$. Intuitively, $\kappa$ controls the extent to which $\epsilon^2$ closes the ``gap'' between $\tau^{\textnormal{im}}$ and unity; i.e.,\ the denominator in the asymptotic MSE bound reads $1 - ac = 1 - \tau^{\textnormal{im}} - \kappa (1- \tau^{\textnormal{im}}) = 1 - \tau^{\textnormal{im}} - \kappa + \kappa\tau^{\textnormal{im}} = (1-\tau^{\textnormal{im}})(1-\kappa) \in (0,1)$.

The asymptotic MSE bound $\forall\kappa\in (0,1)$ is
\allowdisplaybreaks
\begin{align}
\label{specific bound}
\lim \sup_{t\to \infty} \textnormal{MSE}^{\mP}_{t|t} &\leq \frac{b + ad}{1 - ac} \notag \\
&= \frac{b + a\tilde{d}(1 + \frac{1}{\epsilon^2})}{1 - \tau^{\textnormal{im}}(1 +\epsilon^2)}\notag\\
&= \frac{b + a\tilde{d}(1 + \frac{\tau^{\textnormal{im}}}{\kappa(1-\tau^{\textnormal{im}})})}{(1-\tau^{\textnormal{im}})(1-\kappa)}\notag\\
&= \frac{b + a\tilde{d}}{1 - \tau^{\textnormal{im}}}\frac{1}{1-\kappa} + \frac{\tau^{\textnormal{im}} a \tilde{d}}{(1 - \tau^{\textnormal{im}})^2}\frac{1}{\kappa(1-\kappa)}\notag\\
&= A \frac{1}{1-\kappa} + B \frac{1}{\kappa(1-\kappa)},
\end{align}
where the second line uses the definition of $\tilde{d}$ and $\tau^{\textnormal{im}}$ and the third line uses our specification of $\epsilon^2$ in terms of $\kappa$. From the final line, it follows that $A,B \geq 0$ are given by
\begin{equation}
 A:=\frac{b + a \tilde{d}}{1-\tau^{\textnormal{im}}}, \quad B:=\frac{\tau^{\textnormal{im}} a \tilde{d}}{(1-\tau^{\textnormal{im}})^2}.
 \label{eq A and B}
\end{equation}
For future reference,
\begin{equation}
\frac{A}{B}= \frac{\frac{b + a \tilde{d}}{1-\tau^{\textnormal{im}}}}{\frac{\tau^{\textnormal{im}} a \tilde{d}}{(1-\tau^{\textnormal{im}})^2}} = \frac{1 - \tau^{\textnormal{im}}}{\tau^{\textnormal{im}}} \frac{b + a\tilde{d}}{a\tilde{d}}.
\label{eq A/B}
\end{equation}

\textbf{Minimizing the bound.} If $\tau^{\textnormal{im}} = 0$, we have that $\epsilon = \infty$ minimizes the MSE bound, as can be seen in the second line of~\eqref{specific bound}. We therefore proceed with the case $\tau^{\textnormal{im}}>0$, which implies $B>0$ as $\tau^{\textnormal{im}} = ac/(1 +\epsilon^2) \neq 0 \Rightarrow a \neq 0$ and $q>0 \Rightarrow \tilde{d}>0$; see~\eqref{eq A and B}. Equation~\eqref{specific bound} illustrates that as $A,B>0$, we have that $\kappa\in (0,1)$ should not approach either boundary too closely.

Minimizing the bound~\eqref{specific bound}  with respect to $\kappa$ yields the following first-order condition:
\begin{equation*}
0=\frac{A}{(1-\kappa)^2} + \frac{B(2\kappa -1)}{\kappa^2(1-\kappa)^2},
\end{equation*}
where we may multiply by $\kappa^2(1-\kappa)^2\in (0,1)$ to obtain
\begin{equation*}
0=A\, \kappa^2  + 2 B \,\kappa - B,
\end{equation*}
which is a quadratic equation in $\kappa$, the solution of which reads
\begin{equation*}
\kappa_{\pm} = \frac{-2 B \pm \sqrt{4B^2+4 A B} }{2 A}.
\end{equation*}
Because $A$ and $B$ are positive, we have $\kappa_+>0$ and $\kappa_{-}<0$. Only the positive solution is  of interest. Specifically, we have
\allowdisplaybreaks
\begin{eqnarray}
    \kappa_{\star} = \kappa_{+}&=& \frac{-2 B + \sqrt{4B^2+4 A B} }{2 A}
    \notag\\
    &=& \frac{-2 B + \sqrt{4B^2+4 A B} }{2 A}  \times \underbrace{\frac{2 B + \sqrt{4B^2+4 A B}}{2 B + \sqrt{4B^2+4 A B}}}_{=1}
    \notag\\
    &=& \frac{4 A B}{4AB+2A\sqrt{4B^2+4AB}}
    \notag\\
        &=& \frac{4 A B}{4AB+\sqrt{(4AB)^2+16A^3 B}}
        \notag\\
        &=&   \frac{1}{1+\sqrt{1+A/B}},\notag
\end{eqnarray}
which shows that $\kappa_{\star} \in (0, 1/2)$.

Using the expression for $A/B$ in~\eqref{eq A/B} and $\epsilon^2 =  \frac{1-\tau^{\textnormal{im}}}{\tau^{\textnormal{im}}}\kappa$, we obtain
\begin{eqnarray}
    \epsilon_{\star}^2 &=& \frac{1-\tau^{\textnormal{im}}}{\tau^{\textnormal{im}}} \kappa_{\star}  =  \frac{1-\tau^{\textnormal{im}}}{\tau^{\textnormal{im}}}\frac{1}{1+\sqrt{1+A/B}} \notag\\
    &=& \frac{1-\tau^{\textnormal{im}}}{\tau^{\textnormal{im}}+\sqrt{(\tau^{\textnormal{im}})^2+ \tau^{\textnormal{im}} (1 - \tau^{\textnormal{im}})\frac{b + a\tilde{d}}{a\tilde{d}}}}\notag\\
    &=& \frac{1-\tau^{\textnormal{im}}}{\tau^{\textnormal{im}}+\sqrt{(\tau^{\textnormal{im}})^2+ \tau^{\textnormal{im}} (1 - \tau^{\textnormal{im}})(\frac{b}{a\tilde{d}}+1)}}\notag\\
    &=& \frac{1-\tau^{\textnormal{im}}}{\tau^{\textnormal{im}}+\sqrt{(\tau^{\textnormal{im}})^2+ \tau^{\textnormal{im}} (1 - \tau^{\textnormal{im}})\frac{b}{a\tilde{d}} + \tau^{\textnormal{im}} (1 - \tau^{\textnormal{im}})}} \notag \\
    &=& \frac{1-\tau^{\textnormal{im}}}{\tau^{\textnormal{im}}+\sqrt{\tau^{\textnormal{im}}+ \tau^{\textnormal{im}} (1 - \tau^{\textnormal{im}})\frac{b}{a\tilde{d}}}}\notag\\
    &=& \frac{1-\tau^{\textnormal{im}}}{\tau^{\textnormal{im}}+\sqrt{\tau^{\textnormal{im}}+ \tau^{\textnormal{im}} (1 - \tau^{\textnormal{im}})\frac{\sigma^2}{\lambda_{\max}(\mP)\lambda_{\min}(\mP)(\|\mI_k-\mathbf{\Phi}\|\, s_{\omega} + q)^2}}},\notag
\end{eqnarray}
where the final line uses the values of $b$ and $d$ for the ISD filter from Table~\ref{tab: overview}. Coerciveness (the bounds tend to infinity as $\kappa$ approaches either zero or one) and continuity imply that this unique stationary point is, in fact, a global minimum.
\end{proof}

\subsection{Minimizing MSE bound of ISD filter w.r.t.\ learning rate}
\label{sec: proof of optimal learning rate}

Under Assumptions~\ref{Assumption: Bounded moments}--\ref{Assumption: Correct specification}, the (Euclidean) MSE bound for the ISD filter derived from Theorem~\ref{Th: (Non-)asymptotic MSE bounds} contains no free parameters, although some freedom remains in selecting the penalty matrix. If we take the penalty matrix to be a scalar multiple of the identity, the value of this multiple that minimizes the asymptotic  MSE bound can be derived analytically.
\begin{corollary}[Optimal penalty parameter for the ISD filter]\label{Cor: Optimal penalty} Let Assumptions~\ref{ass1}, \ref{Assumption: Bounded moments}(a,b), and \ref{Assumption: Correct specification} hold. Furthermore, assume that (i) the postulated log-likelihood contribution is strongly concave (i.e., $\alpha > 0$), (ii) the penalty matrix is a scalar multiple of the identity matrix (i.e.,\ $\mP = \rho \mI_k$ for some $\rho > 0$), and (iii) $\sigma, \sigma_{\xi} > 0$ such that $b,d> 0$. Then the value of $\rho > 0$ that minimizes the asymptotic (Euclidean) MSE bound for the ISD filter is
    \begin{equation*}
        \rho_{\star}=\frac{\sigma^2\left(1-\|\mathbf{\Phi}_0\|^2\right)-\alpha^2 \sigma_{\xi}^2+\sqrt{\left(\alpha^2 \sigma_{\xi}^2-\sigma^2\left(1-\|\mathbf{\Phi}_0\|^2\right)\right)^2+4 \alpha^2 \sigma^2 \sigma_{\xi}^2}}{2 \alpha \sigma_{\xi}^2} < \infty.
    \end{equation*}
\end{corollary}

\begin{proof}
Using the result of Theorem~\ref{Th: (Non-)asymptotic MSE bounds} and that $\mP = \rho \mI_k$, we have that $a = \left(\frac{\rho}{\rho + \alpha}\right)^2 \in (0,1)$ and $b = a\rho^{-1}\sigma^2$ for the ISD filter with strong concavity ($\alpha>0$). In addition, when the DGP is a known state-space model (Assumption~\ref{Assumption: Correct specification} holds) and again using $\mP = \rho \mI_k$, we have that $c = \|\mathbf{\Phi}_0\|_{\mP}^2 = \|\mP^{1/2} \mathbf{\Phi}_0 \mP^{-1/2}\|^2 = \|\rho^{1/2} \mI_k \mathbf{\Phi}_0 \rho^{-1/2} \mI_k\|^2 = \|\mathbf{\Phi}_0 \|^2$ and $d = \rho \sigma^2_{\xi}$. Substituting these into the asymptotic filter $\mP$-weighted MSE bound in \eqref{eq: asymptotic error bound} and converting to the (Euclidean) MSE bound, we obtain
\begin{eqnarray}
    \limsup _{t \rightarrow\infty} \;\textnormal{MSE}_{t\mid t} &=& \frac{1}{\rho} \limsup _{t \rightarrow\infty} \;\textnormal{MSE}^{\mP}_{t\mid t} \leq \frac{1}{\rho}\frac{b+ad}{1-ac} = \frac{a\sigma^2/\rho^2 +a\sigma_\xi^2}{1-a\|\mathbf{\Phi}_0 \|^2}\notag \\
    &=& \frac{\sigma^2/\rho^2 +\sigma_\xi^2}{a^{-1}-\|\mathbf{\Phi}_0 \|^2} = \frac{\sigma^2/\rho^2 +\sigma_\xi^2}{\left(\frac{\rho + \alpha}{\rho}\right)^2-\|\mathbf{\Phi}_0 \|^2} = \frac{\sigma^2 +\rho^2\sigma_\xi^2}{(\rho + \alpha)^2- \rho^2\|\mathbf{\Phi}_0 \|^2},\notag
\end{eqnarray}
where the second line first multiplies by $a^{-1}/a^{-1} = 1$ and subsequently by $\rho^{2}/\rho^{2} = 1$.

To optimize the bound with respect to $\rho$, we consider the first-order condition:
\allowdisplaybreaks
\begin{eqnarray}
0 &=& \frac{\dd}{\dd \rho} \frac{\sigma^2 +\rho^2\sigma_\xi^2}{(\rho + \alpha)^2- \rho^2\|\mathbf{\Phi}_0 \|^2}  \notag \\
&=& \frac{2\rho\sigma_\xi^2 ((\rho + \alpha)^2- \rho^2\|\mathbf{\Phi}_0 \|^2)  - (\sigma^2 +\rho^2\sigma_\xi^2)(2(\rho + \alpha) - 2\rho \|\mathbf{\Phi}_0 \|^2  )}{((\rho + \alpha)^2- \rho^2\|\mathbf{\Phi}_0 \|^2)^2}\notag \\
&=&  \frac{2\rho\sigma_\xi^2 ((\rho + \alpha)^2- \rho^2\|\mathbf{\Phi}_0 \|^2)  - (\sigma^2 +\rho^2\sigma_\xi^2)(2(\rho + \alpha) - 2\rho \|\mathbf{\Phi}_0 \|^2  )}{\rho^{4} (a^{-1} - c)^2}, \label{Eq_FOC_rho}
\end{eqnarray}
where the second line multiplies with $\frac{1}{\rho^4 \rho^{-4}} = 1$ followed by using the definition of $a$ and $c$. Since $a>0$ (as it is a squared quantity), the contraction condition $ac < 1$ implies that the denominator in \eqref{Eq_FOC_rho} is positive, as $ac < 1 \Rightarrow 1 - ac > 0 \Rightarrow a^{-1} - c > 0 \Rightarrow \rho^{4} (a^{-1} - c)^2 > 0$. This means that to solve the first-order condition, we need to set the numerator equal to 0, i.e.,
\begin{eqnarray}
0 &=& 2\rho\sigma_\xi^2((\rho + \alpha)^2- \rho^2\|\mathbf{\Phi}_0 \|^2)  - (\sigma^2 +\rho^2\sigma_\xi^2)(2(\rho + \alpha) - 2\rho \|\mathbf{\Phi}_0 \|^2)  \notag \\
&=&   2\rho\sigma_\xi^2(\rho + \alpha)^2  - 2\rho^3\sigma_\xi^2  \|\mathbf{\Phi}_0 \|^2  - \sigma^2(2(\rho + \alpha) - 2\rho \|\mathbf{\Phi}_0 \|^2) - 2\rho^2\sigma_\xi^2(\rho + \alpha)  +  2\rho^3\sigma_\xi^2 \|\mathbf{\Phi}_0 \|^2\notag \\
&=&   2\rho\sigma_\xi^2(\rho + \alpha)^2   - 2\sigma^2(\rho + \alpha) + 2\rho\sigma^2\|\mathbf{\Phi}_0 \|^2 - 2\rho^2\sigma_\xi^2(\rho + \alpha) \notag \\
&=& 2\rho^3\sigma_\xi^2 + 2\rho\sigma_\xi^2\alpha^2 + 4\rho^2\sigma_\xi^2 \alpha - 2\rho\sigma^2 - 2\sigma^2\alpha + 2\rho\sigma^2\|\mathbf{\Phi}_0 \|^2 - 2\rho^3\sigma_\xi^2 - 2\rho^2\sigma_\xi^2\alpha \notag\\
&=& 2\rho\sigma_\xi^2\alpha^2 + 4\rho^2\sigma_\xi^2 \alpha - 2\rho\sigma^2 - 2\sigma^2\alpha + 2\rho\sigma^2\|\mathbf{\Phi}_0 \|^2 - 2\rho^2\sigma_\xi^2\alpha \notag \\
&=& \rho^2(2\sigma_\xi^2 \alpha) + \rho(2\sigma_\xi^2\alpha^2 - 2\sigma^2(1- \|\mathbf{\Phi}_0 \|^2))   - 2\sigma^2\alpha \notag\\
&=& A\rho^2 + B \rho + C, \notag
\end{eqnarray}
which yields a quadratic equation in $\rho$ with coefficients $A = 2\sigma_\xi^2 \alpha$, $B = 2\sigma_\xi^2\alpha^2 - 2\sigma^2(1- \|\mathbf{\Phi}_0 \|^2)$ and $C = - 2\sigma^2\alpha$ and solution
\begin{equation*}
\rho_{\pm} = \frac{- B \pm \sqrt{B^2 - 4 A C} }{2 A}.
\end{equation*}
Because $AC = -4 \sigma_\xi^2 \sigma^2 \alpha^2 < 0$ and $A = 2\sigma_\xi^2 \alpha > 0$, we have that $\rho_{-} < 0$ and $\rho_{+}>0$, and the latter is therefore the solution to the minimization problem at hand. Specifically, the final result reads
\begin{eqnarray}
\rho_{\star} &=& \rho_{+} = \frac{- 2\sigma_\xi^2\alpha^2 + 2\sigma^2(1- \|\mathbf{\Phi}_0 \|^2) + \sqrt{4\left(\sigma_\xi^2\alpha^2 - \sigma^2(1- \|\mathbf{\Phi}_0 \|^2)\right)^2  + 16 \sigma_\xi^2 \sigma^2 \alpha^2 } }{4\sigma_\xi^2 \alpha} \notag \\
&=& \frac{\sigma^2(1- \|\mathbf{\Phi}_0 \|^2) - \alpha^2 \sigma_\xi^2 + \sqrt{\left(\alpha^2\sigma_\xi^2 - \sigma^2(1- \|\mathbf{\Phi}_0 \|^2)\right)^2  + 4 \alpha^2\sigma_\xi^2 \sigma^2  } }{2\alpha\sigma_\xi^2},\notag
\end{eqnarray}
where the second line multiplies by $(1/2)/(1/2) = 1$ and rearranges. Because $A = 2\sigma_\xi^2 \alpha > 0$, we have that the numerator of \eqref{Eq_FOC_rho} is a convex parabola with largest root $\rho_{\star}$, which means that it becomes negative for $\rho \in (0, \rho_{\star})$ and positive for $\rho \in (\rho_{\star}, \infty)$. Combined with the fact that the denominator of \eqref{Eq_FOC_rho} is always positive, we have that the bound is decreasing in $\rho$ on the interval $(0,\rho_{\star})$ and increasing on the interval $(\rho_{\star}, \infty)$. In sum, this means that the stationary point $\rho_{\star}$ is indeed the global minimum.
\end{proof}

\subsection{Minimized MSE bound of ISD filter can be tight}
\label{Proof Example 4}

Example~\ref{Example: MSE bounds ISD filter AR(1) + noise model} below illustrates a case in which the minimized asymptotic MSE bound for the ISD filter is tight, as it equals the steady-state variance of the Kalman filter. The optimal learning-rate selection reveals a connection with the Kalman gain.

\begin{example}[Tight MSE bounds of ISD filter]\label{Example: MSE bounds ISD filter AR(1) + noise model}
Consider an AR(1) model with additive observation noise, defined by the observation equation $y_t = \vartheta_{t} + \varepsilon_t,$ where $ \varepsilon_t\sim\textnormal{i.i.d.}\, \rN(0, \sigma_{\varepsilon}^{ 2})$ with $\sigma_\varepsilon>0$, and linear first-order state dynamics $\vartheta_{t+1}=(1-\phi_{0}) \omega_{0} + \phi_{0} \vartheta_{t}+\xi_{t},$ where $\xi_t \sim\textnormal{i.i.d.}\, \rN(0, \sigma_{\xi}^{ 2})$ with $\sigma_\xi>0$.
Let the learning rate be positive and constant, $\mH_t = \mP_t^{-1} = \rho^{-1} = \eta>0$, for all $t$. As in \citeauthor{kalman1960new}'s classic setting, let  Assumption~\ref{Assumption: Correct specification} hold; that is, the true density $p^0(\cdot|\cdot)$ coincides with the postulated density $p(\cdot|\cdot)$, 
and the parameters $\omega_{0}$ and $\phi_{0}$ are known to the researcher. Then, the asymptotic MSE bound
        \begin{equation*}
            \limsup _{t \rightarrow\infty} \textnormal{MSE}_{t \mid t} \leq \frac{\eta^2 \sigma_\varepsilon^{ 2}  + \sigma^2_{\xi} \sigma_\varepsilon^{ 4}}{2 \eta \sigma_\varepsilon^{ 2} + \eta^2 + (1 - \phi_{0}^2)\sigma_\varepsilon^{ 4}},
        \end{equation*}
        is minimized when the learning rate $\eta$ is chosen as
        \begin{equation*}
            \eta_{\star} = \frac{\sigma_{\xi}^2-\sigma_{\varepsilon}^2\left(1-\phi_{0}^2\right)+\sqrt{\sigma_{\xi}^4+\sigma_{\varepsilon}^4\left(1-\phi_{0}^2\right)^2+2 \sigma_{\xi}^2 \sigma_{\varepsilon}^2\left(1+\phi_{0}^2\right)}}{2}.
        \end{equation*}
        For the local-level model, where $\phi_{0} = 1$, the minimizer $\eta_{\star}$ simplifies to
        \begin{equation*}
        \eta_{\star} = \frac{\sigma_{\xi}^2 +\sqrt{\sigma_{\xi}^4+4 \sigma_{\xi}^2 \sigma_{\varepsilon}^2}}{2} =\frac{\sigma_{\varepsilon}^2}{2}\left(\frac{\sigma^2_{\xi}}{\sigma^2_{\varepsilon}}+\sqrt{\frac{\sigma^4_{\xi}}{\sigma^4_{\varepsilon}} +4 \frac{\sigma^2_{\xi}}{\sigma^2_{\varepsilon}}}\right).
        \end{equation*}
       As it turns out, $\eta_{\star}$ equals the steady-state variance of prediction errors in the Kalman filter, as given in~\citet[p.\ 37]{durbin2012time} and~\citet[p.\ 175]{harvey1990forecasting}, where $\sigma^2_{\xi} / \sigma^2_{\varepsilon}$ is the signal-to-noise ratio.
       Substituting the minimizer $\eta_{\star}$ into the ISD update, we obtain the exact form of Kalman's level update with the (optimal) Kalman gain. Since the Kalman filter is optimal for this model, our minimized MSE bound is tight. 
\end{example}

\begin{proof} For readability, we omit all superscripts $\text{im}$. Under Assumptions~\ref{ass1}--\ref{Assumption: Correct specification}, the asymptotic (Euclidean) MSE bound for the ISD filter is obtained by substituting $a = \left(\frac{\rho}{\rho + \alpha}\right)^2$, $b = a\rho^{-1}\sigma^2$, $c=\|\mathbf{\Phi}_{0}\|^2$, and $d = \rho \sigma^2_{\xi}$ in equation~\eqref{eq: asymptotic error bound}, and multiplying by $\eta$ to obtain
\begin{equation}\label{eq: proof ISD MSE bound correct specification penalty simplification}
    \limsup _{t \rightarrow\infty} \textnormal{MSE}_{t \mid t} \leq \frac{\sigma^2+\sigma^2_{\xi}\rho^2}{\alpha^2 + 2 \alpha \rho + (1 - \|\mathbf{\Phi}_{0}\|^2)\rho^2} = \frac{\sigma^2 \eta^2 + \sigma^2_{\xi} }{\alpha^2 \eta^2+ 2 \alpha \eta + (1 - \|\mathbf{\Phi}_{0}\|^2)},
\end{equation}

\noindent where $\eta:=1/\rho$. In the context of Example~\ref{Example: MSE bounds ISD filter AR(1) + noise model}, $\mathbf{\Phi}_{0}=\phi_{0}$,  $\alpha = -\nabla^2 \ell (y_t|\theta_{t|t-1}) = 1/{\sigma_\varepsilon^2}$ and $\sigma^2 = \sup_{t} \underset{y_t}{\mathbb{E}}(\|\nabla \ell (y_t |\vartheta_t)\|^2) = 1/{\sigma_\varepsilon^2}$. To see this, consider the postulated log-likelihood function
\begin{equation*}
\ell(y_t|\theta_{t|t-1}) = -\frac{1}{2} \log(2\pi\sigma_\varepsilon^2) - \frac{(y_t - \theta_{t|t-1})^2}{2\sigma_\varepsilon^2}.
\end{equation*}
\noindent Taking the first derivative with respect to $\theta_{t|t-1}$, the score is
\begin{equation*}
\nabla \ell(y_t|\theta_{t|t-1}) = \frac{y_t - \theta_{t|t-1}}{\sigma_\varepsilon^2}.
\end{equation*}
\noindent The second derivative with respect to $\theta_{t|t-1}$ is
\begin{equation*}
\nabla^2 \ell(y_t|\theta_{t|t-1}) = -\frac{1}{\sigma_\varepsilon^2}.
\end{equation*}

\noindent Thus, $\alpha = 1/{\sigma_\varepsilon^2}$. Now, for $y_t \sim \mathrm{N}(\vartheta_t, \sigma_\varepsilon^2)$, the score evaluated at the true parameter is
\begin{equation*}
\nabla \ell(y_t|\vartheta_t) = \frac{y_t - \vartheta_t}{\sigma_\varepsilon^2},
\end{equation*}
and taking the squared norm, we obtain
\begin{equation*}
\|\nabla \ell(y_t|\vartheta_t)\|^2 = \frac{(y_t - \vartheta_t)^2}{\sigma_\varepsilon^4}.
\end{equation*}
Taking the expectation with respect to $y_t$, we get:
\begin{equation*}
\underset{y_t}{\mathbb{E}}\left(\|\nabla \ell(y_t|\vartheta_t)\|^2\right) = \frac{\underset{y_t}{\mathbb{E}}[(y_t - \vartheta_t)^2]}{\sigma_\varepsilon^4} = \frac{\sigma_\varepsilon^2}{\sigma_\varepsilon^4} = \frac{1}{\sigma_\varepsilon^2}.
\end{equation*}
Thus, $\sigma^2 = \frac{1}{\sigma_\varepsilon^2}$.
Next, we substitute $\mathbf{\Phi}_{0}=\phi_{0}$, $\alpha=1/\sigma_{\varepsilon}^2$ and $\sigma^2 = 1/\sigma_\varepsilon^2$ in equation~\eqref{eq: proof ISD MSE bound correct specification penalty simplification}, and subsequently multiply the numerator and denominator by $\sigma_{\varepsilon}^4>0$, to obtain
\begin{equation}
\limsup _{t \rightarrow\infty} \textnormal{MSE}_{t \mid t} \leq  \frac{\sigma_\varepsilon^{ 2} \eta^2 + \sigma^2_{\xi} \sigma_\varepsilon^{ 4}}{\eta^2 + 2  \sigma_\varepsilon^{ 2}\eta + (1 - \phi_{0}^2)\sigma_\varepsilon^{ 4}}.
\label{bound to be minimized}
\end{equation}
If $\eta=0$, we obtain an upper bound equal to $\sigma_\xi^2/(1-\phi_0^2)$, which is the unconditional variance of the true state, $\vartheta_t$. If $\eta=\infty$, we obtain an upper bound equal to $\sigma_\varepsilon^2$, which is the conditional variance of the observations. We hope to find an (optimal) learning rate $\eta>0$ that minimizes the upper bound, yielding a lower value than either $\eta=0$ or $\eta=\infty$.

To this end, we compute the derivative (with respect to $\eta$) of the bound~\eqref{bound to be minimized} and equate it to zero:
\begin{equation*}
\frac{\left[2 \sigma_{\varepsilon}^2 \eta\right]\left[2 \sigma_{\varepsilon}^2 \eta+\eta^2+\left(1-\phi_{0}^2\right) \sigma_{\varepsilon}^4\right]-\left[\sigma_{\varepsilon}^2 \eta^2+\sigma_{\xi}^2 \sigma_{\varepsilon}^4\right]\left[2 \sigma_{\varepsilon}^2+2 \eta\right]}{\left( \eta^2 +2\sigma_{\varepsilon}^2\eta  + (1 - \phi_{0}^2)\sigma_{\varepsilon}^4\right)^2}=0.
\end{equation*}
We will find that there exists a unique value of $\eta>0$ that solves this equation. Moreover, this unique value $\eta>0$ will turn out to deliver a lower MSE bound than either $\eta=0$ or $\eta=\infty$. Because the bound is continuously differentiable in $\eta\geq 0$, this means that we will have found the global minimum.

Given that the denominator above is positive,
finding the stationary point is equivalent to solving
\begin{equation*}
4 \sigma_{\varepsilon}^4 \eta^2+2 \sigma_{\varepsilon}^2 \eta^3+2 \sigma_{\varepsilon}^6 \eta\left(1-\phi_{0}^2\right)  -  \left(2 \sigma_{\varepsilon}^4 \eta^2+2 \sigma_{\varepsilon}^2 \eta^3+2 \sigma_{\xi}^2 \sigma_{\varepsilon}^6+2 \sigma_{\xi}^2 \sigma_{\varepsilon}^4 \eta\right)=0.
\end{equation*}
Canceling out $2 \sigma_{\varepsilon}^2 \eta^3$, using that $4 \sigma_{\varepsilon}^4 \eta^2-2 \sigma_{\varepsilon}^4 \eta^2=2 \sigma_{\varepsilon}^4 \eta^2$, and that $2 \sigma_{\varepsilon}^6 \eta(1-\phi_{0}^2) - 2 \sigma_{\xi}^2 \sigma_{\varepsilon}^4 \eta = \eta \,(2 \sigma_{\varepsilon}^6 \eta(1-\phi_{0}^2)-2 \sigma_{\xi}^2 \sigma_{\varepsilon}^4)$, combining the terms we get a quadratic equation in $\eta$:
\begin{equation*}
2 \sigma_{\varepsilon}^4 \eta^2+\eta\left(2 \sigma_{\varepsilon}^6\left(1-\phi_{0}^2\right)-2 \sigma_{\xi}^2 \sigma_{\varepsilon}^4\right)-2 \sigma_{\xi}^2 \sigma_{\varepsilon}^6=0.
\end{equation*}
Dividing each term by $2\sigma_{\varepsilon}^4>0$ leads to
\begin{equation*}
\eta^2+\eta\left(\sigma_{\varepsilon}^2\left(1-\phi_{0}^2\right)-\sigma_{\xi}^2\right)-\sigma_{\xi}^2 \sigma_{\varepsilon}^2=0.
\end{equation*}
Solving for $\eta_{\star}$ using the quadratic formula, we obtain
\begin{equation*}
\eta_{\pm}=\frac{\sigma_{\xi}^2-\sigma_{\varepsilon}^2\left(1-\phi_{0}^2\right) \pm \sqrt{D}}{2},
\end{equation*}
where $D=(\sigma_{\varepsilon}^2(1-\phi_{0}^2)-\sigma_{\xi}^2)^2+4 \sigma_{\xi}^2 \sigma_{\varepsilon}^2=\sigma_{\xi}^4+\sigma_{\varepsilon}^4(1-\phi_{0}^2)^2+2 \sigma_{\xi}^2 \sigma_{\varepsilon}^2(1+\phi_{0}^2)$. Because $\sqrt{D}>\sigma_{\xi}^2-\sigma_{\varepsilon}^2\left(1-\phi_{0}^2\right)$, we have $\eta_{+}>0$ and $\eta_{-}<0$. Only the positive solution is of interest. Specifically, we have
\begin{equation*}
\eta_{\star} = \eta_{+}=\frac{\sigma_{\xi}^2-\sigma_{\varepsilon}^2\left(1-\phi_{0}^2\right)+\sqrt{\sigma_{\xi}^4+\sigma_{\varepsilon}^4\left(1-\phi_{0}^2\right)^2+2 \sigma_{\xi}^2 \sigma_{\varepsilon}^2\left(1+\phi_{0}^2\right)}}{2}.
\end{equation*}
In the case of a local-level model (i.e., $ \phi_{0}= 1 $), we have $ a \times c = a \times \phi_{0} = \left(\frac{\rho}{\rho + \alpha}\right)^2 < 1$ as $\alpha = 1/{\sigma_\varepsilon^2}$. Thus, the optimal learning rate reduces to
\begin{equation}
\eta_{\star}=\frac{\sigma_{\xi}^2+\sqrt{\sigma_{\xi}^4+4 \sigma_{\xi}^2 \sigma_{\varepsilon}^2}}{2}
=\frac{\sigma_{\varepsilon}^2}{2}\left(\frac{\sigma^2_{\xi}}{\sigma^2_{\varepsilon}}+\sqrt{\frac{\sigma^4_{\xi}}{\sigma^4_{\varepsilon}} +4 \frac{\sigma^2_{\xi}}{\sigma^2_{\varepsilon}}}\right),
\label{optimal learning rate}
\end{equation}
which is the steady-state covariance associated with  predictions in the Kalman filter, where $\sigma^2_{\xi}/\sigma^2_{\varepsilon}$ is the signal-to-noise ratio. 

Substituting $\eta_\star$ from equation~\eqref{optimal learning rate} back into the bound~\eqref{bound to be minimized}, which was to be minimized, we can confirm (after some algebra) that the unique stationary point yields a value that does not exceed the value at either endpoint (i.e., at $\eta=0$ or $\eta=\infty$). By continuous differentiability of the MSE bound in $\eta\geq 0$, the stationary point yields the global minimum as desired.
\end{proof}

\setcounter{equation}{0}
\section{Detailed discussion and further numerical results}
\label{Appendix C. Further numerical results}

\subsection{Detailed discussion of ISD and ESD updates~\eqref{eq: Implicit optimization problem}--\eqref{Explicit optimization problem}}
\label{A: discussion overshooting}

In optimization~\eqref{eq: Implicit optimization problem}, the optimal value of the objective function (i.e.,\ when evaluated at the argmax) must exceed the (suboptimal) value at any other point (e.g.,\ at the predicted parameter). This fact yields
$\ell(\vy_t|\vtheta_{t|t}^\textnormal{im})-\frac{1}{2}\|\vtheta_{t|t}^{\textnormal{im}}-\vtheta_{t|t-1}^\textnormal{im}\|_{\mP_t}^2 \geq \ell(\vy_t|\vtheta_{t|t-1}^\textnormal{im}),$
where the left-hand side is the optimized value. After rearrangement, this implies
$$\ell(\vy_t|\vtheta_{t|t}^\textnormal{im})-\ell(\vy_t|\vtheta_{t|t-1}^\textnormal{im})\; \geq\; 1/2\|\vtheta_{t|t}^{\textnormal{im}}-\vtheta_{t|t-1}^\textnormal{im}\|_{\mP_t}^2 \;\geq \; 0,$$
which yields two desirable consequences: (i) the fit is improved at every time step, i.e.,\ $\ell(\vy_t|\vtheta_{t|t}^\textnormal{im})-\ell(\vy_t|\vtheta_{t|t-1}^\textnormal{im})\geq 0$,
and (ii) the stepsize is bounded, i.e.,\ $\|\vtheta_{t|t}^{\textnormal{im}}-\vtheta_{t|t-1}^\textnormal{im}\|_{\mP_t}<\infty$, as long as
$\vtheta\mapsto \ell(\vy_t|\vtheta)$ is upper bounded (almost surely in $\vy_t$) and $\ell(\vy_t|\vtheta
_{t|t-1}^\textnormal{im})\neq -\infty$.
Hence the boundedness of the implicit update derives \emph{not} from the boundedness of the gradient, but from the upper boundedness of the objective function itself.

In contrast, the solution~\eqref{Explicit parameter update step} to the linearized update~\eqref{Explicit optimization problem} may be prone to ``overshooting''; i.e.,\ unless the learning rate is very small, the undesirable situation $\ell(\vy_t|\vtheta^{\textnormal{ex}}_{t|t})<\ell(\vy_t|\vtheta^{\textnormal{ex}}_{t|t-1})$ may regularly occur. In Section~\ref{sec:Error bounds for score-driven filters} we find that for the explicit method to asymptotically achieve bounded filtering errors over time,  we require that, almost surely in $\vy_t$, the driving mechanism $ \mH_t \nabla \ell(\vy_t|\vtheta^\textnormal{ex}_{t|t-1})$ is Lipschitz in $\vtheta^\textnormal{ex}_{t|t-1}$. This additional condition, which is not needed for the implicit method, is required to prevent the explicit method from repeatedly overshooting and, possibly, diverging.

\subsection{Details for Section~\ref{subsec:Least-squares recovery.}}
\label{subsec:Simulation study 1: Tracking under default settings}

\begin{figure}[h]
    \centering
        \includegraphics[width=0.5\textwidth]{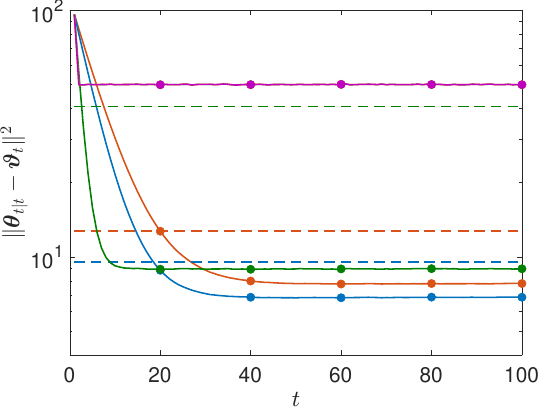}
            \vskip\baselineskip
        \begin{subfigure}{1\textwidth}
            \centering
            \includegraphics[width=\textwidth]{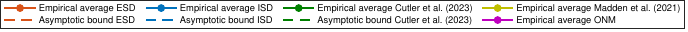}
            \label{fig:legend3}
        \end{subfigure}
        \vspace{-10mm}
        \caption{Semilog plot of guaranteed bounds and empirical tracking errors for least-squares recovery with respect to iteration $t$. Empirical averages are computed over $10{,}000$ trials. Default parameter values: $\alpha=1, \beta=1, \sigma=10, \sigma_{\xi}=1, \eta=\eta_{\star}$, $k=50$, $n=100$.}
        \label{fig:Tracking when alpha=beta=1}
\end{figure}

Figure~\ref{fig:Tracking when alpha=beta=1} presents the MSE performance of the ISD and ESD filters alongside three recent competitors in the setting of Section~\ref{subsec:Least-squares recovery.}. In this case, however, we follow~\cite{cutler2023stochastic} by setting $\alpha=\beta=1$. As a result, the ``momentum'' term in the ONM algorithm vanishes, reducing it to the stochastic gradient descent method from~\cite{madden2021bounds}. As opposed to when $\alpha=1, \beta=40$, the learning rate in the~\cite{cutler2023stochastic} algorithm no longer remains fixed at its cap, $1/(2\beta)$. While~\citeauthor{cutler2023stochastic}'s (\citeyear{cutler2023stochastic}) algorithm performs well for small $t$, the ISD and ESD filters perform empirically better (and have better performance guarantees) for large $t$.

\subsection{Details for Section~\ref{subsec:Koopman DGPs Monte Carlo}}
\label{app: DGPs}

\begin{landscape}
\begin{table}[h!]
\caption{\label{tab:DGPs} Overview of data-generating processes in simulation studies.}
\begin{footnotesize}
\begin{threeparttable}
\begin{tabular}{l@{\hspace{0.15cm}}l@{\hspace{0.15cm}}c@{\hspace{0.15cm}}c@{\hspace{0.15cm}}c@{\hspace{0.15cm}}c@{\hspace{0.15cm}}c}
  \toprule
\multicolumn{2}{l}{\bf{DGP}} & \bf{Link function}  & \bf{Density} & \bf{Score} & \bf{Negative Hessian}&  \bf{Information}
 \\
Type & Distribution &   & $p({y}_t|\theta_t)$ & $\nabla \ell(y_t|\theta_t)$ & $-\nabla^2 \ell(y_t|\theta_t)$ &
\\
  \cmidrule(r{10pt}){1-2}  \cmidrule(lr){3-3} \cmidrule(lr){4-4} \cmidrule(lr){5-5} \cmidrule(lr){6-6} \cmidrule(lr){7-7}

  Count & Poisson & $\lambda_t=\exp(\theta_t)$ &$\displaystyle\lambda_t^{y_t}\, \exp(-\lambda_t)/y_t! $ & $y_t-\lambda_t$ & $\lambda_t$ & $\lambda_t$
\\
Count  & Negative bin.& $\lambda_t=\exp(\theta_t)$ & $\displaystyle \frac{\Gamma(\kappa+y_t)\left(\frac{\kappa}{\kappa+\lambda_t}\right)^\kappa\left(\frac{\lambda_t}{\kappa+\lambda_t}\right)^{y_t}}{\Gamma(\kappa)\Gamma(y_t+1)}$ &$\displaystyle y_t-\frac{\lambda_t(\kappa+y_t)}{\kappa+\lambda_t}$ & $\displaystyle \frac{\kappa \lambda_t(\kappa+y_t)}{(\kappa+\lambda_t)^2}$ & $\displaystyle \frac{\kappa\,\lambda_t}{\kappa+\lambda_t}$
\\
Intensity & Exponential & $\lambda_t=\exp(\theta_t)$ & $\displaystyle\lambda_t\, \exp(-\lambda_t y_t)$ &  $1-\lambda_t\,y_t$ & $y_t \lambda_t$ & $1$
\\
Duration & Gamma & $\beta_t=\exp(\theta_t)$ & $\displaystyle
\frac{y_t^{\kappa-1}\exp(-y_t/\beta_t)}{\Gamma(\kappa)\beta_t^\kappa }
$ & $\displaystyle \frac{y_t}{\beta_t}-\kappa$& $\displaystyle \frac{y_t}{\beta_t}$ & $\kappa$
\\
Duration & Weibull  & $\beta_t=\exp(\theta_t)$ & $\displaystyle\frac{\kappa\, \left(y_t/\beta_t \right)^{\kappa-1} }{\beta_t \exp\{(y_t/\beta_t)^\kappa\} }
$& $\displaystyle \kappa\left(\frac{y_t}{\beta_t}\right)^\kappa-\kappa$& $\displaystyle \kappa^2 \left(\frac{y_t}{\beta_t}\right)^\kappa$&$\kappa^2$
\\
Volatility & Gaussian & $\sigma^2_t=\exp(\theta_t)$ &$ \displaystyle
\frac{\exp\{-y_t^2/(2\sigma_t^2)\}}{ \{2\pi \sigma_t^2\}^{1/2} }$ & $\displaystyle \frac{y_t^2}{2\sigma_t^2}-\frac{1}{2}$
&$\displaystyle \frac{y_t^2}{2\sigma_t^2}$ &$\displaystyle\frac{1}{2}$
\\
Volatility  & Student's $t$& $\sigma^2_t=\exp(\theta_t)$ &  $\displaystyle\frac{\Gamma\left(\frac{\nu+1}{2}\right)\left(1+\frac{y_t^2}{(\nu-2)\sigma_t^2}\right)^{-\frac{\nu+1}{2}}}{\sqrt{(\nu-2)\pi}\Gamma\left(\nu/2\right)\sigma_t}$ &
$\displaystyle \frac{\omega_t \, y_t^2}{2\sigma_t^2}-\frac{1}{2}$
& $\displaystyle \frac{\nu-2}{\nu+1}\,\frac{\omega_t^2\,y_t^2}{2\sigma_t^2}$
& $\displaystyle \frac{\nu}{2\nu+6}$ \\
&&&& $\displaystyle \omega_t:=\frac{\nu+1}{\nu-2+y_t^2/\sigma_t^2}$ & &
\\
Dependence & Gaussian & $\displaystyle\rho_t=\frac{1-\exp(-\theta_t)}{1+\exp(-\theta_t)}$ & $\displaystyle\frac{\exp\left\{-\frac{y_{1t}^2+y_{2t}^2-2\rho_t y_{1t}y_{2t} }{2(1-\rho_t^2)} \right\}}{2\pi \sqrt{1-\rho_t^2}}$ & $
\displaystyle \frac{\rho_t}{2}+
\frac{1}{2}\frac{ z_{1t}\,z_{2t}}{1-\rho_t^2} $
& $\displaystyle 0\nleq \frac{1}{4}\frac{z_{1t}^2+z_{2t}^2}{1-\rho_t^2}- \frac{1-\rho_t^2}{4}$
& $\displaystyle \frac{1+\rho_t^2}{4}$
\\
&& $\displaystyle$ && $z_{1t}:=y_{1t}-\rho_t y_{2t}$
\\
&&&& $z_{2t}:=y_{2t}-\rho_t y_{1t}$
\\
Dependence & Student's $t$ & $\displaystyle\rho_t=\frac{1-\exp(-\theta_t)}{1+\exp(-\theta_t)}$ & $\displaystyle\frac{\nu \left(1+\frac{y_{1t}^2+y_{2t}^2-2\rho_t y_{1t}y_{2t} }{(\nu-2)(1-\rho_t^2)}\right)^{-\frac{\nu+2}{2}}}{2\pi(\nu-2) \sqrt{1-\rho_t^2}}$ & $\displaystyle \frac{\rho_t}{2}+
\frac{\omega_t}{2}\frac{ z_{1t}\,z_{2t}}{1-\rho_t^2} $
& $\displaystyle 0\nleq \frac{\omega_t}{4}\frac{z_{1t}^2+z_{2t}^2}{1-\rho_t^2}-\frac{1-\rho_t^2}{4}-\frac{1}{2}\frac{\omega_t^2}{\nu+2}\frac{z_{1t}^2 \, z_{2t}^2}{(1-\rho_t^2)^2}$
& $\displaystyle \frac{2+\nu(1+\rho_t^2)}{4(\nu+4)}$
\\
&& $\displaystyle$ && $z_{1t}:=y_{1t}-\rho_t y_{2t}$ &  \multirow{2}{*}{$\displaystyle \omega_t:=\frac{\nu+2}{\nu-2+\frac{y_{1t}^2+y_{2t}^2-2\rho_t y_{1t}y_{2t} }{1-\rho_t^2}}$}
\\
&&&& $z_{2t}:=y_{2t}-\rho_t y_{1t}$
\\
\\
  \bottomrule
\end{tabular}
\begin{tablenotes}
\item Note: The table contains nine distributions and link functions taken from~\citet{koopman2016predicting}. To aid the comparison (but at the expense of some consistency), we retain most of their parameter notation, even though some symbols are also used in the main text for different quantities. We assume that the observation density is known, i.e., $p^0(\cdot|\vartheta_t)=p(\cdot|\theta_t)$ (hence, $\vartheta_t=\theta_t$). In each case, the state-transition equation is $\vartheta_t=(1-\phi_0)\omega_0+\phi_0\vartheta_{t-1}+\xi_t$ where $\xi_t\sim \text{i.i.d.}(0,\sigma_\xi^2)$ is distributed as either a Gaussian or Student's $t$ variate.
\end{tablenotes}
\end{threeparttable}
\end{footnotesize}
\end{table}
\end{landscape}

\begin{table}[ht]
\centering
\caption{Static (hyper-)parameters for DGPs in Table~\ref{tab:DGPs} in the low-volatility setting} \label{tab:dgpparametervalues}
\begin{threeparttable}

\begin{tabular}{@{}llcccccl@{}}
\toprule
\textbf{Type}       & \textbf{Distribution} & $\omega_{0}$ & $\phi_{0}$ & $\sigma_{\xi}$ & \textbf{Shape} \\ \midrule
Count               & Poisson               & 0.00        & 0.97  &      0.15            &                \\
Count               & Negative binomial     & 0.00        & 0.97  &      0.15            & $\kappa = 4$      \\ \midrule
Intensity           & Exponential           & 0.00        & 0.97  &      0.15            &                \\ \midrule
Duration            & Gamma                 & 0.00        & 0.97  &     0.15            & $\kappa = 1.5$    \\
Duration            & Weibull               & 0.00        & 0.97  &    0.15            & $\kappa = 1.2$    \\ \midrule
Volatility          & Gaussian              & 0.00        & 0.97  &    0.15            &                \\
Volatility          & Student's $t$         & 0.00        & 0.97  &    0.15            & $\nu = 6$       \\ \midrule
Dependence              & Gaussian              & 0.00        & 0.97  &     0.15            &                \\
Dependence              & Student's $t$         & 0.00        & 0.97  &    0.15            & $\nu = 6$       \\ \bottomrule
\end{tabular}
\end{threeparttable}
\end{table}

{
\begin{table}
\centering
\caption{\label{tab: MSE Gaussian} Out-of-sample MSE of $\{\theta^j_{t|t-1}\}$ for $j\in\{\textnormal{im},\textnormal{ex}\}$ relative to true states $\{\vartheta_t\}$ under Gaussian state increments.}
\begin{footnotesize}
\begin{threeparttable}
\begin{tabular}{l@{\hspace{0.3cm}}l@{\hspace{0.1cm}}c@{\hspace{0.1cm}}c@{\hspace{0.1cm}}c@{\hspace{0.1cm}}c@{\hspace{0.1cm}}c@{\hspace{0.1cm}}c@{\hspace{0.1cm}}c@{\hspace{0.1cm}}c@{\hspace{0.1cm}}c@{\hspace{0.1cm}}c@{\hspace{0.1cm}}c@{\hspace{0.1cm}}c}
\toprule
\multirow{2}{*}{\bf DGP} &
&   \multicolumn{2}{c}{\multirow{2}{*}{\bf Assumption~\ref{ass1}}}
& \multicolumn{2}{c}{\multirow{1}{*}{\bf MSE}}
&  \multicolumn{2}{c}{\bf Low volatility}
& \multicolumn{2}{c}{\bf Medium volatility}
& \multicolumn{2}{c}{\bf High volatility}
\\
  & & & &	\multicolumn{2}{c}{\multirow{1}{*}{\bf bound?}}   &   \multicolumn{2}{c}{($\sigma_\xi=0.15$)} & \multicolumn{2}{c}{($\sigma_\xi=0.30$)} & \multicolumn{2}{c}{($\sigma_\xi=0.60$)}
\\

\cmidrule(r{5pt}l{5pt}){1-2} \cmidrule(r{5pt}l{5pt}){3-4} \cmidrule(r{5pt}l{5pt}){5-6} \cmidrule(r{5pt}l{5pt}){7-8} \cmidrule(r{5pt}l{5pt}){9-10} \cmidrule(r{5pt}l{5pt}){11-12}

Type &	Distribution &
\multicolumn{1}{c}{$\alpha$} & \multicolumn{1}{c}{$\beta$} & \multicolumn{1}{c}{ISD} & \multicolumn{1}{c}{ESD} & \multicolumn{1}{c}{ISD} & \multicolumn{1}{c}{ESD} & \multicolumn{1}{c}{ISD} & \multicolumn{1}{c}{ESD} & \multicolumn{1}{c}{ISD} & \multicolumn{1}{c}{ESD}	\\

\cmidrule(r{5pt}l{5pt}){1-2} \cmidrule(r{5pt}l{5pt}){3-4} \cmidrule(r{5pt}l{5pt}){5-6} \cmidrule(r{5pt}l{5pt}){7-8} \cmidrule(r{5pt}l{5pt}){9-10} \cmidrule(r{5pt}l{5pt}){11-12}

Count & 	Poisson
&  \multicolumn{1}{c}{0}
& \multicolumn{1}{c}{$\infty$}
& \cmark
& \xmark
& \multicolumn{1}{c}{$0.146$}
& \multicolumn{1}{c}{$0.148\phantom{^\dagger}$}
& \multicolumn{1}{c}{$0.405$}
& \multicolumn{1}{c}{$\infty$}
&  \multicolumn{1}{c}{$1.64$ }
& \multicolumn{1}{c}{$\infty$ }

\\
Count &	Neg. Binomial
& 	\multicolumn{1}{c}{0}
& \multicolumn{1}{c}{$\infty$}
& \cmark
& \xmark
& \multicolumn{1}{c}{$0.158$}
& \multicolumn{1}{c}{$0.160\phantom{^\dagger}$}
& \multicolumn{1}{c}{$0.430$}
& \multicolumn{1}{c}{$0.452$}
& \multicolumn{1}{c}{$1.64$ }
& \multicolumn{1}{c}{$1.85$ }
\\

Intensity  &	Exponential
& \multicolumn{1}{c}{0}
& \multicolumn{1}{c}{$\infty$}
& \cmark
& \xmark
& \multicolumn{1}{c}{$0.146$}
& \multicolumn{1}{c}{$0.150\phantom{^\dagger}$}
& \multicolumn{1}{c}{$0.368$}
& \multicolumn{1}{c}{$0.423$}
&\multicolumn{1}{c}{$1.03$}
& \multicolumn{1}{c}{$	2.24$}
\\

Duration  &	Gamma
& \multicolumn{1}{c}{0}
& \multicolumn{1}{c}{$\infty$}
& \cmark
& \xmark
& \multicolumn{1}{c}{$0.157$}
& \multicolumn{1}{c}{$0.163\phantom{^\dagger}$}
& \multicolumn{1}{c}{$0.473$}
& \multicolumn{1}{c}{$0.546$}
& \multicolumn{1}{c}{$1.29$}
& \multicolumn{1}{c}{$2.24$}

\\

Duration  &	Weibull
& \multicolumn{1}{c}{0}
& \multicolumn{1}{c}{$\infty$}
& \cmark
& \xmark
& \multicolumn{1}{c}{$0.125$}
& \multicolumn{1}{c}{$0.128\phantom{^\dagger}$}
& \multicolumn{1}{c}{$0.306$}
& \multicolumn{1}{c}{$0.329$}
& \multicolumn{1}{c}{$0.80$}
& \multicolumn{1}{c}{$0.96$}

\\
Volatility  &	Gaussian
& \multicolumn{1}{c}{0}
& \multicolumn{1}{c}{$\infty$}
& \cmark
& \xmark
& \multicolumn{1}{c}{$0.192$}
& \multicolumn{1}{c}{$0.198\phantom{^\dagger}$}
& \multicolumn{1}{c}{$0.503$}
& \multicolumn{1}{c}{$	0.609$}
& \multicolumn{1}{c}{$1.46$}
& \multicolumn{1}{c}{$4.68$}

\\

Volatility  &	Student's \emph{t}
& \multicolumn{1}{c}{$0$}
& \multicolumn{1}{c}{$\frac{\nu+1}{8}$}
& \cmark
& \cmark
& \multicolumn{1}{c}{$0.226$}
& \multicolumn{1}{c}{$0.227\phantom{^\dagger}$}
& \multicolumn{1}{c}{$0.610$}
& \multicolumn{1}{c}{$0.617$}
& \multicolumn{1}{c}{$1.55$}
& \multicolumn{1}{c}{$	1.60$}
\\

Dependence  &	Gaussian
& \multicolumn{1}{c}{$-\frac{1}{4}$}
& \multicolumn{1}{c}{$\infty$}
& \cmark
& \xmark
& \multicolumn{1}{c}{$0.239$}
& \multicolumn{1}{c}{$0.231^\dagger$}
& \multicolumn{1}{c}{$0.596$}
& \multicolumn{1}{c}{$\infty$}
& \multicolumn{1}{c}{$1.56$}
&  \multicolumn{1}{c}{$\infty$}
\\

Dependence  &	Student's \emph{t}
& \multicolumn{1}{c}{$-\frac{1}{4}$}
& \multicolumn{1}{c}{$\frac{\nu+1}{4}$}
& \cmark
& \cmark
& \multicolumn{1}{c}{$0.251$}
& \multicolumn{1}{c}{$0.252\phantom{^\dagger}$}
& \multicolumn{1}{c}{$0.620$}
& \multicolumn{1}{c}{$0.625$}
& \multicolumn{1}{c}{$1.50$}
& \multicolumn{1}{c}{$1.53$}

\\
\bottomrule
 \end{tabular}
\begin{tablenotes}
\item \emph{Note}: MSE = mean squared error. ISD = implicit score driven. ESD = explicit score driven. {$\dagger=$} For the Gaussian dependence model, the ESD filtered path diverged for a single (out-of-sample) path in the low-volatility setting. For simplicity, we ignore this path and report a finite MSE.
\end{tablenotes}
\end{threeparttable}
\end{footnotesize}
\end{table}
}

\subsection{Computing the ISD update}
\label{subsec:Computing the implicit-gradient update}

Standard Newton-Raphson iterates for solving the ISD update~\eqref{eq: Implicit optimization problem} with $\mP_t=\mP$ read
\begin{equation}
\label{NR iterates}
\vtheta_{t|t}^\textnormal{im} \; \leftarrow \; \vtheta_{t|t}^\textnormal{im}
\,+\, \big[\mP - \nabla^2 \ell(\vy_t|\vtheta^\textnormal{im}_{t|t})\big]^{-1} \;\big[\, \nabla \ell(\vy_t|\vtheta^\textnormal{im}_{t|t}) \,-\, \mP\, (\vtheta^\textnormal{im}_{t|t}- \vtheta^\textnormal{im}_{t|t-1})\, \big] ,
\end{equation}
where $\nabla^2:=\nabla \nabla^{\top}=(\dd/\dd\vtheta)(\dd/\dd\vtheta)^{\top}$
denotes the Hessian operator.
The algorithm may be initialized with $\vtheta^\textnormal{im}_{t | t} \leftarrow \vtheta^\textnormal{im}_{t | t-1}$. The inverse exists as $\mP-\nabla^2 \ell(\vy_t|\vtheta^\textnormal{im}_{t|t})$ is positive definite because of Assumption~\ref{ass1}(b). Adding a standard line search is typically helpful.

For high-dimensional problems, it may be beneficial to employ an algorithm that avoids large-matrix inversions, such as the Broyden–Fletcher–Goldfarb–Shanno (BFGS) algorithm~(e.g., \citealp{liu1989limited}). When computational efficiency is critical, algorithm~\eqref{NR iterates} may be terminated after a single NR iteration, in which case the output (after one iteration) reads $\vtheta_{t | t-1}^\textnormal{im} + [ \mP - \nabla^2 \ell(\vy_t | \vtheta^\textnormal{im}_{t | t-1})]^{-1} \nabla \ell(\vy_t | \vtheta^\textnormal{im}_{t \mid t-1}) $. This ``1NR'' version is similar to the \textit{explicit} update~\eqref{Explicit parameter update step} in being computationally inexpensive; however, it is based on a quadratic (rather than linear) approximation of $\ell(\vy_t | \vtheta)$ around the prediction, which is advantageous when $\vtheta\mapsto \ell(\vy_t|\vtheta)$ exhibits strong curvature. On the other hand, additional iterations typically provide additional precision; hence, depending on the available computer power, researchers may decide to execute more or fewer iterations of algorithm~\eqref{NR iterates}.

\subsection{Details for Section~\ref{subsec:Dynamic Poisson distribution with misspecified link function.}}
\label{subsec:Estimated learning rates}

Here, we provide additional implementation details for the ESD and ISD filters, as well as the MSE bounds for the latter from Section~\ref{subsec:Dynamic Poisson distribution with misspecified link function.}.

\textbf{ESD filter.} For the ESD filters with \emph{exponential} link functions, $\mu^\textnormal{ex}_{t|t} = \,\exp(\theta^\textnormal{ex}_{t|t})$, we define the learning rates as $\mH_t^{\textnormal{ex}} = \eta^{\textnormal{ex}}  \, \mathcal{I}(\theta_{t|t-1}^{\textnormal{ex}})^{-\zeta}$, where $\mathcal{I}(\theta) = \exp(\theta)$ represents the Fisher information and $\zeta \in \{0, 1/2, 1\}$ determines the scaling of the score function. The ESD update becomes
\begin{equation}
    \theta^{\textnormal{ex}}_{t|t} = \theta^{\textnormal{ex}}_{t|t-1} + \eta^{\textnormal{ex}}  \, \exp(-\zeta\theta^{\textnormal{ex}}_{t|t-1}) (y_t-\exp(\theta^{\textnormal{ex}}_{t|t-1})).\notag
\end{equation}
Combined with the prediction step~\eqref{Parameter prediction step}, this leads to the standard score-driven prediction-to-prediction recursion (e.g., \citealp[p.\ 779]{creal2013generalized}).

\textbf{ISD filter.} The ISD filter with \emph{exponential} link function, $\mu^\textnormal{im}_{t|t} = \,\exp(\theta^\textnormal{im}_{t|t})$,  and static learning rate, $\mH_t^{\textnormal{im}} = \eta^{\textnormal{im}}$ for all $t$, has update
\begin{align}\label{eq:Poisson correctly specified ISD update}
    \theta_{t \mid t}^{\textnormal{im}}&=\underset{\theta\in \Theta}{\operatorname{argmax}} \; \left\{
    \ell \left(y_t \mid \theta \right) - \frac{1}{2\eta^{\textnormal{im}}}\left(\theta-\theta^\textnormal{im}_{t \mid t-1}\right)^2\right\} \notag \\
    &= \underset{\theta\in \Theta}{\operatorname{argmax}} \; \left\{
    y_t \, \theta\, -\exp(\theta) - \log(y_t!) - \frac{1}{2\eta^{\textnormal{im}}}\left(\theta-\theta^\textnormal{im}_{t \mid t-1}\right)^2\right\},\notag
\end{align}
which is solved numerically using a standard Newton{-}Raphson algorithm (see Appendix~\ref{subsec:Computing the implicit-gradient update}).
The ISD filter with \emph{quadratic} link function, $\mu^\textnormal{im}_{t|t} = \,(\theta^{\textnormal{im}}_{t|t})^2$, and static learning rate, has an implicit update that can be solved analytically as
\begin{align}
    \theta_{t \mid t}^{\textnormal{im}}&=\underset{\theta\in \Theta}{\operatorname{argmax}} \; \left\{
    \ell \left(y_t \mid \theta \right) - \frac{1}{2\eta^{\textnormal{im}}}\left(\theta-\theta^\textnormal{im}_{t \mid t-1}\right)^2\right\}, \notag \\
    &= \underset{\theta\in \Theta}{\operatorname{argmax}} \; \left\{
    y_t \log(\theta^{2}) - \theta^{2} - \log(y_t!) - \frac{1}{2\eta^{\textnormal{im}}}\left(\theta-\theta^\textnormal{im}_{t \mid t-1}\right)^2\right\}.\notag
\end{align}
Taking first-order conditions with respect to $\theta$, and evaluating at $\theta=\theta_{t \mid t}^{\textnormal{im}}$, yields
\begin{equation*}
    \frac{2y_t}{\theta_{t|t}^\textnormal{im}} - 2\theta_{t|t}^\textnormal{im} - \frac{1}{\eta^{\textnormal{im}}}\left(\theta^\textnormal{im}_{t|t} -\theta^\textnormal{im}_{t \mid t-1}\right)=0.
\end{equation*}
Next, we assume $\theta_{t|t}^\textnormal{im}>0$, which is without loss of generality,
 as the case $\theta_{t|t}^\textnormal{im}=0$ (which occurs when $y_t=\theta^\textnormal{im}_{t|t-1}=0$) will be automatically covered below.
As $\eta^{\textnormal{im}}>0$, we can then multiply both sides by $\eta^{\textnormal{im}}\theta_{t|t}^\textnormal{im}>0$ to obtain a quadratic equation in $\theta_{t|t}^\textnormal{im}$ as follows:
\begin{equation*}
    -(1+2\eta^{\textnormal{im}})(\theta_{t|t}^{\textnormal{im}})^2 +\theta^\textnormal{im}_{t|t-1}\theta_{t|t}^\textnormal{im} + 2 \eta^{\textnormal{im}} y_t=0.
\end{equation*}
Solving for $\theta_{t|t}^\textnormal{im}$ by the usual formula yields two potential solutions:
\begin{equation*}
    \theta_{t|t}^\textnormal{im} = \frac{-\theta^\textnormal{im}_{t|t-1} \pm \sqrt{(\theta_{t|t-1}^{\textnormal{im}})^2 + 8(1+2\eta^{\textnormal{im}})\eta^{\textnormal{im}} y_t}}{-2(1+2\eta^{\textnormal{im}})}.
\end{equation*}
Multiplying the numerator and denominator by $-1$ and taking only the nonnegative solution, we obtain the ISD update
\begin{equation*}
    \theta^{\textnormal{im}}_{t|t} = \frac{\theta^{\textnormal{im}}_{t|t-1} + \sqrt{(\theta_{t|t-1}^{\textnormal{im}})^2 + 8\eta^{\textnormal{im}}(1+2\eta^{\textnormal{im}})y_t}}{2(1+2 \eta^{\textnormal{im}})}\geq 0.
\end{equation*}
This solution yields $\theta^\textnormal{im}_{t|t}=0$ if and only if $\theta^\textnormal{im}_{t|t-1}=y_t=0$, such that the limiting case is correctly covered.

\textbf{MSE bounds.} The gradient noise (in Assumption~\ref{Assumption: Bounded moments}) of the ISD filter with quadratic link can be computed as
\begin{align}
    \sigma^2 & = \sup_{t}\mathbb{E}\left[\left(\frac{\textnormal{d}\ell(y_t|\theta_t)}{\textnormal{d}\theta_t}\right)^2\mid_{\theta_t = \theta_t^{\star}}\right] \notag \\
    &=\sup_{t}\mathbb{E}\left[\left( \frac{2y_t}{\theta_t}-2 \theta_t \right)^2\mid_{\theta_t =\theta_t^{\star}}\right]\notag \\
    &=\sup_{t}\mathbb{E}\left[\left( \frac{4y_t^2}{\theta_t^2}-8 y_t + 4 \theta_t^2 \right)\mid_{\theta_t =\theta_t^{\star}} \right] \notag \\
    &=\sup_{t}\mathbb{E}\left[ \frac{4y_t^2}{\mu_t}-8 y_t + 4 \mu_t \right]\notag \\
    &= \sup_{t}\mathbb{E}\left[4(\mu_t+1)-8 \mu_t + 4\mu_t\right]\notag \\
    &= 4,\notag
\end{align}
where in the first line we used that $\ell(y_t|\theta_t) = y_t \log(\mu_t) - \mu_t - \log(y_t!)= y_t \log(\theta_t^{2}) - \theta_t^{2} - \log(y_t!)$, using $\mu_t = \theta_t^2$, hence $\dd\ell(y_t|\theta_t)/\dd\theta_t= 2 y_t \theta_t/\theta_t^{2} - 2 \theta_t = 2y_t/\theta_t - 2 \theta_t$. In the fourth line we used that the pseudo-true state is identified as $\theta_t^{\star}=\sqrt{\mu_t}=\exp(\vartheta_t/2) \in (0, \infty)$ for all $t$, and thus $\theta_t^{\star 2} = \exp(\vartheta_t) = \mu_t$. In the fifth line, we used that $\mathbb{E}[y_t^2/ \mu_t] = \mathbb{E}[\underset{y_t}{\mathbb{E}}[y_t^2]/\mu_t]=\mathbb{E}[(\mu_t^2 + \mu_t)/\mu_t]=\mathbb{E}[\mu_t+1]$ and $\mathbb{E}[y_t] = \mathbb{E}[\underset{y_t}{\mathbb{E}}[y_t]]=\mathbb{E}[\mu_t]$, both using the tower property. The MSE bound of the ISD filter with quadratic link function additionally depends on $q^2=\sup_{t} \mathbb{E}[\|\theta_t^{\star}-\theta_{t-1}^{\star}\|^2] <\infty$ and $s_{\omega}^2=\sup_{t} \mathbb{E} [\|\theta_t^{\star}-\omega\|^2]<\infty$, which are assumed to be given. Here, we compute these quantities analytically assuming $\omega_0$, $|\phi_0|<1$ and $\sigma_{\xi}$ are known.
\begin{align}
    q^2 &= \sup_{t}\mathbb{E} \left[\left(\theta_t^{\star}-\theta_{t-1}^{\star}\right)^2\right]\notag\\
    &= \sup_{t}\mathbb{E} \left[\left( \sqrt{\mu_t} - \sqrt{\mu_{t-1}} \right)^2\right]\notag\\
    &= \sup_{t}\mathbb{E} \left[ {\mu_t} +{\mu_{t-1}} - 2 \sqrt{\exp(\vartheta_t+\vartheta_{t-1})}\right]\notag\\
    &=  \sup_{t}\left(2\exp \left(\omega_{0} + \frac{\sigma_{\xi}^2}{2(1-\phi_{0}^{2})}\right) - 2 \mathbb{E} \left[\sqrt{\exp(\vartheta_t+\vartheta_{t-1})}\right]\right) 
    \notag\\
    &= 2\exp \left(\omega_{0} + \frac{\sigma_{\xi}^2}{2(1-\phi_{0}^{2})}\right)- 2 \exp\left(\omega_{0} + \frac{(1+\phi_{0})\sigma_{\xi}^2}{4(1-\phi_{0}^{2})}\right)\notag\\
    &= 2\exp \left(\omega_{0} + \frac{\sigma_{\xi}^2}{2(1-\phi_{0}^{2})}\right)- 2 \exp\left(\omega_{0} +\frac{\sigma_{\xi}^2}{4(1 - \phi_{0})} \right).\notag \\
    s_{\omega}^2&=\sup_{t} \mathbb{E} \left[\|\theta_t^{\star}-\omega\|^2\right] \notag\\
    &= \sup_{t}\mathbb{E} \left[ \mu_t + \omega^{2} -2 \omega \sqrt{\mu_t}\right]\notag\\
    &=  \sup_{t}\mathbb{E} \left[\mu_{t} -2 \omega \sqrt{\mu_{t}} \right] + \omega^{2}
    \notag\\
    &=  \exp\left(\omega_0 + \frac{\sigma_{\xi}^2}{2(1-\phi_0^{2})}\right) + \omega^{2} -2 \omega\exp \left(\frac{\omega_0}{2} + \frac{\sigma_{\xi}^2}{8(1-\phi_0^{2})}\right).\notag
\end{align}
As $\{\vartheta_t\}$ follows stationary AR(1) dynamics $\vartheta_t = \omega_0 (1 - \phi_0) + \phi_0 \vartheta_{t-1} + \xi_t$, where $\xi_t \sim \mathrm{N}(0, \, \sigma_\xi^2)$, the unconditional distribution of $\vartheta_t$ is $\mathrm{N}(\omega_0 , \, \sigma_{\xi}^2/(1 - \phi_0^2))$. Moreover, $X$ being normally distributed implies that $\mathbb{E}[\exp(X)] = \exp(\mathbb{E}(X) + \textnormal{var}(X)/2)$, where $\textnormal{var}(X)$ denotes the variance of $X$. Taken together, this yields the following three identities:
\begin{align}
    \mathbb{E}[\exp(\vartheta_t)] &= \exp\left(\omega_0 + \frac{\sigma_\xi^2}{2(1 - \phi_0^2)}\right)\label{eq:lognormal moment},\\
    \mathbb{E}[\sqrt{\exp(\vartheta_t)}] &= \exp\left(\frac{\omega_0}{2} + \frac{\sigma_\xi^2}{8(1 - \phi_0^2)}\right),\notag \\
    \mathbb{E}[\sqrt{\exp(\vartheta_t + \vartheta_{t-1})}] &= \exp\left(\omega_0 + \frac{(1+\phi_0)\sigma_\xi^2}{4(1 - \phi_0^2)}\right)\label{eq:square root sum exponents}.
\end{align}
To derive equation~\eqref{eq:square root sum exponents}, we write $\sqrt{\exp(\vartheta_t + \vartheta_{t-1})} = \exp\left(S\right)$, where $S := (\vartheta_t + \vartheta_{t-1})/2$ is (unconditionally) Gaussian. By (weak) stationarity, its mean is
\begin{equation*}
\mathbb{E}[S] = \frac{1}{2}(\mathbb{E}[\vartheta_t] + \mathbb{E}[\vartheta_{t-1}]) = \omega_0,
\end{equation*}
while its variance is
\begin{equation*}
\textnormal{var}(S) = \frac{1}{4} \, \textnormal{var}(\vartheta_t + \vartheta_{t-1}) = \frac{1}{4} \left(2\,\textnormal{var}(\vartheta_t) + 2\,\text{Cov}(\vartheta_t, \vartheta_{t-1})\right),
\end{equation*}
where $\text{Cov}(\vartheta_t, \vartheta_{t-1}) = \phi_0\,\textnormal{var}(\vartheta_t) = {\phi_0 \sigma_\xi^2}/({1 - \phi_0^2})$. Hence,
\begin{equation*}
\textnormal{var}(S) = \frac{1}{4} \left(2 + 2\phi_0\right) \frac{\sigma_\xi^2}{1 - \phi_0^2} = \frac{(1+\phi_0)\sigma_\xi^2}{2(1 - \phi_0^2)}.
\end{equation*}
Using $\mathbb{E}[\exp(S)] = \exp(\mathbb{E}(S) + \textnormal{var}(S)/2)$ yields equation~\eqref{eq:square root sum exponents}.

\medskip

The gradient noise $\sigma^2$ in Assumption~\ref{Assumption: Bounded moments}(b) for the ISD filter with an exponential link function follows from $\nabla \ell(y_t \mid \theta_t^\star) = y_t - \exp(\theta_t^\star) = y_t - \mu_t$, which uses the fact that $\theta_t^\star = \vartheta_t$. Since $y_t \mid \mu_t \sim \mathrm{Poisson}(\mu_t)$,
and noting that $\mathbb{E}[(y_t - \mu_t)^2]$ is the centralized second moment, i.e., the unconditional variance of $y_t$,
we have
$$
\sigma^2 = \sup_t \, \mathbb{E}[(y_t - \mu_t)^2] = \sup_t \, \mathbb{E}[\mu_t] = \sup_t \,\mathbb{E}[\exp(\vartheta_t)] = \exp\left(\omega_0 + \frac{\sigma_\xi^2}{2(1 - \phi_0^2)}\right),
$$
where the last equality follows from equation~\eqref{eq:lognormal moment}.

\begin{figure}
    \centering
    \begin{subfigure}{0.5\textwidth}
        \centering
        \includegraphics[width=\textwidth]{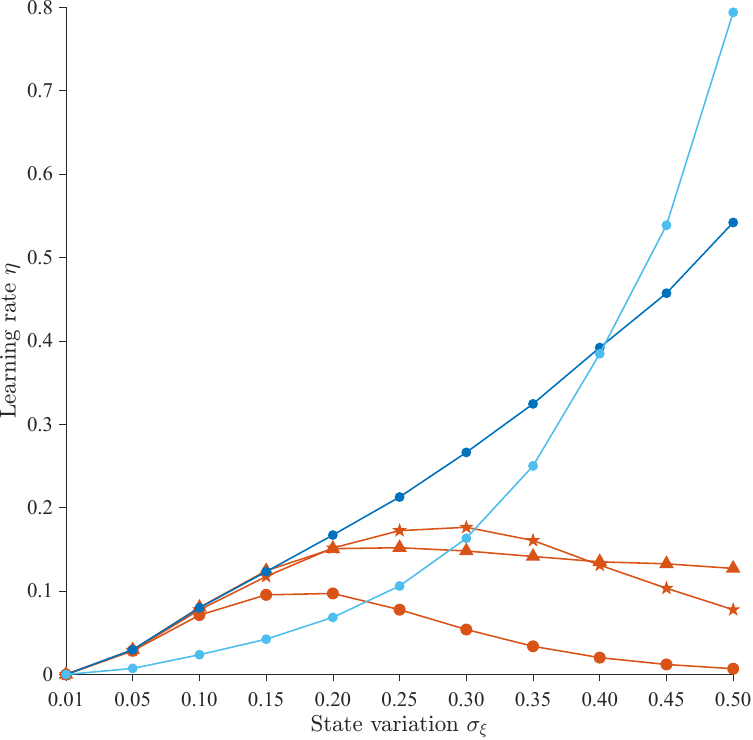}
        \label{fig:Learning_rates}
        \vspace{-4mm}
    \end{subfigure}

    \begin{subfigure}{\textwidth}
        \centering
        \includegraphics[width=\textwidth]{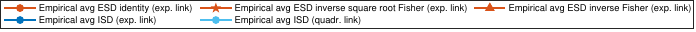}
        \label{fig:legend_Poisson_learning_ratesv1}
    \end{subfigure}
    \vspace{-12mm}
    \caption{Plot of the maximum likelihood-estimated learning rates for the ESD filter with identity scaling ($\zeta = 0$), inverse square root Fisher scaling ($\zeta = 1/2$), and inverse Fisher scaling ($\zeta = 1$), and the ISD filter, all using exponential link functions, along with the ISD filter using a quadratic link function.}
    \label{fig:Learning rates.}
\end{figure}

\noindent \textbf{Figure}~\ref{fig:Learning rates.} plots the maximum likelihood-estimated learning rates of the ISD and ESD filters. The learning rates of the ISD filters both increase monotonically with the state variation, which is intuitive, as more sensitivity of the filter is needed to track more volatile states. The learning rates of the ESD filters, on the other hand, peak before declining, likely as an attempt to prevent divergence of the filter when the true states are volatile.

\putbib[paper-ref]
\end{bibunit}

\end{document}